\documentclass[%
superscriptaddress,
pof,
 amsmath,
 amssymb,
preprint,%
 longbibliography,  
]{revtex4-2}

\usepackage{graphicx}
\usepackage{dcolumn}
\usepackage{bm}
\usepackage[mathlines]{lineno}

\usepackage[utf8]{inputenc}
\usepackage[T1]{fontenc}
\usepackage{mathptmx}
\usepackage{etoolbox}

\usepackage{enumerate}  
\usepackage{enumitem}   
\usepackage{tikz}       
\usepackage[linesnumbered,ruled,vlined]{algorithm2e} 
\usepackage{hyperref}   
\setlist{nosep}         
\usepackage{makecell}   
\usepackage{booktabs} 
\usepackage{multirow} 

\makeatletter
\def\@email#1#2{%
 \endgroup
 \patchcmd{\titleblock@produce}
  {\frontmatter@RRAPformat}
  {\frontmatter@RRAPformat{\produce@RRAP{*#1\href{mailto:#2}{#2}}}\frontmatter@RRAPformat}
  {}{}
}%
\makeatother
\begin{document}

\preprint{AIP/123-QED}

\title[Sample title]{Temperature transformation recovering the compressible law of the wall for turbulent channel flow}

\author{Youjie Xu}
\thanks{youjie.xu@tum.de}
\affiliation{Chair of Aerodynamics and Fluid Mechanics, TUM School of Engineering and Design, Technical University of Munich, Boltzmannstraße 15, 85748 Garching, Germany}

\author{Steffen J. Schmidt}
\affiliation{Chair of Aerodynamics and Fluid Mechanics, TUM School of Engineering and Design, Technical University of Munich, Boltzmannstraße 15, 85748 Garching, Germany}

\author{Nikolaus A. Adams}
\affiliation{Chair of Aerodynamics and Fluid Mechanics, TUM School of Engineering and Design, Technical University of Munich, Boltzmannstraße 15, 85748 Garching, Germany}
\affiliation{Munich Institute of Integrated Materials, Energy and Process Engineering, Technical University of Munich, Lichtenbergstraße 4a, 85748 Garching, Germany}

\date{\today}

\begin{abstract}
  Velocity and temperature distributions are both crucial for modeling compressible wall-bounded turbulent flows. The compressible law of the wall for velocity has been extensively examined through velocity transformations. However, a well-established temperature transformation remains an open issue. We propose new Van Driest type (VD-type) and semi-local type (SL-type) temperature transformation for compressible turbulent channel flow. Our approach is based on an analysis of the momentum and energy balance equations in the overlap layer. It accounts for the influences of mixing length model, the work of the body force, and the turbulent kinetic energy (TKE) flux. The proposed transformations are evaluated using data from direct numerical simulations and wall-resolved large eddy simulations of compressible turbulent channel flow. The SL-type transformation provides better data collapse than the VD-type in the viscous sublayer and buffer layer. With a suitable mixing length model, the SL-type transformed temperature agrees well with the incompressible temperature profile or the extended law of the wall. For the isothermal wall, the integral mean error over the entire boundary layer remains below \(2\%\) for most cases, with root mean square value of about \(1.7\%\). The results highlight the importance of mitigating the energy imbalance in the transformation. This work identifies the multi-layer structure of the turbulent TKE flux, which in turn enables approximate models and corresponding simplified yet effective temperature transformations. Applications of the proposed approach in near-wall modeling and inverse transformation, as well as its potential extension to more general configurations, are also discussed.
\end{abstract}

\maketitle
\section{\label{section:introduction}Introduction}
Wall-bounded turbulent flow plays a crucial role in various applications, including aircraft aerodynamics \citep{Slotnick2014,Goc2021}, atmospheric flows \citep{Stoll2020}, wind farm optimization \citep{Sørensen2011}, etc. It is well known that, in high-Reynolds-number incompressible turbulent boundary layers, the mean streamwise velocity follows the law of the wall (LoW), typically expressed as \citep{Pope2000turbulent}:
\begin{equation}\label{eq:LoW_Uplus}
  U^+ =
  \begin{cases}
  y^+, & \text{viscous sublayer}, \\[1.2ex]
  \dfrac{1}{\kappa} \ln(y^+) + B, & \text{logarithmic layer.}
  \end{cases}
\end{equation}
Here, \( U^+ = \bar{u}/u_\tau \), \( u_\tau = \sqrt{\tau_w/\bar\rho_w} \), and \( y^+ = {\bar \rho_w u_\tau y}/{\bar \mu_w} \), where \( \tau_w \), \( \bar\rho_w \), and \( \bar\mu_w \) are the mean shear stress, density, and dynamic viscosity at the wall, respectively. \( \kappa \) represents the von Kármán constant, and \( B \) is an integration constant. 

Reynolds analogy suggests that heat and momentum transfer exhibit similar behavior when the Prandtl number approaches unity. The underlying logic is that fluid motions transport the momentum and heat flux simultaneously \citep{tennekes_lumley_1972}. This implies the existence of the LoW for temperature above a non-adiabatic wall, expressed as \citep{Bradshaw1995, Huang2023}:
\begin{equation}\label{eq:LoW_Tplus}
  T^+ =
  \begin{cases}
  Pr_w \, y^+, & \text{viscous sublayer}, \\[1.2ex]
  \dfrac{Pr_t}{\kappa} \ln(y^+) + B_T, & \text{logarithmic layer.}
  \end{cases}
\end{equation}

Here, \( T^+ = (\bar T - \bar T_w)/T_\tau \), and \( T_\tau = q_w/(\bar\rho_w c_p u_\tau) \). The terms \( \bar T_w \) and \( q_w \) represent the wall temperature and heat flux removed from the wall. \( \bar \rho_w \) and \( u_\tau \) retain their definitions from Eq.~\eqref{eq:LoW_Uplus}. \(Pr_w\) is the molecular Prandtl number at the wall, and \( Pr_t \) is the turbulent Prandtl number. \( B_T \) is the counterpart of \(B\) in Eq.~\eqref{eq:LoW_Uplus} and typically depends on \(Pr\).

Eqs.\eqref{eq:LoW_Uplus} and \eqref{eq:LoW_Tplus} are applicable to incompressible flows. Many existing studies support the LoW for velocity distribution \citep{Lee2015, Hoyas2022, Liakopoulos2024} and temperature distribution \citep{Kader1981, Carvin1988, Lee2014, Pirozzoli2016a, Alcantara-Avila2021}. However, their accuracy decrease with increasing Mach number due to aerodynamic heating effects and the coupling between velocity and temperature fields. According to \citet{Morkovin1962}, the difference between compressible and incompressible turbulence for moderate Mach numbers can be accounted for by taking into consideration of variations in fluid properties. Based on this idea, numerous velocity transformations have been established to transform the compressible turbulent velocity profile into its incompressible counterpart \citep{VanDriest1951, Zhang2012, Brun2008, Trettel2016, Patel2016,Griffin2021,Volpiani2020,Lee2023, Younes2023, Zhu2024}, thereby validating the compressible LoW for velocity.

In order to describe the temperature profile in compressible boundary layers, two primary strategies have been explored in the literature. The first one involves developing a temperature-velocity relation (TV-relation), where the mean temperature is expressed as a function of the mean velocity. The second strategy focuses on establishing a temperature transformation, analogous to the idea of velocity transformation.

Examples of TV-relation include the formulations from the last century \citep{Crocco1932, Busemann1931, Walz1962} and recent developments \citep{Duan2011a,Zhang2014, Cheng2024a, Zhu2025a}. These relations are built upon the Strong Reynolds Analogy \citep{Morkovin1962} and exhibit good performance across a wide range of flows \citep{Modesti2016, Duan2010}. However, most of these TV-relations require the velocity and temperature at the boundary layer edge as input \citep{Modesti2016, Cheng2024b, Chen2025}. For internal flows, such as compressible turbulent channel and pipe flows, the centerline velocity and temperature are not known \emph{a priori}. Regarding this issue, \citet{Song2022} proposed an approach to determine the centerline temperature for turbulent channel and pipe flows. However, this method is limited to the classical isothermal wall configuration. For more complex configurations, such as compressible turbulent channel flows with mixed isothermal/adiabatic wall conditions \citep{Lusher2022}, the commonly used TV-relations encounter several challenges. First, it is often difficult to determine the boundary layer edge \citep{Cheng2024b}, as the mean flow field is no longer symmetric about the centerline. In \citet{Lusher2022} and \citet{Huang2023}, the boundary layer edge is defined as the location where the mean velocity approaches its maximum value. But it remains unclear whether this position can be regarded as the thermal boundary layer edge. Second, neither the centerline temperature nor the temperature at the position of maximum velocity can be predicted by the approach of \citet{Song2022}. More importantly, even when these temperatures are available, applying them in the commonly used TV-relations still leads to noticeable discrepancies, which can be validated using the DNS data of \citet{Lusher2022}. Regarding this issue, the integral TV-relation proposed more recently by \citet{Zhu2025a} provides a more accurate solution.



In contrast to the TV-relation, the temperature transformation is formulated through an incremental wall-normal integration without relying on velocity and temperature values at the boundary layer edge. The resulting formulation is consistent with the classical LoW \citep{Coles1956}. However, such a temperature scaling law has not been well established. Preliminary results for this strategy have been reported in several studies \citep{Huang1994, Brun2008, Patel2017, Chen2022, Huang2023, Cheng2024b, Modesti2024}.

A straightforward approach to construct a temperature transformation for compressible turbulent flows is to follow the philosophy of VD-type velocity transformation \citep{VanDriest1951}. For example, \citet{Brun2008} derived a VD-type transformation of the total temperature expressed as \( \tilde T^+_{i,VD} = \int_0^{\tilde T^+_i} \sqrt{\bar{\rho}/\bar{\rho}_w} \, d{\tilde T^+_i} = \frac{Pr_t} {\kappa} \text{log} \ (y^+) + B_T\). However, it does not collapse the buffer layer very well. To address this, they further proposed an integral length scaling and a corresponding integral temperature transformation, which account for variation in both density and dynamic viscosity. This transformation reduces the scatter of the intercept significantly, leading to better agreement with experimental values.

\citet{Patel2017} investigated flows over non-adiabatic walls under low-Mach-number conditions and proposed an extended VD-type temperature transformation. Their results demonstrate a good collapse of the transformed temperature profile. More recently, \citet{Modesti2024} developed a temperature transformation that accounts for variable fluid properties under low-Mach-number conditions. However, since aerodynamic heating was not considered in both studies, the performance of these transformations in compressible flows cannot be guaranteed. Furthermore, they encounter a singularity issue when applied to adiabatic wall boundary conditions. To address this problem, \citet{Chen2022} proposed using local heat flux instead of wall heat flux, which leads to unified temperature transformations applicable to both isothermal and adiabatic walls. It is worth noting that high-order statistics, especially the turbulent TKE flux, is retained in the local heat flux, which improves the performance of their transformations.

\citet{Huang2023} proposed VD-type and SL-type temperature transformations, which apply to both isothermal and adiabatic wall conditions, with the SL-type transformation showing superior performance. More recently, \citet{Cheng2024b} proposed three Mach number invariant functions and a new SL-type transformation that demonstrates good performance above adiabatic wall in turbulent channel flow and isothermal wall in turbulent boundary layer flow.

The studies of \citet{Chen2022}, \citet{Huang2023}, and \citet{Cheng2024b} demonstrate the possibility of recovering the temperature LoW in compressible turbulent flows. The most important lesson of their studies is that variations in fluid properties and aerodynamic heating effects should be taken into consideration. Incorporating the effect of high-order statistics can also enhance the performance of the transformation. However, it should be noted that there is still room for improvement in previous transformations. For instance, the slope of the logarithmic profile remains unsatisfactory in the results of \citet{Chen2022}. At relatively low Reynolds numbers, the logarithmic profiles under the transformation by \citet{Huang2023} are less pronounced. In compressible turbulent channel flow, the transformation by \citet{Cheng2024b} is effective for adiabatic wall, but its performance for isothermal walls is less satisfactory.

Considering the Reynolds analogy, the log-law for temperature and velocity distribution would share the same fundamental arguments. The log-law for velocity is supported by the arguments of Prandtl and Millikan \citep{Pirozzoli2014}. Prandtl's reasoning relies on the assumptions of linear variation of mixing length (\( l_m = \kappa y \)) and uniform (constant) shear stress in the near wall region. Millikan's argument is based on asymptotic matching of the LoW in the inner layer and the velocity-defect law in the outer layer \citep{Coles1956,Pope2000turbulent,wilcox2006turbulence,Pirozzoli2014,Luchini2017}. 

When focusing on compressible turbulent channel flow, there are three aspects that could be improved. First, it has been shown that the mixing length model \( l_m = \kappa y \) is inaccurate, and the parabolic form \( l_m = \kappa y \sqrt{1 - {y}/{h}} \) is a more suitable choice for turbulent channel flow \citep{Pirozzoli2014}. Here, \(h\) is the half-channel height. Second, the assumption of constant shear stress in the logarithmic region is also problematic. The driving force (external body force or pressure gradient) and its work on the fluid should be considered in the energy equation if we expect to obtain satisfactory results in the outer layer. Third, the turbulent TKE flux, typically neglected in the transformations proposed by \citet{Huang2023}, could also be included, as demonstrated in the transformation by \citet{Chen2022} and \citet{Cheng2024b}.

In light of these considerations, we propose new VD-type and SL-type temperature transformations that account for the effects of mixing length model, body force, and the turbulent TKE flux. These transformations are applicable to both isothermal and adiabatic wall boundary conditions in compressible turbulent channel flow. In addition, mathematical models for the turbulent TKE flux and corresponding simplified transformations are provided.

The manuscript is organized as follows: Sec.~\ref{sec:compressible_law_of_the_wall} gives the detailed derivation of the temperature transformation. Sec.~\ref{sec:performance_of_temperature_transformation} evaluates  the performance of the proposed transformation. Sec.~\ref{sec:simplified_temperature_transformations} provides insights into the effect of the introduced parameters. Simplified transformations and their performance are presented. In Sec.~\ref{sec:application_and_limitation}, applications of the transformation to near-wall modeling and to the inverse transformation, as well as its potential extension to more general flow configurations, are discussed. Finally, concluding remarks are given in Sec.~\ref{sec:Conclusion}.

\section{Compressible law of the wall for temperature}
\label{sec:compressible_law_of_the_wall}
The derivation is based on compressible turbulent channel flow and can be applied in a similar manner to compressible turbulent pipe flow, as shown in Appendix~\ref{sec:app_A}. Our method follows \citet{Chen2022} and \citet{Huang2023}, but differs in three key differences. First, the transformation is defined based on the momentum and energy balance equations in the overlap layer, with an additional requirement imposed on the mixing length \(l_m \) to realize the linear law within the viscous sublayer. Second, we do not neglect the body force and its work on the fluid. Third, we account for the turbulent TKE flux in the energy balance equation. Considering these differences, we provide a complete derivation in this section for clarity and readability.

Throughout this study, \( x \), \( y \), and \( z \) denote the streamwise, wall-normal, and spanwise directions, with corresponding velocity components denoted by \( u \), \( v \) and \( w \). For generalization, \( u_i(i=1,2,3) \) represents the velocity components. Reynolds averaging is expressed as \( \phi = \bar\phi + \phi^\prime \), whereas Favre averaging is given by \( \phi = \tilde \phi + \phi^{\prime\prime} \), where \(\tilde\phi = \overline{\rho \phi}/\bar \rho \). Quantities at the wall are denoted by the subscript \( w \), while the superscripts \( + \) and \( * \) represent wall scaling and semi-local scaling. The subscripts VD and SL denote Van Driest type and semi-local type transformations.

\subsection{Governing equations}\label{sec:governing_equation}
For simplicity, we focus on compressible turbulent channel flow with periodic boundary conditions in streamwise and spanwise directions, and no-slip condition at the two walls. The governing equations for mass, momentum and energy conservations are:
\begin{equation}\label{eq:mass_conservation}
  \frac{\partial \rho}{\partial t} + \frac{\partial {\rho u_j}}{\partial x_j} = 0,
\end{equation}
\begin{equation}\label{eq:momentum_conservation}
  \frac{\partial \rho u_i}{\partial t} 
  + \frac{\partial {\rho u_j u_i}}{\partial x_j} 
  = -\frac{\partial p}{\partial x_i} + \frac{\partial \tau_{ij}}{\partial x_j}
  + f_x \delta_{i1},
\end{equation}
\begin{equation}\label{eq:energy_conservation}
  \frac{\partial }{\partial t} \bigg [\rho \left(c_v T + \frac{u_i u_i}{2}\right) \bigg]
  + \frac{\partial }{\partial x_j} \bigg [\left(\rho c_v T + \frac{\rho u_i u_i}{2} + p \right) u_j \bigg]
  = \frac{\partial \tau_{ij} u_i}{\partial x_j}
  - \frac{\partial q_i}{\partial x_i}
  + f_x u_1,
\end{equation}
with the viscous stress \(\tau_{ij}\) and heat flux vector \(q_i\) given by:
\begin{equation}\label{eq:tau_ij}
\tau_{ij} = \mu \left(\frac{\partial u_i}{\partial x_j} 
+ \frac{\partial u_j}{\partial x_i} 
- \frac{2}{3} \delta_{ij} \frac{\partial u_k}{\partial x_k}\right), \ \ q_i = - \lambda \frac{\partial T}{\partial x_i}.
\end{equation}

Here, \(\rho, p, T\) represent density, pressure, temperature, respectively. \(\mu\) and \(\lambda\) are the dynamic viscosity and molecular thermal conductivity. \( f_x \) is the external body force in streamwise direction. \(\delta_{ij}\) denotes Kronecker delta notation. The ideal gas material model \( p = \rho R T \) is used to close the governing equations, where \( R \) is gas constant. The specific heat capacities at constant volume and constant pressure are given by \( c_v = R/(\gamma - 1)\) and \( c_p = \gamma R/(\gamma - 1)\), respectively, where the specific heat ratio is \(\gamma = 1.4\).

As pointed out by \citet{Huang1995}, the flow is driven by an external streamwise body force \( f_x \) in order to avoid non-zero streamwise gradients of mean density and pressure. Hence \( f_x \) acts as an "effective pressure gradient" to maintain a prescribed mass flow rate \( m = \int_0^{2h}{\bar\rho \tilde u}dy/2h \), which is in practice more relevant to explain the physics of a fully developed flow. The mean gradient of the actual, thermal dynamic pressure \(d\bar p/dx\) is zero \citep{Lusher2022}. Two types of body force are frequently implemented: volume-based, with \( -\left( \frac {\partial p} {\partial x}\right)_{eff} = f_x \), and density-based, with \( -\left( \frac {\partial p} {\partial x}\right)_{eff} = \rho f_x \). 
To derive the temperature transformation, \( f_x \) is chosen to be volume-based in this section. Variations for different configurations are introduced in Sec.~\ref{sec:performance_on_DNS_GV2024} and \ref{sec:performance_above_isoWall_of_LC2022}.

In the temperature transformation, the primary goal is to establish the relationship between the mean temperature gradient and the heat flux. To achieve this, a momentum and energy balance analysis is required.

\subsection{Momentum and energy balance}\label{sec:momentum_balance}
\begin{figure*}
  \includegraphics[width=\linewidth]{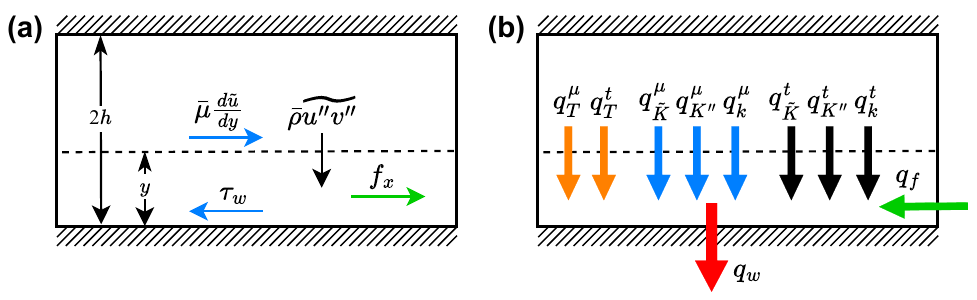}
  \caption{Momentum (a) and energy (b) balance in a statistically steady turbulent channel flow. \(h\) is the channel half-height. The black dashed line indicates a reference \( y \)-plane in the lower half-channel (\(y/h\) = 0 to 1). Th expressions for each heat flux component are provided in Eqs.~\eqref{eq:q_mu_T} to \eqref{eq:q_f}.}
\label{fig:momentum_energy_balance}
\end{figure*}

Consider the control volume from the wall to a reference \( y \)-plane, for statistically steady flows, time-derivative and convective terms in wall-parallel directions vanish, leading to the momentum and energy balance, as shown in Fig.~\ref{fig:momentum_energy_balance}. 

Integrating Eqs.~\eqref{eq:momentum_conservation} and \eqref{eq:energy_conservation} from the wall to a reference \( y \)-plane, we obtain the momentum and energy balance equations:
\begin{equation}\label{eq:global_momentumn_balance}
  \bar\mu \frac{d\tilde u }{dy} - 
  \bar\rho \widetilde{u^{\prime\prime} v^{\prime\prime}} + 
  \int_0^{y} {f_x(\eta)} d{\eta} = 
  \tau_w.
\end{equation}

\begin{equation}\label{eq:global_energy_balance}
  \overline{\lambda \frac{dT}{dy}} 
  - \overline{\rho c_p v^{\prime\prime} T^{\prime\prime}} 
  + \overline{\tau_{2i} u_i} 
  - \overline{\rho v^{\prime\prime} \frac{1}{2} u_i u_i} 
  + {\int_0^y {f_x\tilde u} d{y}} 
  = q_w.
\end{equation}
Here, \(\bar{\mu} d\tilde{u}/dy\) is the mean viscous stress, \(-\bar\rho \widetilde{u^{\prime\prime} v^{\prime\prime}}\) represents the mean turbulent momentum flux, \( \int_0^y f_x(\eta) \, d\eta \) corresponds to the total streamwise body force, and \(\tau_w\) is the mean wall shear stress. In Eq.~\eqref{eq:global_energy_balance}, \(q_w\) is the mean heat flux removed from the wall. The energy fluxes across the \( y \)-plane can be categorized into two primary mechanisms: molecular diffusion and turbulent diffusion. Each of these mechanisms involves the diffusion of both thermal energy and kinetic energy.

Before proceeding, we first decompose the instantaneous kinetic energy into three components, analogous to the approach of \citet{Huang1995, Huang2023} and \citet{Chen2022}:
\begin{equation}\label{eq:K_decomposition}
  K 
  = \frac{1}{2}{u_i u_i} 
  = \frac{1}{2} \tilde u_i \tilde u_i
    + {\tilde u_i u^{\prime\prime}_i} 
    + \frac{1}{2} u^{\prime\prime}_i u^{\prime\prime}_i.
\end{equation}

Since the mean flow in spanwise and wall-normal directions is negligible, Eq.~\eqref{eq:K_decomposition} simplifies to \(K = \tilde K + K^{\prime\prime} + k\), where \(\tilde K = \frac{1}{2}{\tilde u_i \tilde u_i} 
\approx \frac{1}{2}{\tilde u \tilde u}\), \(K^{\prime\prime} = {\tilde u_i u^{\prime\prime}_i} 
\approx {\tilde u u^{\prime\prime}}\), and \(k = \frac{1}{2}{u^{\prime\prime}_i u^{\prime\prime}_i}\).

In Eq.~\eqref{eq:global_energy_balance}, the first two terms represent the molecular and turbulent diffusion of thermal energy (or equivalently molecular and turbulent heat conduction).
\begin{equation}\label{eq:q_mu_T}
q^\mu_T = \overline{\lambda \frac{dT}{dy}} 
\approx \bar\lambda \frac{d\tilde T}{dy}, 
\end{equation}
\begin{equation}\label{eq:q_t_T}
q^t_T = -\overline{\rho c_p v^{\prime\prime} T^{\prime\prime}}. 
\end{equation}

The third term in Eq.~\eqref{eq:global_energy_balance} represents the molecular diffusion of kinetic energy, which can be split into three parts:
\begin{equation}\label{eq:q_mu_K}
  \overline{\tau_{2i} u_i} = \overline{\mu \frac{dK}{dy}} 
  = q^\mu_{\tilde K} + q^\mu_{K^{\prime\prime}} + q^\mu_k
\end{equation}
with
\begin{equation}\label{eq:q_mu_Ktilde}
q^\mu_{\tilde K} = \overline {\mu \frac{d{\tilde K}}{dy}} 
\approx \bar\mu \frac{d\tilde u}{dy} \tilde u,
\end{equation}
\begin{equation}\label{eq:q_mu_Kprime}
q^\mu_{K^{\prime\prime}}
= \overline{\mu \frac{d{K^{\prime\prime}}}{dy}}
= \overline{\mu^\prime \frac{ d{(\tilde u u^{\prime\prime})} }{dy}},
\end{equation}
\begin{equation}\label{eq:q_mu_k}
q^\mu_k
= \overline{\mu \frac{d{k}}{dy}}
\approx \bar\mu \frac{ d\overline{(u^{\prime\prime}_i u^{\prime\prime}_i/2)} }{dy}.
\end{equation}

The fourth term in Eq.~\eqref{eq:global_energy_balance} represents the turbulent diffusion of kinetic energy, which is also split into three parts:
\begin{equation}\label{eq:q_t_K}
  - \overline{\rho v \frac{1}{2}{u_i u_i}} = q^t_{\tilde K} + q^t_{K^{\prime\prime}} + q^t_k
\end{equation}
with
\begin{equation}\label{eq:q_t_Ktilde}
  q^t_{\tilde K} = -\overline{\rho v^{\prime\prime}\tilde K}\approx 0,
\end{equation}
\begin{equation}\label{eq:q_t_Kprime}
q^t_{K^{\prime\prime}} = -\overline{\rho v^{\prime\prime}K^{\prime\prime}}
\approx -\overline{\rho v^{\prime\prime} u^{\prime\prime}} \tilde u,
\end{equation}
\begin{equation}\label{eq:q_t_k}
q^t_k 
= -\overline{\rho v^{\prime\prime}k}
= -\overline{\rho v^{\prime\prime} \frac{1}{2} u^{\prime\prime}_i u^{\prime\prime}_i}.
\end{equation}

The last term on the left hand side of Eq.~\eqref{eq:global_energy_balance} is associated with the work of the body force. Considering the overall balance between the body force and wall shear stress in the channel, and assuming a uniform body force per unit volume, we have \(f_x = \tau_w/h\). Hence, 
\begin{equation}\label{eq:q_f}
q_f = 
{\int_0^y {f_x \, \tilde u(\eta)} d{\eta}} = 
\tau_w \tilde u^i_b \frac {y}{h},
\end{equation}
\begin{equation}\label{eq:ub_i}
\tilde u^i_b = \frac{1}{y}\int_0^y {\tilde u(\eta)}d\eta,  \ \ \  y \in (0, h].
\end{equation}

Here, we introduce the integral bulk velocity, \( \tilde u^i_b \), which is connected to the mean velocity profile. Note that the expression of \( \tilde u^i_b \) should be modified correspondingly if the body force implemented in the solver is density-based, as is the case in Sec.~\ref{sec:performance_on_DNS_GV2024}. 

On the right hand side of Eq.~\eqref{eq:global_energy_balance}, \(q_w\) represents the heat flux removed from the channel. Invoking \(f_x\) in Eq.~\eqref{eq:global_momentumn_balance} and substituting Eqs.~\eqref{eq:q_mu_T} to \eqref{eq:q_f} into Eq.~\eqref{eq:global_energy_balance}, we obtain the momentum and energy balance equations:
\begin{equation}\label{eq:linear_shear_stress}
  \bar\mu \frac{d{\tilde u}}{dy} - 
  \bar\rho \widetilde{u^{\prime\prime} v^{\prime\prime}} = \tau_w \left(1 - \frac{y}{h} \right), 
\end{equation}

\begin{equation}\label{eq:global_energy_balance_2}
  \underbrace{\bar\lambda \frac{d\tilde T}{dy}}_{q^\mu_T}
  \underbrace{-\overline{\rho c_p v^{\prime\prime} T^{\prime\prime}}}_{q^t_T}
  + \underbrace{\bar\mu \frac{d\tilde u}{dy} \tilde u}_{q^\mu_{\tilde K}}
  + \underbrace{\overline{\mu^\prime \frac{d{(\tilde u u^{\prime\prime})}}{dy}}}_{q^\mu_{K^{\prime\prime}}}
  + \underbrace{\bar\mu \frac{ d\overline{(u^{\prime\prime}_i u^{\prime\prime}_i/2)} }{dy}}_{q^\mu_k}
  \underbrace{-\overline{\rho v^{\prime\prime} u^{\prime\prime}}\tilde u}_{q^t_{K^{\prime\prime}}}
  \underbrace{-\overline{\rho v^{\prime\prime} \frac{1}{2} u^{\prime\prime}_i u^{\prime\prime}_i}}_{q^t_k}
  + \underbrace{\tau_w \tilde u^i_b \frac {y}{h}}_{q_f}
  = q_w.
\end{equation}

\subsection{Temperature transformation}\label{sec:temperature_transformation}
Regarding the transformation to account for compressibility, two approaches have been commonly applied in previous studies: wall scaling \citep{VanDriest1951} and semi-local scaling \citep{Huang1995, Trettel2016, Patel2016}, typically referred to as VD-type and SL-type transformations, respectively. The VD-type transformation is motivated by overlap layer balance and neglects viscous effects, while the SL-type transformation considers both viscous and turbulent effects. In this study, we demonstrate that both VD-type and SL-type temperature transformations can be derived directly from the overlap layer, with an additional constraint on the mixing length to ensure the linear law in the viscous sublayer.

\begin{figure*}
  \includegraphics[width=\linewidth]{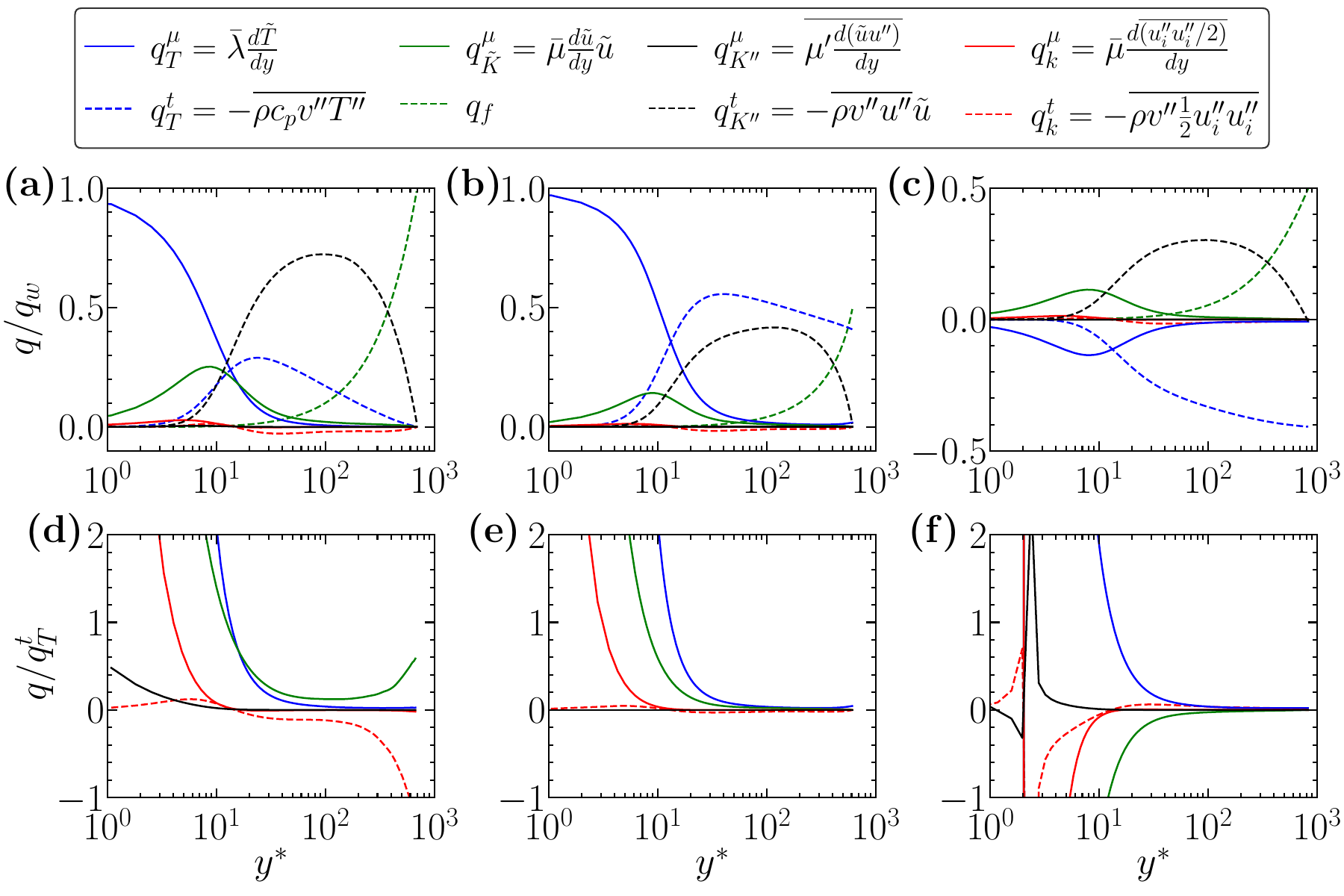}
  \caption{Energy budget in compressible turbulent channel flow. (a, d): isothermal wall for the classical isothermal setup (case JXF-M1.5Re17000 in Table \ref{table:WRLES_JAXFluids}); (b, e): isothermal wall side for the mixed isothermal/adiabatic configuration (case LC2022-iF2 in Table \ref{table:DNS_LC2022_isothermal_wall}); (c, f): adiabatic wall side for the mixed isothermal/adiabatic configuration (case LC2022-aF2 in Table \ref{table:DNS_LC2022_adiabatic_wall}). Here, \(M_b\) and \(Re_b\) represent the bulk Mach number and bulk Reynolds number, respectively, as defined in Sec.~\ref{sec:performance_of_temperature_transformation}. The heat flux in panel (c) is normalized using \(q_w\) from the corresponding isothermal wall side. Negative values indicate heat flux away from the wall (see Fig.~\ref{fig:momentum_energy_balance}).}
\label{fig:energy_budget}
\end{figure*}

In the study of \citet{Huang2023}, \(q^\mu_{K^{\prime\prime}}\) is not considered. Both \(q^\mu_k\) and \(q^t_k\) are neglected in the energy balance equations because their magnitudes are significantly smaller than \(q^\mu_{\tilde K}\) and \(q^t_{K^{\prime\prime}}\), respectively. \citet{Chen2022} and \citet{Cheng2024b} reported similar treatment in the near wall region. However, since we focus on the relationship between \( {d\tilde T}/{dy} \) and \( q^t_T \) in the overlap layer, the latter should be the basis for comparison when determining which terms in Eq.~\eqref{eq:global_energy_balance_2} can be neglected.

In Fig.~\ref{fig:energy_budget}, the magnitudes of each term in Eq.~\eqref{eq:global_energy_balance_2}, normalized by \(q_w\) and \(q^t_T\), are plotted for three types of wall-boundary conditions in compressible turbulent channel flow: (1) the classical setup with both walls isothermal, (2) the isothermal wall side in mixed isothermal/adiabatic configuration, and (3) the adiabatic wall side in the same mixed thermal configuration. As \(q_w = 0\) for adiabatic wall, the heat flux in panel (c) is normalized using \(q_w\) at the corresponding isothermal wall side.

Outside the viscous sublayer and buffer layer, the flux components \(q^\mu_T, q^\mu_{\tilde K}, q^\mu_{K^{\prime\prime}}, q^\mu_k\) and \(q^t_k\) are significantly smaller than \(q_w\) across all three types of thermal wall-boundary configurations. However, when compared to \(q^t_T\), both \(q^\mu_{\tilde K}\) and \(q^t_k\) exhibit comparable magnitudes in this region for the classical isothermal configuration, as seen in panel (d). Their roles become more and more significant when approaching the channel centerline. Therefore, \(q^\mu_{\tilde K}\) and \(q^t_k\) should be retained for this configuration.

In contrast, under the mixed thermal configuration, \(q^t_T\) is consistently directed from the adiabatic wall toward the cold wall, in agreement with the findings of \citet{Zhu2025b}. Its magnitude remains dominant throughout the outer layer, as shown in panels (e) and (f). Consequently, the smaller flux components \(q^\mu_T, q^\mu_{K^{\prime\prime}}, q^\mu_k\), and \(q^t_k\) can be reasonably neglected. In principle, \(q^\mu_{\tilde K}\) may also be neglected when compared to \(q^t_T\). However, this term can be further simplified in combination with Eq.~\eqref{eq:linear_shear_stress}, which improves the overall accuracy. For this reason, \citet{Huang2023} retained it in the energy balance equation, and we follow the same approach in this work.

Based on above observations, we retain \(q^\mu_{\tilde K}\) and \(q^t_k\) for all three types of thermal wall-boundary conditions, leading to the following simplified energy balance equation:
\begin{equation}\label{eq:log_layer_energy_balance}
  -\overline{\rho c_p v^{\prime\prime} T^{\prime\prime}}
  = q_w - \tau_{tot} \, \tilde u + \overline{\rho v^{\prime\prime} \frac{1}{2} u^{\prime\prime}_i u^{\prime\prime}_i} - \tau_w \tilde u^i_b \frac {y}{h}.
\end{equation}

Here, \(\tau_{tot} = \bar\mu {d\tilde u}/{dy}- \overline{\rho v^{\prime\prime} u^{\prime\prime}}\) represents the total shear stress. Invoking Prandtl's mixing length hypothesis, \(\nu_t = l_m^2 d \tilde u/dy\), along with the turbulent Prandtl number \(Pr_t\), we obtain:
\begin{equation}\label{eq:q_t_T_lm}
  -\overline{\rho c_p v^{\prime\prime} T^{\prime\prime}}
  = \frac{\bar\rho \nu_t c_p}{Pr_t}\frac{d\tilde T}{dy}.
\end{equation}

In the overlap layer, the viscous stress can be neglected. In other words, the total shear stress is approximately equal to the Reynolds stress. Using Boussinesq's assumption, we have: 
\begin{equation}\label{eq:lm_du_dy}
  l_m\frac {d \tilde u}{dy} = 
  \sqrt {\frac{\tau_{tot}} {\bar \rho}}.
\end{equation}

Note that the general form of \(l_m\) is applied in Eqs.~\eqref{eq:q_t_T_lm} and \eqref{eq:lm_du_dy}. The influence of mixing length model will be presented in subsequent sections. Following \citet{Huang2023}, we define the friction Mach number as \(M_\tau = {u_\tau}/{\sqrt{\gamma R \tilde T_w}}\), the non-dimensional heat flux as \(B_q = {-q_w}/{(\bar\rho_w c_p u_\tau \tilde T_w)}\), and the non-dimensional temperature difference as \(\theta^+ = (\tilde T_w - \tilde T)/{\tilde T_w}\). Substituting these definitions and Eqs.~\eqref{eq:linear_shear_stress}, \eqref{eq:q_t_T_lm}, \eqref{eq:lm_du_dy} into Eq.~\eqref{eq:log_layer_energy_balance} yields: 
\begin{equation}\label{eq:log_layer_energy_balance_3}
  \frac
  {l_m\sqrt{\tau^+_{tot}}} {Pr_t}
  \sqrt{\frac{\bar\rho} {\bar\rho_w}}
  \frac{d\theta^+}{dy} = B_q + \left(\tau^+_{tot} + \frac {\tilde u^i_b}{\tilde u} \frac{y}{h}\right) {(\gamma - 1)M_\tau^2 u^+} - \frac{\overline{\rho v^{\prime\prime} 
  {u^{\prime\prime}_i u^{\prime\prime}_i/2}}}{\bar\rho_w c_p u_\tau \tilde T_w},
\end{equation}

Here, \(u^+ = \tilde u/u_\tau\) and \(\tau^+_{tot} = \tau_{tot}/\tau_w\). Furthermore, we define the following three parameters:
\begin{equation}\label{eq:params_f1f2f3}
  \psi_1 = \frac{l_m\sqrt{\tau^+_{tot}}}{\kappa y}, \,\,\,
  \psi_2 = \tau^+_{tot} + \frac {\tilde u^i_b}{\tilde u} \frac{y}{h}, \,\,\,
  \psi_3 = \frac {-\overline{\rho v^{\prime\prime} u^{\prime\prime}_i u^{\prime\prime}_i/2}}{\bar\rho_w c_p u_\tau \tilde T_w}
\end{equation}

Substituting these definitions, we obtain a simplified equation:
\begin{equation}\label{eq:energy_f1f2f3}
\frac
{\psi_1}
{B_q + \psi_2 {(\gamma - 1)M_\tau^2 u^+} + \psi_3}
\sqrt{\rho^+}
{d\theta^+} = \frac{Pr_t}{\kappa}\frac{dy}{y}.
\end{equation}

Here, \(\rho^+ = \bar\rho/\bar\rho_w\). Note that for turbulent channel flow driven by a volume-based body force, \(\tau^+_{tot} = 1 - y/h\). When the driving force is density-based or a mixed thermal wall condition is applied, the expressions for \(\psi_1\) and \(\psi_2\) differ slightly and will be presented in Sec.~\ref{sec:performance_on_DNS_GV2024} and \ref{sec:performance_above_isoWall_of_LC2022}, respectively.

Based on Eq.~\eqref{eq:energy_f1f2f3}, the VD-type and SL-type transformations can be established by applying wall scaling and semi-local scaling, respectively.

\vspace{12pt}
\begin{itemize}
\item \,\, \textbf{VD-type temperature transformation}
\end{itemize}

Scaling the \( y \)-coordinate in Eq.~\eqref{eq:energy_f1f2f3} with wall quantities, we obtain:
\begin{equation}\label{eq:energy_VD}
\frac {\psi_1} {B_q + \psi_2 {(\gamma - 1)M_\tau^2 u^+} + \psi_3}
\sqrt{\rho^+}
{d\theta^+} = \frac{Pr_t}{\kappa}\frac{dy^+}{y^+}.
\end{equation}

Here, \(y^+ = \sqrt{\tau_w \bar\rho_w}y/{\bar\mu_w}\). Following the VD-type transformation \citep{VanDriest1951, Huang1994}, we define the VD-type temperature transformation as:
\begin{equation}\label{eq:VD_Tplus}
T^+_{VD} = \int_0^{\theta^+} 
\frac {\psi_1} {B_q + \psi_2 {(\gamma - 1)M_\tau^2 u^+} + \psi_3}
\sqrt{\rho^+}
{d\theta^+}.
\end{equation}

According to Eq.~\eqref{eq:energy_VD}, \(T^+_{VD}\) is expected to exhibit a logarithmic profile in the overlap layer:
\begin{equation}\label{eq:VD_Tplus_loglaw}
T^+_{VD} = \frac{Pr_t}{\kappa} \ \log (y^+) + B_{T,VD}.
\end{equation}

Here, \(B_{T,VD}\) is an integration constant, which is usually a function of Prandtl number. 

\vspace{12pt}
\begin{itemize}
\item \,\, \textbf{SL-type temperature transformation}
\end{itemize}

Scaling the \( y \)-coordinate in Eq.~\eqref{eq:energy_f1f2f3} with local quantities, we obtain:
\begin{equation}\label{eq:energy_SL}
\frac{\psi_1}{B_q + \psi_2 {(\gamma - 1)M_\tau^2 u^+} + \psi_3}
\sqrt{\rho^+}
\left(1 + \frac{1}{2}\frac{y^+}{\rho^+}\frac{d\rho^+}{dy^+} - \frac{y^+}{\mu^+}\frac{d\mu^+}{dy^+}\right)
{d\theta^+} = \frac{Pr_t}{\kappa}\frac{dy^*}{y^*}.
\end{equation}

Here \(\mu^+ = \bar\mu/\bar\mu_w\), \(y^* = \sqrt{\tau_w \bar\rho}y/{\bar\mu}\). The SL-type temperature transformation is defined as:
\begin{equation}\label{eq:SL_Tplus}
T^+_{SL} = \int_0^{\theta^+} 
\frac{\psi_1}{B_q + \psi_2 {(\gamma - 1)M_\tau^2 u^+} + \psi_3}
\sqrt{\rho^+}
\left(1 + \frac{1}{2}\frac{y^+}{\rho^+}\frac{d\rho^+}{dy^+} - \frac{y^+}{\mu^+}\frac{d\mu^+}{dy^+}\right)
{d\theta^+}.
\end{equation}

According to Eq.~\eqref{eq:energy_SL}, \(T^+_{SL}\) is expected to exhibit a logarithmic profile in the overlap layer:
\begin{equation}\label{eq:SL_Tplus_loglaw}
T^+_{SL} = \frac{Pr_t}{\kappa} \ \log (y^*) + B_{T,SL}.
\end{equation}

Here, \(B_{T,SL}\) is an integration constant, which is usually a function of Prandtl number.

In the present study, we adopt the von Kármán constant \(\kappa = 0.41\)  following \citet{Pope2000turbulent}, although recent studies have reported slightly different values \citep{Nagib2008, Lee2015, She2017, Liakopoulos2024}. For the turbulent Prandtl number, we use \( Pr_t = 0.85 \), which has been reported by \citet{Lusher2022} to be appropriate for both isothermal and adiabatic walls in the logarithmic layer. The same value is also reported in \citet{Coles1956}. Additionally, \citet{Huang2023} proposed the relation \(Pr_t = 1.05 - 0.2 \ \text{tanh}^3(-y^*/17)\), which also yields \(Pr_t \approx 0.85\) in the logarithmic region.

Regarding the three parameters given in Eqs.~\eqref{eq:VD_Tplus} and \eqref{eq:SL_Tplus}, \(\psi_1\) can be computed using a suitable \(l_m\), while \(\psi_2\) and \(\psi_3\) are obtained from the simulation results. Note that the influence of the body force is incorporated through the ratio \({\tilde u^i_b}/{\tilde u}\). Representative distributions of \({\tilde u^i_b}/{\tilde u}\) are shown in Fig.~\ref{fig:ub_i} in Appendix~\ref{sec:app_B}. Additionally, the high-order term in \(\psi_3\) makes the transformations unclosed. In Sec.~\ref{sec:simplified_temperature_transformations}, several approximate models for this term are introduced, which lead to simplified yet satisfactory transformations. The reader is referred to Eqs.~\eqref{eq:SL_Tplus_modeled_1} to \eqref{eq:SL_Tplus_modeled_3} and Eqs.~\eqref{eq:SL_Tplus_pipe_modeled_1} to \eqref{eq:SL_Tplus_pipe_modeled_3}.

We acknowledge that the above derivation is entirely based on the logarithmic region, where the viscous shear stress and heat flux are neglected. Therefore, the energy balance relations in Eqs.~\eqref{eq:energy_VD} and \eqref{eq:energy_SL} are only valid in the logarithmic layer. However, both \(T^+_{VD}\) and \(T^+_{SL}\) defined in Eqs.~\eqref{eq:VD_Tplus} and \eqref{eq:SL_Tplus} are formulated to include the entire half-channel height. This may cause potential physical inconsistency below the overlap layer, which will be addressed in next section. 

\subsection{Distribution in the viscous sublayer}\label{sec:viscous_sublayer}
The LoW includes both the linear law for the viscous sublayer and the log-law for the overlap layer \citep{Coles1956, Chen2022, Guo2022}. To ensure that the linear law is satisfied in the viscous sublayer, additional constraints should be imposed. Before proceeding, we emphasize that \(T^+_{VD}\) and \(T^+_{SL}\) are not redefined in the viscous sublayer. Rather, their distributions defined in Eqs.~\eqref{eq:VD_Tplus} and \eqref{eq:SL_Tplus} are evaluated in this region.

Considering the global energy balance equation, Eq.~\eqref{eq:global_energy_balance_2}, it can be verified that the molecular heat conduction \(q^\mu_T\) dominates in the viscous sublayer, while \(q^t_k\) and \(q_f\) are negligible. Consequently, Eq.~\eqref{eq:global_energy_balance_2} reduces to:
\begin{equation}\label{eq:sublayer_energy_balance_1}
  \frac{\bar\mu c_p}{Pr}\frac{d\tilde T}{dy}
  = q_w - \tau_{tot} \, \tilde u. 
\end{equation}

The right hand side of Eq.~\eqref{eq:log_layer_energy_balance} is approximately equal to the right hand side of Eq.~\eqref{eq:sublayer_energy_balance_1}. In this case, the denominator in Eqs.~\eqref{eq:VD_Tplus} and \eqref{eq:SL_Tplus} reduces to:
\begin{equation}\label{eq:sublayer_energy_balance_2}
  B_q + \psi_2 {(\gamma - 1)M_\tau^2 u^+} + \psi_3 
  \approx 
  \frac{\bar\mu}{Pr \bar\rho_w u_\tau} \frac{d\theta^+}{dy} .
\end{equation}

Invoking Eq.~\eqref{eq:sublayer_energy_balance_2} to Eqs.~\eqref{eq:VD_Tplus} and \eqref{eq:SL_Tplus}, and applying corresponding scaling, we obtain:
\begin{equation}\label{eq:sublayer_Tplus}
T^+_{VD} \approx Pr \int_0^{y^+} \psi_1 \frac{\sqrt{\rho^+}}{\mu^+} dy^+ , \,\,\,\,\, 
T^+_{SL} \approx Pr \int_0^{y^*} \psi_1 dy^* .
\end{equation}

As stated earlier, Eq.~\eqref{eq:sublayer_Tplus} does not redefine the transformations in the viscous sublayer. Rather, it represents the equivalent distributions of the proposed transformations, Eqs.~\eqref{eq:VD_Tplus} and \eqref{eq:SL_Tplus}, in this region. Furthermore, it also accounts for the different near-wall behaviors of the two transformations over isothermal and adiabatic walls (see Sec.~\ref{sec:performance_of_temperature_transformation}).

To ensure the linear law \citep{Chen2022, Guo2022} in the viscous sublayer (\(0 \leq y^+ \leq 5\)), the proposed transformations should meet additional requirements such that:
\begin{equation}\label{eq:linear_law_Tplus}
  T^+_{VD} \approx Pr \, y^+ , \,\,\,\,\, T^+_{SL} \approx Pr \, y^* .
\end{equation}

For the SL-type transformation, Eqs.~\eqref{eq:sublayer_Tplus}  and \eqref{eq:linear_law_Tplus} imply:
\begin{equation}\label{eq:sublayer_f1}
  \psi_1 = \frac{l_m\sqrt{\tau^+_{tot}}}{\kappa y} \approx 1,
\end{equation}
which indicates
\begin{equation}\label{eq:sublayer_lm}
  l_m \approx \frac{\kappa y} {\sqrt{\tau^+_{tot}}} \,. 
\end{equation}

Typically, \(\tau^+_{tot} \approx 1\) in this region for turbulent channel, pipe, and zero-pressure-gradient boundary layer flows. Hence, the requirement of \(\psi_1 \approx 1\) is equivalent to \(l_m \approx {\kappa \, y}\) in the viscous sublayer. 

For the VD-type transformation, this requirement still holds. However, due to the presence of \(\rho^+\) and \(\mu^+\) in Eq.~\eqref{eq:sublayer_Tplus}, the performance of the VD-type temperature transformation is generally not as good as the SL-type in the viscous sublayer and buffer layer.

It should be noted that, the eddy viscosity is typically damped in the viscous sublayer, as shown in previous transformations \citep{Huang2023, Hasan2023} and near-wall modeling \citep{Larsson2016, Yang2018, Chen2022b}. The damped eddy viscosity corresponds to a damped \(l_m\). According to \citet{She2017}, \(l_m\) has a multi-layer structure, with \(l_m \sim y^{3/2}\) in the viscous sublayer, \(l_m \sim y^2\) in the buffer layer, and \(l_m \sim y\) in the logarithmic layer. However, Eq.~\eqref{eq:sublayer_lm} implies that wall damping should not be employed in our transformation. These observations may lead to inconsistency in the mixing length \(l_m\), which is a limitation of the present transformation. In this regard, we provide further discussions in Sec.~\ref{sec:application_and_limitation}.

\subsection{Mixing length model}\label{sec:mixing_length_model}
The above analysis indicate that the desired mixing length model in our transformations should meet two requirements: (1) it follows \(l_m \approx {\kappa \, y}\) in the viscous sublayer, and (2) it is capable of modeling the Reynolds stress \(\bar\rho \widetilde{u^{\prime\prime}v^{\prime\prime}}\) in the overlap layer. Various mixing length models have been developed in the literature. In the following, we introduce three representative models that illustrate different characteristics of the transformation.

The first model is the most widely utilized linear formulation based on Prandtl's hypothesis of linear variation of mixing length:
\begin{equation}\label{eq:lm_linear}
  l_m^L = \kappa y \,.
\end{equation}

It satisfies the first requirement while is not well-suited for the second. As a result, \(\psi_1 = \sqrt{\tau^+_{tot}}\), and the transformed temperature follows the linear law in the viscous sublayer, while exhibiting a logarithmic behavior in the overlap layer at sufficiently high Reynolds numbers.

The second model is a special case for the channel and pipe configurations \cite{Pirozzoli2014, Kundu2018}. The studies by \citet{Pirozzoli2014} and \citet{Galbraith1977} indicate that the linear variation of total shear stress and the velocity log-law directly yields the parabolic form:
\begin{equation}\label{eq:lm_parabolic}
  l_m^P = \kappa y \sqrt{\tau^+_{tot}} \quad \text{with} \quad \tau^+_{tot} = 1 - {y}/{h}.
\end{equation}

It satisfies both requirements and leads to \(\psi_1 = {\tau^+_{tot}}\). Consequently, the transformed temperature follows a linear distribution in the viscous sublayer and exhibits a pronounced logarithmic profile in the overlap layer. Moreover, the transformed temperatures coincides with their incompressible counterparts at comparable characteristic Reynolds number.

The third model is the enhanced mixing length formulation proposed by \citet{Xu2025}, motivated by the idea of extending the logarithmic profile in the outer layer:
\begin{equation}\label{eq:lm_new}
  \frac{l_m^E}{h} = 
  \begin{cases}
    \displaystyle \kappa \frac{y}{h}\sqrt{\tau^+_{tot}}  & \text{for } y/h \in [0, \eta_{mix}], \\[3ex]
    \displaystyle \frac{K_{mix}(1 - r^{M_{mix}})}{M_{mix}(1+r^2_{core})^{1/4}} \left[1 + \left(\frac{r_{core}}{r}\right)^2 \right]^{1/4}  & \text{for } y/h \in (\eta_{mix}, 1], \\
  \end{cases}
\end{equation}
\begin{equation}\label{eq:fitted_eta_mix}
  \eta_{mix} = 0.060 + 0.340\exp{(-Re^*_\tau/595)},
\end{equation}
\begin{equation}\label{eq:fitted_K_mix}
  K_{mix} = 0.416 + 0.172 \exp{(-Re^*_\tau/373)},
\end{equation}
\begin{equation}\label{eq:fitted_M_mix}
  M_{mix} = 3.104 + 0.871 \exp{(-Re^*_\tau/3144)}.
\end{equation}

Here, \(\tau^+_{tot} = 1 - y/h\), \( r = 1 - y/h\), \( r_{core} = 0.45\), and \( Re^*_\tau = {\bar\rho_c \sqrt{\tau_w/\bar \rho_c} \, h}/{\bar\mu_c} \). Here, the subscript \(c\) denoting quantities at the channel centerline. In the first part, the parabolic form is applied, which aligns Eq.~\eqref{eq:sublayer_lm} in the viscous sublayer. The second part is a revised form of the model by \citet{She2017}, originally developed for incompressible turbulent channel flow. In our tests, their model performs well at \(Re_\tau = 1000\), but its accuracy deteriorates for \( Re_\tau < 400 \) and \( Re_\tau > 4000 \), where \( Re_\tau = \rho_w u_\tau h/\bar \mu_w \) is the friction Reynolds number. Outside this range the performance in the outer layer degrades. Although \citet{Zhu2024} proposed a revised version by applying semi-local scaling, the \(l_m\) distribution in the outer layer does not significantly differ. In this regard, Eq.~\eqref{eq:lm_new} introduces three parameters, \(\eta_{mix}\), \(K_{mix}\), and \(M_{mix}\), to more effectively account for Reynolds and Mach number effects in the outer layer. Using this model, \(\psi_1\) exhibits a compound profile. A representative distribution of \(\psi_1\) is shown in Fig.~\ref{fig:f1f2f3_influence_wrles}.

In addition to meeting the first requirement, the enhanced model \(l_m^E\) also satisfies the second over a significantly wider region beyond the overlap layer. In this study, it is used to demonstrate that, with a suitable mixing length model, the transformed temperature would exhibit an extended logarithmic profile.

Moreover, to illustrate the damping effects, we consider the linear model with a Van Driest damping function \citep{VanDriest1956, She2017}: 
\begin{equation}\label{eq:lm_linear_damp}
  l_m^{LD} = \kappa y \left[1 - \text{exp}\left(-y^*/A^+\right) \right] \,\,\, \text{with} \,\,\,  A^+ = 27
\end{equation}

It is used to demonstrate that the mixing length indeed should remain undamped in the proposed transformation. Note that the value of \(A^+\) is different from those used in the eddy viscosity models \cite{Yang2018,Griffin2023}. 

\section{Performance of temperature transformation}
\label{sec:performance_of_temperature_transformation}
In this section, we evaluate the transformation in compressible turbulent channel flow. Three types of data are employed: 
\begin{itemize}
  \item \,\, DNS from \citet{Gerolymos2023, Gerolymos2024a, Gerolymos2024b} with the isothermal wall boundary condition and a density-based body force. 
  \item \,\, WRLES from our own computations with the isothermal wall boundary condition and a volume-based body force. 
  \item \,\, DNS from \citet{Lusher2022} with the mixed adiabatic/isothermal wall boundary condition and a volume-based body force.
\end{itemize}

Details of these data are provided in Table \ref{table:DNS_GV2024} to Table \ref{table:DNS_LC2022_adiabatic_wall}. Additionally, we compare the performance of our transformations with those proposed by \citet{Chen2022}, \citet{Huang2023}, and \citet{Cheng2024b}, as given by Eq.~\eqref{eq:VD_Tplus_Chen2022} to \eqref{eq:SL_Tplus_Cheng2024} in Appendix \ref{sec:app_C}.

As the compressible transformations are designed to map the compressible temperatures to their incompressible counterparts, the transformed temperature profiles should be compared with the incompressible distributions. In the present study, the analytical profile proposed by \citet{Pirozzoli2024} is used, hereafter denoted as \(T^+_{IC}\). It shows excellent agreement with the DNS results of incompressible temperature distribution \citep{Pirozzoli2023, Pirozzoli2024}. Details are provided in Appendix \ref{sec:app_D}. Furthermore, to show the logarithmic behavior of the transformed temperature, the inner layer distribution of \(T^+_{IC}\) is  employed, given by \citep{Pirozzoli2024}:
\begin{equation}\label{eq:IC_exLoW}
  T^+_{LoW} = \int_0^{y^+} \frac{Pr}{1 + Pr \, \alpha_t^+} \, dy^+ 
  \quad \text{where} \quad 
  \alpha^+_t = \frac{(k_\theta \, y^+)^3}{(k_\theta \, y^+)^2 + C_\theta^2}.
\end{equation}

Here, \(k_\theta = \kappa/Pr_t\) and \(C_\theta = 11\). Note that in the study of \citet{Pirozzoli2024}, \(k_\theta = 0.459\) and \(C_\theta = 10\). In fact, these two quantities are directly related to the values of \(\kappa\) and \(Pr_t\). As introduced in Sec.~\ref{sec:compressible_law_of_the_wall}, \(\kappa = 0.41\) and \(Pr_t = 0.85\) are applied in the present study, leading to slightly different values of \(k_\theta\) and \(C_\theta\). Nevertheless, these differences do not result in significant discrepancies with the LoW. In the present study, we use \(T^+_{LoW}\) throughout the half-channel as reference.

Before proceeding, we define the bulk Mach number \( M_b = u_b/\sqrt{\gamma R \tilde T_w} \), the bulk Reynolds number \( Re_b = \rho_b u_b h/\bar \mu_w \), and the friction coefficient \(C_f = \tau_w/(\frac{1}{2} \rho_b \, u_b^2)\), where \( \rho_b = \int_0^h \bar\rho dy/h \), \( u_b = \int_0^h \overline{\rho u} dy/(\rho_b h)\). All other terms are as defined in section Sec.~\ref{sec:compressible_law_of_the_wall}.

\subsection{Performance above the isothermal wall with a density-based driving force}\label{sec:performance_on_DNS_GV2024}
In the DNS of \citet{Gerolymos2023, Gerolymos2024a, Gerolymos2024b}, hereafter referred to as GV2024, the authors investigated the statistics of total and static temperature in compressible turbulent channel flow, along with the effects of Mach number on pressure fluctuations. Both bottom and top walls are isothermal boundary condition. This dataset cover a wide range of Mach and Reynolds numbers, making it well-suited for evaluating the performance of our transformations. Table \ref{table:DNS_GV2024} lists the critical information of the data. Since the \( Re^*_\tau \) is relatively low in many of their simulations, only those with \( Re^*_\tau>140 \) are considered to mitigate strong low-Reynolds-number effects \citep{Modesti2016, Griffin2023}.

\begin{table}[t]
  \caption{\label{table:DNS_GV2024}DNS of \citet{Gerolymos2023, Gerolymos2024a, Gerolymos2024b} for compressible turbulent channel flow with isothermal wall boundary conditions on both the bottom and top walls. The case name in the first column follows the same nomenclature of this database. For instance, "MCLx0p32" refers to Mach number at channel center line, \( M_{CLx} = \bar u_{CL}/\bar a_{CL} = 0.32 \), where \( \bar u_{CL} \) and \( \bar a_{CL} \) are the mean streamwise velocity and sound speed at the channel center line, respectively.}
  \begin{ruledtabular}
  \footnotesize
  \begin{tabular}{l c c c c c c}
    Case & \( M_b \) & \( Re_b \) & \( Re_\tau \) & \( Re^*_\tau \) & \( M_\tau \) & \( -B_q \)\\[3pt]
    \hline
    GV2024-MCLx0p32	& 0.28	& ~2197	& ~145	& 143	& 0.0181	& 0.0020 \\
    GV2024-MCLx0p79	& 0.71	& ~2508	& ~168	& 151	& 0.0437	& 0.0123 \\
    GV2024-MCLx1p50	& 1.51	& ~3100	& ~228	& 151	& 0.0812	& 0.0490 \\
    GV2024-MCLx0p35	& 0.30	& ~2786	& ~180	& 177	& 0.0191	& 0.0023 \\
    GV2024-MCLx1p99	& 2.39	& ~6909	& ~555	& 245	& 0.1005	& 0.0963 \\
    GV2024-MCLx0p83	& 0.75	& ~4479	& ~282	& 251	& 0.0430	& 0.0129 \\
    GV2024-MCLx1p47	& 1.49	& ~5468	& ~377	& 254	& 0.0757	& 0.0451 \\
    GV2024-MCLx0p80	& 0.72	& ~6266	& ~378	& 340	& 0.0402	& 0.0117 \\
    GV2024-MCLx1p98	& 2.39	& ~9962	& ~768	& 341	& 0.0964	& 0.0922 \\
    GV2024-MCLx1p51	& 1.56	& ~7813	& ~523	& 342	& 0.0750	& 0.0468 \\
    GV2024-MCLx1p50	& 1.57	& 25216	& 1479	& 965	& 0.0660	& 0.0414 \\
    GV2024-MCLx0p81	& 0.74	& 21092	& 1100	& 985	& 0.0356	& 0.0106 \\
  \end{tabular}
  \end{ruledtabular}
\end{table}

It is important to note that the flow in this dataset is driven by density-based body force \( \rho f_x \) rather than volume-based force \( f_x \) \citep{Gerolymos2014}. Consequently, the density profile, \( {\bar \rho}/{\rho_b} \), should be considered when calculating \(\psi_2\). Following \citet{Huang1995}, the total shear stress profile in Eq.~\eqref{eq:linear_shear_stress} and the definition of \( \tilde u^i_b \) in Eq.~\eqref{eq:ub_i} should be computed as follows:
\begin{equation}\label{eq:tau_y_new}
  \tau^+_{tot} = 1 - \frac{1}{h} \int_0^y \frac {\bar \rho(\eta)}{\rho_b} d\eta,
\end{equation}
\begin{equation}\label{eq:ub_i_new}
  \tilde u^{i}_{b,\rho} = \frac{1}{y}\int_0^y {\frac {\bar \rho}{\rho_b} \tilde u(\eta)}d\eta.
\end{equation}

Subsequently, we obtain:
\begin{equation}\label{eq:f1_rho}
  \psi_{1,\rho} = \frac{l_m}{\kappa y} \sqrt{1 - \frac{1}{h} \int_0^y \frac {\bar \rho(\eta)}{\rho_b} d\eta}, \,\,\,
\end{equation}
\begin{equation}\label{eq:f2_rho}
  \psi_{2,\rho} = 1-\frac{1}{h} \int_0^y \frac {\bar \rho(\eta)}{\rho_b} d\eta + \frac {\tilde u^{i}_{b,\rho}}{\tilde u} \frac{y}{h}.
\end{equation}

\subsubsection{Influence of mixing length model}\label{sec:influence_of_lm_model}
\begin{figure*}
  \includegraphics[width=\linewidth]{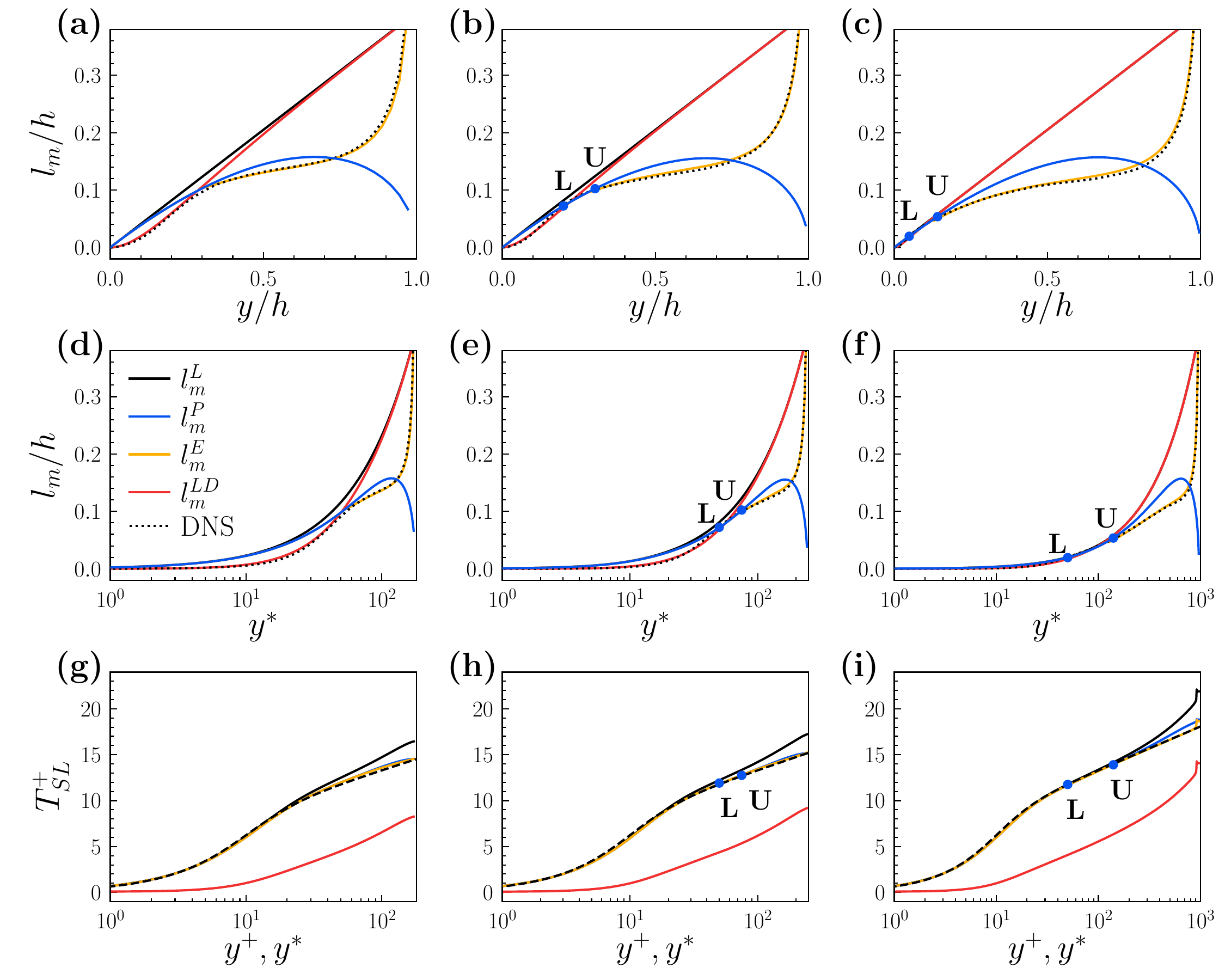}
  \caption{Distribution of mixing length model and its influence on the transformed temperature. \(l_m^L\), \(l_m^P\), \(l_m^E\), and \(l_m^{LD}\) correspond to models given by Eqs.~\eqref{eq:lm_linear}, \eqref{eq:lm_parabolic} , \eqref{eq:lm_new}, and \eqref{eq:lm_linear_damp}, respectively. Cases: GV2024-MCLx0p35 \textbf{(a, d, g)}, GV2024-MCLx1p99 \textbf{(b, e, h)}, GV2024-MCLx1p50 \textbf{(c, f, i)}. DNS data from \citet{Gerolymos2023, Gerolymos2024a, Gerolymos2024b} are employed. Black dotted lines: theoretical \(l_m\) from DNS using \(l_m = {{(-\widetilde{u^{\prime\prime} v^{\prime\prime}})}^{1/2}}/{(d\tilde{u}/dy)}\). Points \(L\) and \(U\): approximate lower and upper bounds of the logarithmic region using \(l_m^P\). Black dashed lines: \(T^+_{LoW}\) from Eq.~\eqref{eq:IC_exLoW}.}
  \label{fig:lm_Tplus_yplus}
\end{figure*}

We compare the distribution of \(l_m\) and its influence on the transformed temperature, as shown in Fig.~\ref{fig:lm_Tplus_yplus}. Three flow conditions are considered, representing cases from weakly to strongly compressible and from weakly to highly turbulent flows.

In the viscous sublayer, only the model with damping function correctly follows the true \(l_m\) value. However, the resulting \(T^+_{SL}\) exhibits a lower magnitude due to the damping, as indicated by the red solid lines in panels (g, h, i). On the contrary, the other three models provide the correct linear distributions of \(T^+_{SL}\), which agree well with the incompressible temperature profile.

In the overlap region, all these models produces logarithmic profile in \(T^+_{SL}\), although the range varies in different flow conditions. Particularly, for turbulent channel flow, the parabolic model \(l_m^P\) gives better performance than the linear model \(l_m^L\) in the overlap region, yielding a more pronounced logarithmic profile. The enhanced model \(l_m^E\) produces \(T^+_{SL}\) with an extended logarithmic profile that reaches nearly to the channel centerline. Particularly in panel (i), \(T^+_{SL}\) computed using \(l_m^E\) agrees well with \(T^+_{LoW}\).

It should be noted that, the damped \(l_m^{LD}\) also produces logarithmic profile, but its magnitude is significantly lower. Such underprediction of \(T^+_{SL}\) does not imply that the common practice of applying damping function in the viscous sublayer is incorrect. Rather, it suggests that the \(l_m\) should remain undamped for the present transformation. Further details are provided in  Sec.~\ref{sec:temperature_transformation}.

Comparing the \(l_m^P\) and corresponding \(T^+_{SL}\) profiles between points \(L\) and \(U\), it is evident that the approximate overlap region between the model predicted \(l_m\) and theoretical values also corresponds to the approximate logarithmic region of \(T^+_{SL}\). The \(l_m^E\) profile aligns with the DNS values in most of the outer layer, and hence provides the broadest range of logarithmic temperature profile. These observations are consistent with the findings of \citet{Xu2025} regarding the transformed velocity profile in compressible turbulent channel flows.

Above analysis leads to three conclusions: (1) At high Reynolds numbers, applying the commonly used linear model \(l_m^L\) in the transformation recovers the temperature LoW. (2) Damping effects results in lower magnitude of the transformed temperature. (3) The enhanced model \(l_m^E\) provides a better prediction of \(l_m\) in the outer layer, resulting in a more pronounced logarithmic profile.

In the following, we evaluate the performance of the proposed transformations across various flow conditions using \(l_m\) without damping effects. 

\subsubsection{Transformed temperature profile}\label{sec:tranformed_temperature}
\begin{figure*}
  \includegraphics[width=\linewidth]{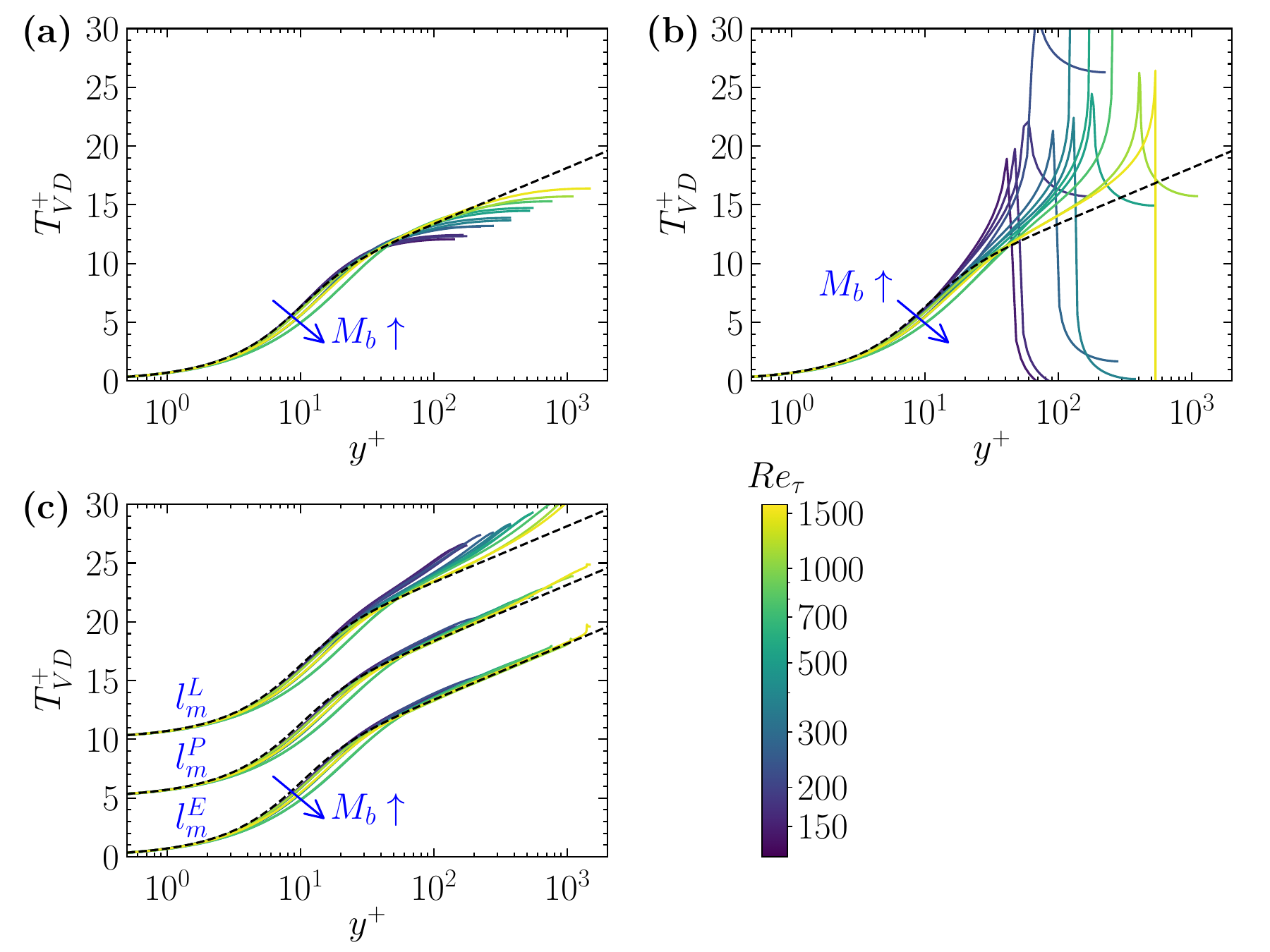}
  \caption{Temperature profiles above the isothermal wall under the VD-type transformation of (a) \citet{Chen2022}, (b) \citet{Huang2023}, and (c) the present transformation given by Eq.~\eqref{eq:VD_Tplus}, using DNS data from \citet{Gerolymos2023, Gerolymos2024a, Gerolymos2024b}. All panels show the same 12 cases listed in Table \ref{table:DNS_GV2024} and share the same color bar. In panel (c), results from \(l_m^P\) and \(l_m^L\) are shifted upward by 5 and 10 units, respectively. Black dashed lines: \(T^+_{LoW}\) from Eq.~\eqref{eq:IC_exLoW}.}
  \label{fig:GV2024_VD_Tplus_yplus}
\end{figure*}

\begin{figure*}
  \includegraphics[width=\linewidth]{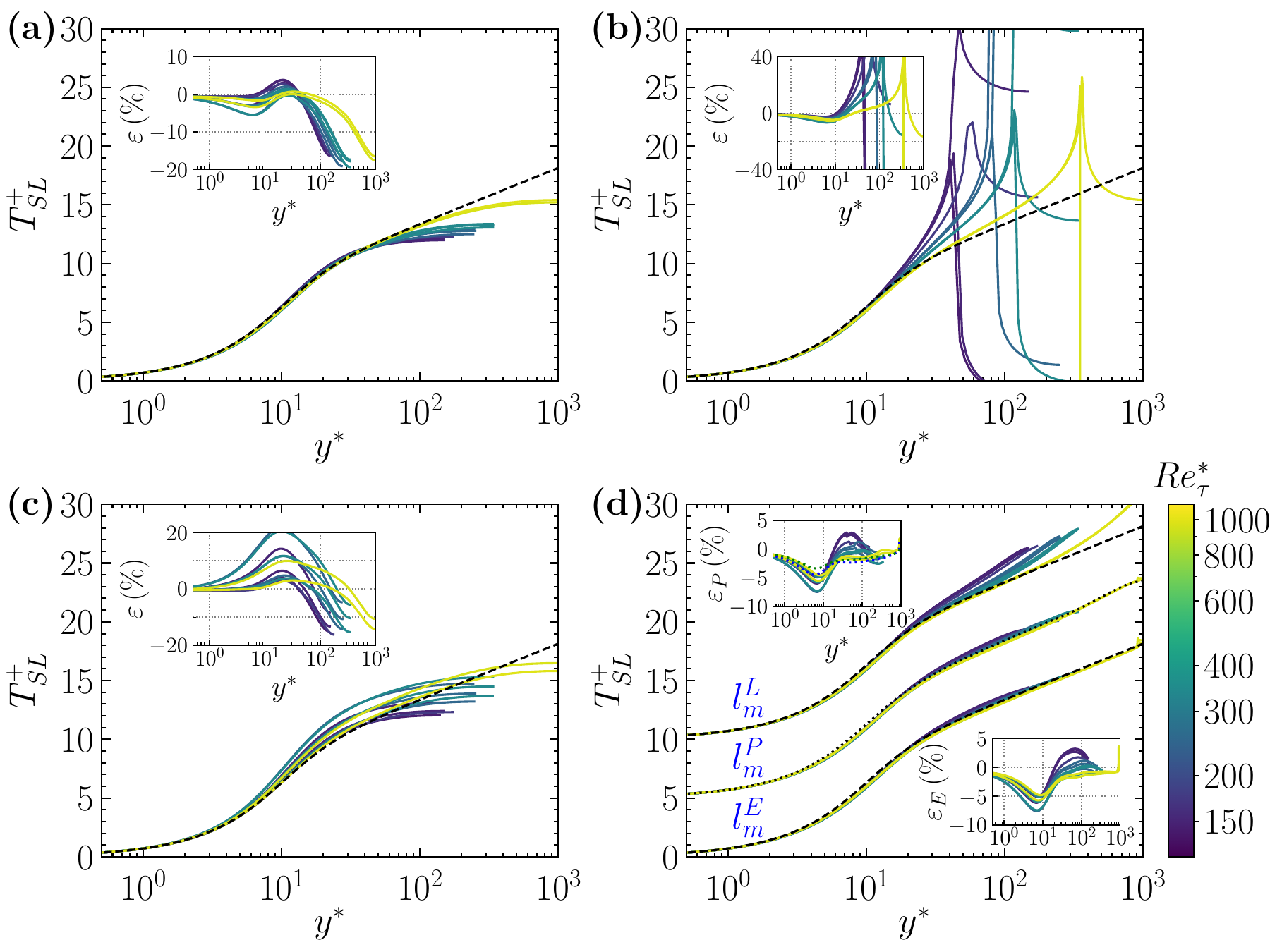}
  \caption{Temperature profiles above the isothermal wall under the SL-type transformation of (a) \citet{Chen2022}, (b) \citet{Huang2023}, (c) \citet{Cheng2024b}, and (d) the present transformation given by Eq.~\eqref{eq:SL_Tplus}, using DNS data from \citet{Gerolymos2023, Gerolymos2024a, Gerolymos2024b}. All panels show the same 12 cases listed in Table \ref{table:DNS_GV2024} and share the same color bar. In panel (d), results from \(l_m^P\) and \(l_m^L\) are shifted upward by 5 and 10 units, respectively. Black dashed lines: \(T^+_{LoW}\) from Eq.~\eqref{eq:IC_exLoW}. Black dotted line: the analytical incompressible profile \(T^+_{IC}\), corresponding to case GV2024-MCLx0p81. The insets show the local relative error.}
  \label{fig:GV2024_SL_Tplus_yplus}
\end{figure*}

The temperature profiles under the VD-type and SL-type transformations are shown in Figs.~\ref{fig:GV2024_VD_Tplus_yplus} and \ref{fig:GV2024_SL_Tplus_yplus}, along with those of \citet{Chen2022}, \citet{Huang2023} and \citet{Cheng2024b} for comparison. Our transformation outperforms the others in both slope and magnitude. As the Reynolds number increases, transformations by \citet{Chen2022}, \citet{Huang2023}, and \citet{Cheng2024b} yield improved results, suggesting that the log-law may be achieved at sufficiently high Reynolds numbers. In contrast, our transformations produce a logarithmic profile even at relatively low Reynolds numbers when \(l_m^P\) and \(l_m^E\) are applied. The linear model \(l_m^L\) achieves the log-law only at high Reynolds numbers.

Since the VD-type transformation uses wall scaling, it cannot account for variations of fluid properties within the near-wall region. Consequently, as the Mach number increases, the discrepancies in the viscous sublayer and buffer layer increase correspondingly, as indicated by the arrows in Fig.~\ref{fig:GV2024_VD_Tplus_yplus}. In contrast, the SL-type transformation more effectively collapses the temperature profiles in the viscous sublayer and buffer layer by accounting for density and viscosity variations, thereby recovering the LoW. Analogous to velocity transformation, it is expected that the compressible temperature profiles can be mapped onto their incompressible counterparts using the SL-type transformation. To demonstrate this, the insets in Fig.~\ref{fig:GV2024_SL_Tplus_yplus} show the local relative percentage error, which is defined as:
\begin{equation}\label{eq:local_error_profile}
  \varepsilon = \frac{T^+_{SL} - T^+_{ref}}{T^+_{ref}} \times 100\%,
\end{equation}
where \(T^+_{ref}\) represents the "incompressible" reference distribution. In principle, \(T^+_{ref}\) corresponds to the true incompressible temperature profile at the same \(Re^*_\tau\). However, due to limited availability of incompressible DNS data at the same \(Re^*_\tau\) considered in the present study, the analytical profile \(T^+_{IC}\) is employed as the reference. For the case employing \(l_m^E\), \(T^+_{ref}\) corresponds to \(T^+_{LoW}\), as indicated in Fig.~\ref{fig:GV2024_SL_Tplus_yplus}(d).

The black dotted line in Fig.~\ref{fig:GV2024_SL_Tplus_yplus} (d) represents \(T^+_{IC}\) distribution corresponding to case GV2024-MCLx0p81. Our transformation using \(l_m^P\) agrees well with the incompressible profiles across the entire half-channel. The insets show \(\varepsilon\) over a wider range of flow conditions. Compared with the other three transformations, the present transformations exhibit relatively lower errors, especially in the outer layer. 

It should be noted that the agreement is affected by low-Reynolds-number effects. At higher Reynolds numbers, such as the two cases GV2024-MCLx1p50 and GV2024-MCLx0p81 with \(Re^*_\tau \approx 1000\), \(\varepsilon\) in our transformations remain below \(2\%\) over most of the outer layer. However, \(\varepsilon\) is relatively larger in the buffer layer, reaching approximately \(6\%\) for these two cases. We note that the analytical \(T^+_{IC}\) and \(T^+_{LoW}\) are slightly higher than the incompressible DNS profile in the buffer layer (by up to 2\%), which partially amplifies the discrepancies observed in this region. To illustrate this, we additionally computed \(\varepsilon\) of the two cases with respective to the incompressible DNS at \(Re_\tau \approx 1000\). The results are presented by the blue and green dotted lines in the upper-left inset of panel (d). As seen, when true incompressible DNS data are used, the maximum \(\varepsilon\) remain below \(5\%\) in the buffer layer. This observation holds throughout the manuscript for all applied transformations. In contrast, the transformation by \citet{Chen2022} shows closer agreement with the reference in the buffer layer.

Additionally, the SL-type transformation using \(l_m^E\) collapses the entire outer layer, producing an extended logarithmic profile. These results for temperature transformation are consistent with the extended logarithmic behavior observed for velocity transformation in our previous study \citep{Xu2025}. This indicates a similarity between velocity and temperature statistics. The local relative errors obtained using \(l_m^E\) with respect to \(T^+_{LoW}\) are similar to those obtained using \(l_m^P\). As \(T^+_{LoW}\) is a fixed distribution representative of high Reynolds numbers, the errors in the outer layer are expected to be larger at low Reynolds numbers.

To quantitatively evaluate the logarithmic profile, we compute the log-law intercept \(B_T\). Following the approach of \citet{Trettel2016}, \(B_T\) can be determined by computing the integral average of the profile within the logarithmic region.
\begin{equation}\label{eq:loglaw_BT}
  B_T = \frac{1}{y^+_u -  y^+_l} \int_{y^+_l}^{y^+_u} \left( T^+ - \frac{Pr_t}{\kappa} \log(y^+) \right) dy^+
\end{equation}
where \(T^+\) denotes the transformed temperature, which may correspond to either \(T^+_{VD}\) or \(T^+_{SL}\), and \(y^+_l\) and \(y^+_u\) represent the lower and upper bounds of the logarithmic layer, respectively. This region is typically located within the range \(y^+ > 30\) and \(y/h < 0.3\) \citep{Pope2000turbulent}. For compressible turbulent flows, two modifications are applied. First, the buffer layer is observed to be thicker than in the incompressible flows, leading to an outward shift of the logarithmic layer \citep{Trettel2016,Hasan2023}. Second, \(y^*\) is employed for the SL-type transformation, as previously adopted in \citet{Guo2022}. Accordingly, we set \(y^+_l = 50, y^+_u = y^+\left.\right|_{y = 0.3h}\) for the VD-type transformation and \(y^*_l = 40, y^*_u = y^*\left.\right|_{y = 0.3h}\) for the SL-type transformation.

In our transformation, \(B_T\) exhibits a decreasing trend with increasing \(Re^*_\tau\) under the SL-type transformation, resembling the behavior of velocity transformations \citep{Huang1994}. The value of \(B_T\) under the VD-type transformation is slightly larger. In addition, \(B_T\) is also influenced by the strength of wall cooling. Considering the last two cases \((Re^*_\tau \approx 1000)\) in table~\ref{table:DNS_GV2024}, we obtain \(B_T \approx 3.65\) for GV2024-MCLx1p50 and \(B_T \approx 3.68 \) for GV2024-MCLx0p81. These values are consistent with the range \( B_T \approx 3.0\) to \( 4.0 \) reported by \citet{Brun2008} and \(B_T = 3.9\) adopted in \citet{Huang2023} and \citet{kays1980}. According to the DNS data of \citet{Pirozzoli2016a}, the incompressible temperature profile at \(Re_\tau \approx 4000\) yields \(B_T \approx 3.73\) when using \(y^+_l = 50\) and \(y^+_u = 300\), which is close to our results. Slight differences may occur when different values of \(y^+_l\), \(y^+_u\), \(\kappa\), and \(Pr_t\) are employed.

It should be noted that distinct spikes occur in the transformation of \citet{Huang2023}, which are attributed to energy imbalance. In our transformation, the inclusion of \(\psi_2\) and \(\psi_3\) significantly reduces or removes these kinks. The remaining tiny kinks in Figs.~\ref{fig:GV2024_VD_Tplus_yplus} and \ref{fig:GV2024_SL_Tplus_yplus} indicate that a minor energy imbalance still exists, but it can be neglected here. A more detailed discussion about the energy imbalance is presented in Sec.~\ref{sec:simplified_temperature_transformations}.

\subsection{Performance above the isothermal wall with a volume-based driving force}\label{sec:performance_on_WRLES}
In this section, we validate the temperature transformation using datasets obtained from simulations with a volume-based body force, consistent with the derivation in Sec.~\ref{sec:compressible_law_of_the_wall}. While several DNS datasets employ this type of body force \citep{Modesti2016,Trettel2016,Yao2020}, they do not provide the high-order statistics required to compute \(\psi_3\). Therefore, we conduct WRLES of compressible turbulent channel flow. All simulations were performed using JAX-Fluids \citep{Bezgin2023, Bezgin2025a}. The solver has been verified by prior studies \citep{Bezgin2023, Bezgin2025a, Bezgin2025b}.

\begin{table}[t]
  \caption{\label{table:WRLES_JAXFluids} WRLES with JAX-Fluids \citep{Bezgin2023, Bezgin2025a} for compressible turbulent channel flow with isothermal wall boundary conditions. "JXF-M1.5Re3000" refers to a case at \( M_b = 1.5, Re_b = 3000 \). The other cases follow the same nomenclature. The mesh sizes in wall units are denoted by \(\Delta x^+\), \(\Delta y^+_{w}\), \(\Delta y^+_{c}\), and \(\Delta z^+\), with subscript \(w\) and \(c\) representing mesh adjacent to the wall and at the channel center, respectively. Note that all flow variables are stored at cell centers. Therefore, the first off-wall cell center is located at half of \(\Delta y^+_{w}\).}
  \begin{ruledtabular}
  \footnotesize
  \begin{tabular}{l  c c c c c c c c c c c c}
    Case & \(L_x \times L_y \times L_z \) & \( N_x \times N_y \times N_z \) & \( \Delta x^+ \) & \( \Delta y^+_{w} \) & \( \Delta y^+_{c} \) & \( \Delta z^+ \) & \( Re_\tau \) & \( Re^*_\tau \) & \( M_\tau \) & \( C_f (\times 10^{-3}) \) & \( -B_q \) & \(\tilde T_c/\tilde T_w \) \\[3pt]
    \hline
    JXF-M0.7Re11750	 & \(2\pi\!\times\!2\!\times\!\pi\)            & \(256\!\times\!192\!\times\!192\) & 16.0 & 1.01 & 14.1 & 10.7 & 651  & 591 & 0.0359 & 5.677 & 0.0100 & 1.083 \\
    JXF-M0.8Re3000	 & \(3\pi\!\times\!2\!\times\!\frac{4}{3}\pi\) & \(160\!\times\!96\!\times\!96\)   & 11.7 & 0.63 & 8.6  & 8.6  & 198  & 175 & 0.0478 & 7.892 & 0.0153 & 1.111 \\
    JXF-M0.8Re7667	 & \(2\pi\!\times\!2\!\times\!\pi\)            & \(162\!\times\!140\!\times\!140\) & 17.5 & 0.97 & 13.3 & 10.1 & 450  & 397 & 0.0425 & 6.250 & 0.0136 & 1.108 \\
    JXF-M1.5Re3000	 & \(3\pi\!\times\!2\!\times\!\frac{4}{3}\pi\) & \(160\!\times\!96\!\times\!96\)   & 12.9 & 0.69 & 9.4  & 9.5  & 218  & 145 & 0.0797 & 7.700 & 0.0486 & 1.400 \\
    JXF-M1.5Re7667	 & \(2\pi\!\times\!2\!\times\!\pi\)            & \(192\!\times\!160\!\times\!160\) & 16.5 & 0.95 & 13.1 & 9.9  & 505  & 341 & 0.0721 & 6.324 & 0.0435 & 1.385 \\
    JXF-M1.5Re17000  & \(2\pi\!\times\!2\!\times\!\pi\)            & \(320\!\times\!192\!\times\!224\) & 20.1 & 0.74 & 27.0 & 14.3 & 1022 & 696 & 0.0661 & 5.292 & 0.0397 & 1.377 \\
    JXF-M1.7Re10000  & \(2\pi\!\times\!2\!\times\!\pi\)            & \(256\!\times\!192\!\times\!192\) & 16.3 & 1.03 & 14.4 & 10.9 & 664  & 412 & 0.0768 & 5.993 & 0.0526 & 1.487 \\
  \end{tabular}
  \end{ruledtabular}
\end{table}

The working fluid is assumed to be ideal gas with constant ratio of specific heats \( \gamma = 1.4 \) and \( Pr = 0.7 \). The dynamic viscosity follows a power law relationship with temperature, given by \( \mu/\mu_w = (T/T_w)^{0.7} \). A uniform grid is employed in streamwise and spanwise direction, while a stretched grid, following a tangent-hyperbolic function, is used in the wall-normal direction to improve the near wall resolution. We perform implicit large eddy simulation(ILES) in JAX-Fluids, utilizing the Adaptive Local Deconvolution Method (ALDM) developed by \citet{Adams2004a}, \citet{Hickel2007}, and \citet{Hickel2014}. Fourth order central finite-difference is used to compute the dissipative fluxes, while third-order Runge-Kutta (RK3) is employed for time integration. No-slip, isothermal boundary conditions are imposed on the bottom and top walls, with periodic boundary conditions applied in the streamwise and spanwise directions. A summary of the simulation is provided in Table \ref{table:WRLES_JAXFluids}. The mesh resolution employed in the WRLES is coarser than that used in the DNS \citep{Modesti2016,Trettel2016,Yao2020}. Nevertheless, the computed mean velocity and temperature profiles, as well as key flow quantities (\(Re_\tau\), \(Re^*_\tau\), \(M_\tau\), \(C_f\), \(B_q\), and \(\tilde T_c/\tilde T_w\)), are in reasonable agreement with the DNS results. Therefore, the WRLES data are well suited for validating the temperature transformations.

\begin{figure*}
  \includegraphics[width=\linewidth]{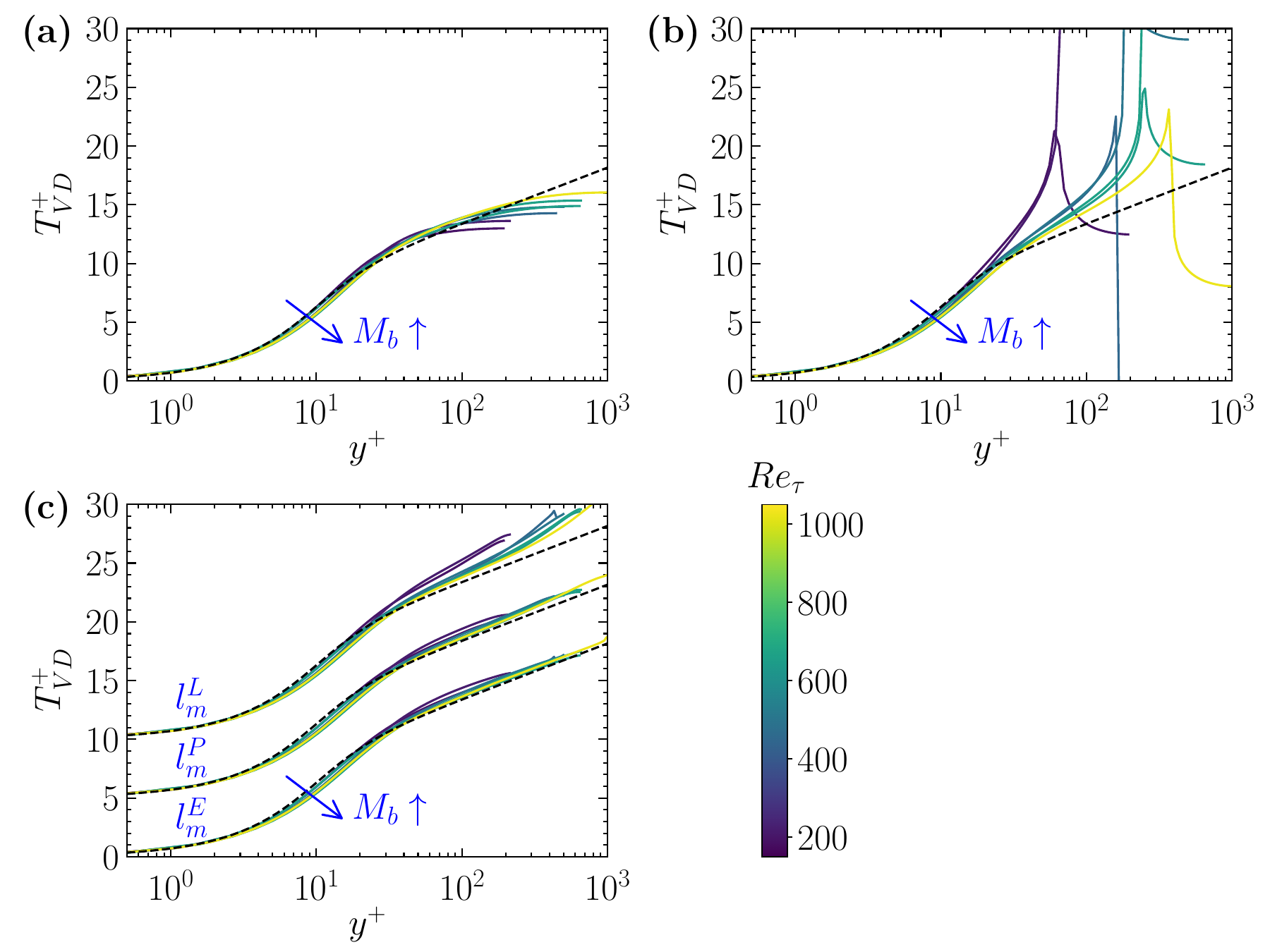}
  \caption{Temperature profiles above the isothermal wall under the VD-type transformation of (a) \citet{Chen2022}, (b) \citet{Huang2023}, and (c) the present transformation given by Eq.~\eqref{eq:VD_Tplus}, using WRLES dataset. All panels show the same 7 cases listed in Table \ref{table:WRLES_JAXFluids} and share the same color bar. In panel (c), results from \(l_m^P\) and \(l_m^L\) are shifted upward by 5 and 10 units, respectively. Black dashed lines: \(T^+_{LoW}\) from Eq.~\eqref{eq:IC_exLoW}.}
  \label{fig:wrles_VD_Tplus_yplus}
\end{figure*}

\begin{figure*}
  \includegraphics[width=\linewidth]{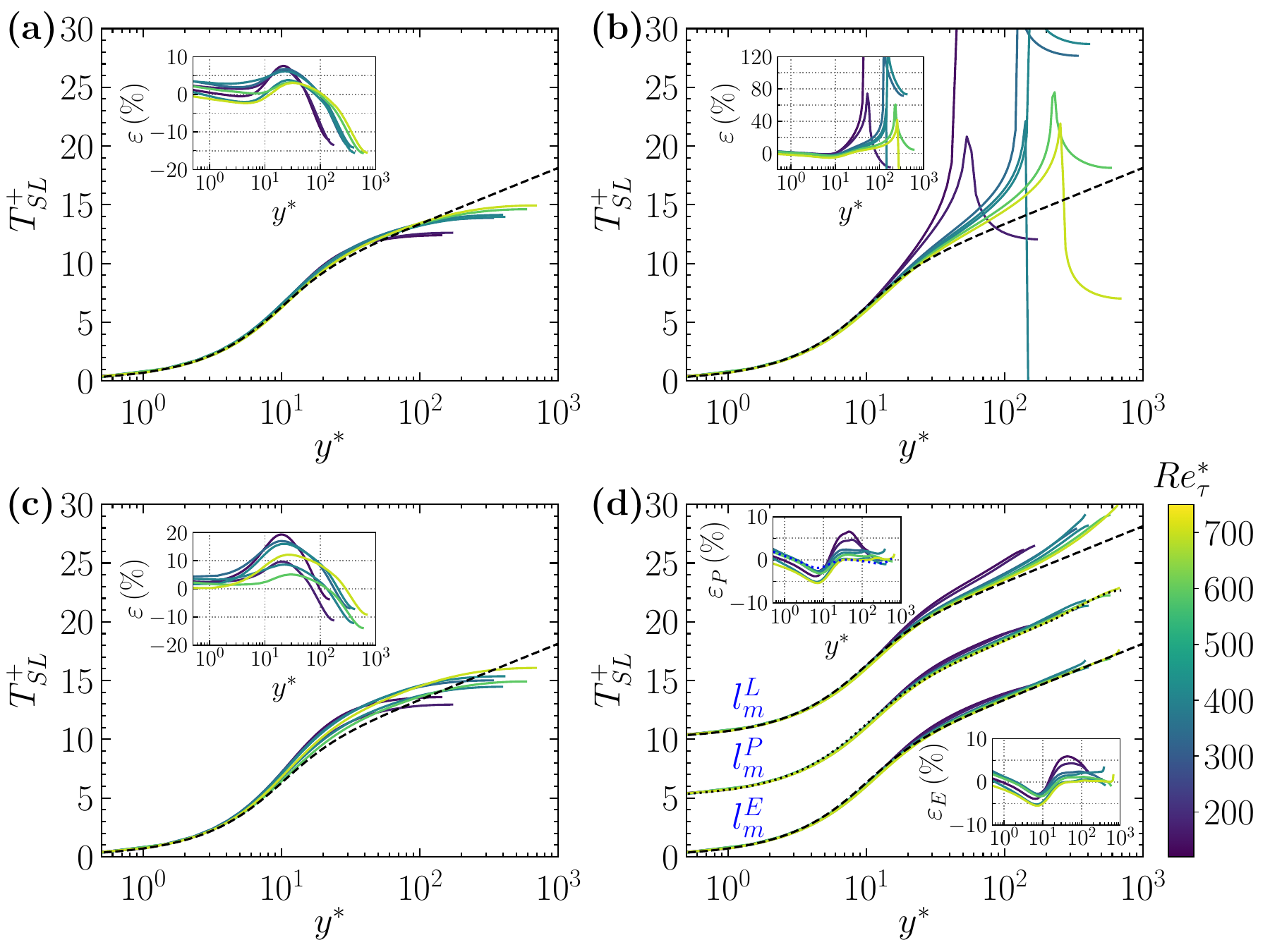}
  \caption{Temperature profiles above the isothermal wall under the SL-type transformation of (a) \citet{Chen2022}, (b) \citet{Huang2023}, (c) \citet{Cheng2024b}, and (d) the present transformation given by Eq.~\eqref{eq:VD_Tplus}, using WRLES dataset. All panels show the same 7 cases listed in Table \ref{table:WRLES_JAXFluids} and share the same color bar. In panel (d), results from \(l_m^P\) and \(l_m^L\) are shifted upward by 5 and 10 units, respectively. Black dashed lines: \(T^+_{LoW}\) from Eq.~\eqref{eq:IC_exLoW}. Black dotted line: the analytical incompressible profile \(T^+_{IC}\), corresponding to case JXF-M1.5Re17000. The insets show the local relative error.}
  \label{fig:wrles_SL_Tplus_yplus}
\end{figure*}

Figs.~\ref{fig:wrles_VD_Tplus_yplus} and \ref{fig:wrles_SL_Tplus_yplus} present the transformed temperature profiles for the VD-type and SL-type, respectively. Although with different driving force, the results closely resemble those of GV2024. As the Mach number increases, the VD-type transformation produces increasingly larger discrepancies in the viscous sublayer and buffer layer, as indicated by the arrows in Fig.~\ref{fig:wrles_VD_Tplus_yplus}. The SL-type transformation provides a better data collapse. In the outer layer, our transformations achieve better agreement with the reference profiles than the other three transformations.

The black dotted line in Fig.~\ref{fig:wrles_SL_Tplus_yplus} (d) represents \(T^+_{IC}\) corresponding to case JXF-M1.5Re17000. The \(T^+_{SL}\) using \(l_m^P\) shows good agreement with \(T^+_{IC}\) across the entire half-channel. A more thorough comparison are provided in the insets. Similar to Fig.~\ref{fig:GV2024_SL_Tplus_yplus}, \(T^+_{SL}\) computed using \(l_m^P\) and \(l_m^E\) are underpredicted in the buffer layer, with a maximum \(\varepsilon\) approaching around \(6\%\). Analogous to Fig.~\ref{fig:GV2024_SL_Tplus_yplus} (d), using true incompressible DNS data as reference results in less pronounced errors, as indicated by the blue dotted line in panel Fig.~\ref{fig:wrles_SL_Tplus_yplus} (d) for the case JXF-M0.7Re11750. The agreement improves in the outer layer, with \(\varepsilon < 1\%\) for case JXF-M0.7Re11750 and \(\varepsilon < 0.5\%\) for case JXF-M1.5Re17000 over most of the outer layer. The other three transformations demonstrate relatively larger \(\varepsilon\) in the outer layer.

\subsection{Performance above the isothermal wall with mixed thermal boundary condition}\label{sec:performance_above_isoWall_of_LC2022}
In this section, we evaluate the performance of our transformations on the isothermal wall side under mixed isothermal/adiabatic boundary conditions. We use the DNS data from \citet{Lusher2022}, hereafter referred to as LC2022. They conducted DNS to study the behavior of turbulent Prandtl number in compressible turbulent channel flows, with no-slip isothermal condition on the bottom wall and and adiabatic condition on the top. This setup creates an asymmetric flow field, where the maximum mean velocity shifted from the channel center toward the adiabatic wall, and the temperature increasing from the isothermal side to the adiabatic wall side. Such a setup is well suited for evaluating the transformations in asymmetric flows. Critical data for the isothermal and adiabatic walls are provided in Table \ref{table:DNS_LC2022_isothermal_wall} and \ref{table:DNS_LC2022_adiabatic_wall}, respectively.

\begin{table}[t]
  \caption{\label{table:DNS_LC2022_isothermal_wall} Flow quantities on the isothermal wall side of compressible turbulent channel flows under the mixed thermal configuration. The values are compatible with that of \citet{Lusher2022} and \citet{Huang2023}.}
  \begin{ruledtabular}
  \footnotesize
  \begin{tabular}{l c c c c c c c}
    Case & \( M_b \) & \( Re_b \) & \( Re_\tau \) & \( Re^*_\tau \) & \( M_\tau \) & \( -B_q \) & \(\bar T_e/\bar T_w\) \\[3pt]
    \hline
    LC2022-iC	  & 2.25  & 9983	& 1358	& 251 & 0.0723 & 0.1187 & 4.098 \\
    LC2022-iD	  & 1.70  & 13846 & 1436	& 426 & 0.0614 & 0.0767 & 2.755 \\
    LC2022-iD2  & 1.78  & 14512 & 1553	& 453 & 0.0630 & 0.0782 & 2.793 \\
    LC2022-iE	  & 3.44  & 20638 & 3789	& 306 & 0.0757 & 0.1886 & 8.197 \\
    LC2022-iE2s & 3.96  & 23770 & 3260	& 496 & 0.0994 & 0.1701 & 4.651 \\
    LC2022-iF2	& 1.86  & 20813 & 2234	& 613 & 0.0620 & 0.0799 & 2.941 \\
    LC2022-iF2s & 1.94  & 21776 & 1964	& 751 & 0.0697 & 0.0677 & 2.232 \\
  \end{tabular}
  \end{ruledtabular}
\end{table}

\begin{figure*}
  \includegraphics[width=\linewidth]{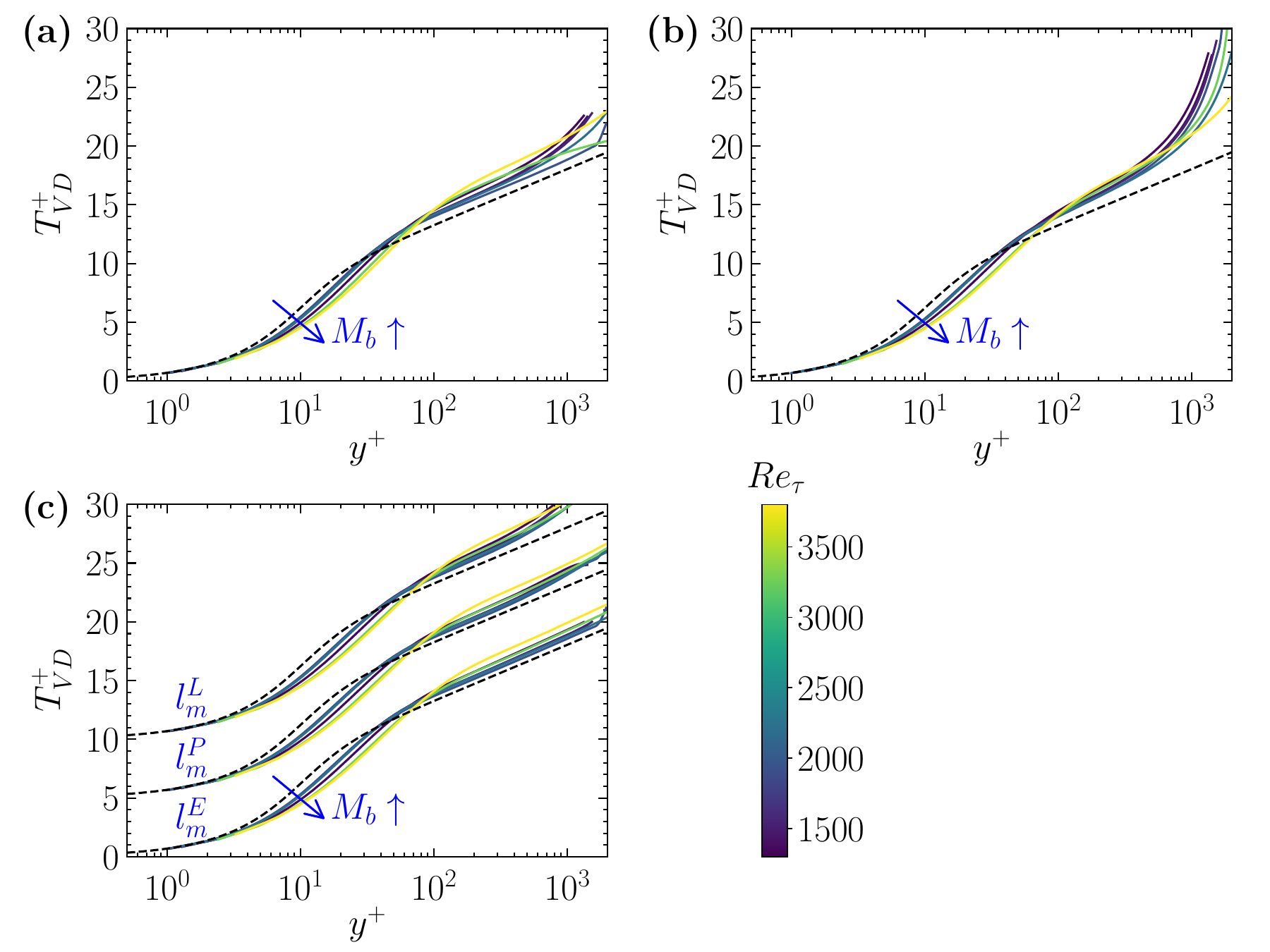}
  \caption{Temperature profiles above the isothermal wall with mixed thermal configuration under the VD-type transformation of (a) \citet{Chen2022}, (b) \citet{Huang2023}, and (c) the present transformation given by Eq.~\eqref{eq:VD_Tplus}, using DNS data from \citet{Lusher2022}. All panels show the same 7 cases listed in Table \ref{table:DNS_LC2022_isothermal_wall} and share the same color bar. In panel (c), results from \(l_m^P\) and \(l_m^L\) are shifted upward by 5 and 10 units, respectively. Black dashed lines: \(T^+_{LoW}\) from Eq.~\eqref{eq:IC_exLoW}.}
  \label{fig:LC2022_VD_Tplus_isoWall}
\end{figure*}

In \citet{Lusher2022}, the boundary layer thickness \(h_e\) is defined as the distance from the maximum velocity location to the corresponding wall, which is larger than the channel half-height above the isothermal wall, and smaller above the adiabatic wall. Accordingly, the values of \(Re_\tau\) and \(Re^*_\tau\) in Table \ref{table:DNS_LC2022_isothermal_wall} and \ref{table:DNS_LC2022_adiabatic_wall} follow this definition. Similar approach is used in the study of \citet{Guo2022} on turbulent channel flow with a cold-wall/hot-wall setup.

Additionally, the mixed thermal configuration leads to asymmetric velocity and temperature fields. Consequently, the wall shear stresses on the isothermal and adiabatic walls are no longer identical. The driving force in \citet{Lusher2022} is given by \(f_i = \bar\tau/h\), where \(\bar\tau = (\tau_i + \tau_a)/2\) is the arithmetic mean shear stress of the isothermal and adiabatic walls. Accordingly, the total shear stress is \(\tau^+_{tot} = 1 - \frac{\bar\tau}{\tau_w} \frac{y}{h}\), and \(\psi_1\) and \(\psi_2\) are computed as:
\begin{equation}\label{eq:f1_mixed_channel}
  \psi_{1,i} = \frac{l_m}{\kappa y}\sqrt{1 - \frac{\bar\tau}{\tau_{w,i}} \frac{y}{h}}, \,\,\,
  \psi_{1,a} = \frac{l_m}{\kappa y}\sqrt{1 - \frac{\bar\tau}{\tau_{w,a}} \frac{y}{h}}.
\end{equation}
\begin{equation}\label{eq:f2_mixed_channel}
  \psi_{2,i} = 1 - \frac{\bar\tau}{\tau_{w,i}}\frac{y}{h} + \frac {\tilde u^{i}_b}{\tilde u} \frac{\bar\tau}{\tau_{w,i}} \frac{y}{h}, \,\,\,
  \psi_{2,a} = 1 - \frac{\bar\tau}{\tau_{w,a}}\frac{y}{h} + \frac {\tilde u^{i}_b}{\tilde u} \frac{\bar\tau}{\tau_{w,a}}\frac{y}{h}.
\end{equation}

\begin{figure*}
  \includegraphics[width=\linewidth]{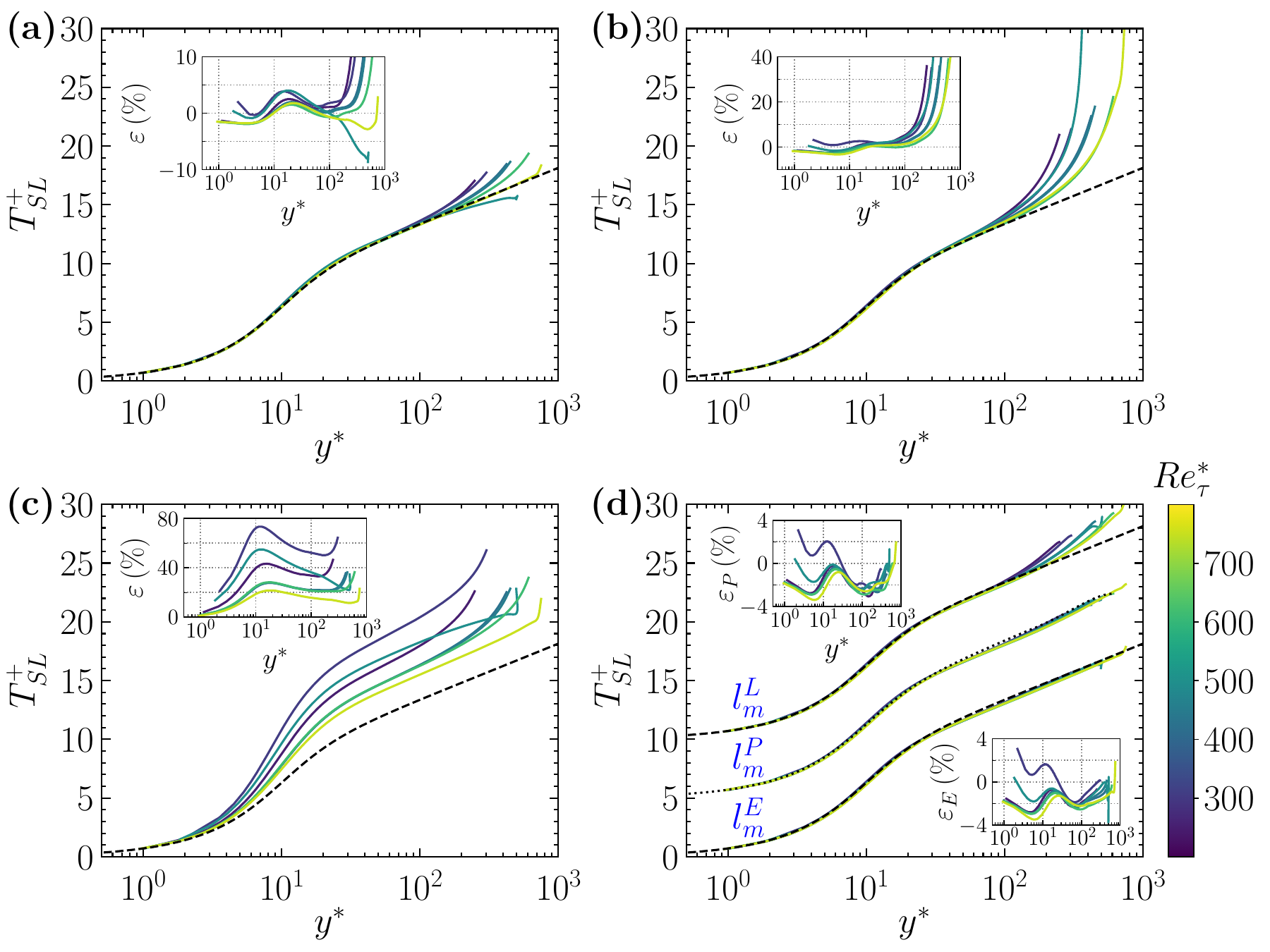}
  \caption{Temperature profiles above the isothermal wall with mixed thermal configuration under the SL-type transformation of (a) \citet{Chen2022}, (b) \citet{Huang2023}, (c) \citet{Cheng2024b}, and (d) the present transformation given by Eq.~\eqref{eq:SL_Tplus}, using DNS data from \citet{Lusher2022}. All panels show the same 7 cases listed in Table \ref{table:DNS_LC2022_isothermal_wall} and share the same color bar. In panel (d), results from \(l_m^P\) and \(l_m^L\) are shifted upward by 5 and 10 units, respectively. Black dashed lines: \(T^+_{LoW}\) from Eq.~\eqref{eq:IC_exLoW}. Black dotted line: the analytical incompressible profile \(T^+_{IC}\), corresponding to case LC2022-iF2. The insets show the local relative error.}
  \label{fig:LC2022_SL_Tplus_isoWall}
\end{figure*}

In addition, since not all datasets from \citet{Lusher2022} provide the necessary information to compute \(\psi_3\), we only test the performance on cases with the required data, labeled as "iC, iD, iD2, iE, iE2s, iF2, iF2s" in Table \ref{table:DNS_LC2022_isothermal_wall}. For more details, the reader can refer to \citet{Lusher2022} and \citet{Huang2023}.

The temperature profiles for the isothermal wall side under VD-type and SL-type transformations are presented in Figs.~\ref{fig:LC2022_VD_Tplus_isoWall} and \ref{fig:LC2022_SL_Tplus_isoWall}. The transformations by \citet{Chen2022}, \citet{Huang2023} and \citet{Cheng2024b} are also plotted for comparison. Regarding this boundary condition, all transformations yield pronounced logarithmic profiles. As the Mach number increases, the increasingly larger discrepancies observed in the viscous sublayer and buffer layer are consistent with the classical isothermal wall. The SL-type transformations provides better data collapse in this region.

In particular, as shown in Fig.~\ref{fig:LC2022_SL_Tplus_isoWall}, the transformations by \citet{Chen2022} and \citet{Huang2023} yield pronounced logarithmic profiles in the overlap layer. For the transformation by \citet{Cheng2024b}, the logarithmic slope remains correct, but the magnitude exhibits relatively large variations across different cases. Such limitation on the isothermal wall was also reported in their study. In our transformations, under the SL-type transformation, all three mixing length models recover the LoW. Analogous to the classical isothermal configuration, the parabolic model \(l_m^P\) retains the wake region while the enhanced model \(l_m^E\) substantially extends the logarithmic profile. 

It should be noted that, our transformation slightly underpredicts \(T^+_{SL}\) in the outer layer when using \(l_m^P\) and \(l_m^E\), as shown in the insets in Fig.~\ref{fig:LC2022_SL_Tplus_isoWall} (d). However, the relative errors remain below \(3.5\%\) throughout the boundary layer. Employing \(l_m^L\) produces smaller \(\varepsilon\) in the overlap layer (not shown here). The other three transformations lead to larger discrepancies in the outer layer.

Finally, the small kinks and bends near the boundary layer edge are likely associated with the complex flow field in this mixed thermal configuration. We neglect this issue in the present study, without affecting the overall conclusion.

\subsection{Performance above the adiabatic wall with mixed thermal boundary condition}\label{sec:performance_above_adiaWall_of_LC2022}

\begin{table}[t]
  \caption{\label{table:DNS_LC2022_adiabatic_wall} Flow quantities on the adiabatic wall side of compressible turbulent channel flows under the mixed thermal configuration. The values are compatible with that of \citet{Lusher2022} and \citet{Huang2023}.}
  \begin{ruledtabular}
  \footnotesize
  \begin{tabular}{l c c c c c c c}
    Case & \( M_b \) & \( Re_b \) & \( Re_\tau \) & \( Re^*_\tau \) & \( M_\tau \) & \( -B_q \) & \( \bar T_e/\bar T_w \)\\[3pt]
    \hline
    LC2022-aC	  & 2.25 & 9983  & 119 & 184  & 0.0654 & 0 & 0.697 \\
    LC2022-aD	  & 1.70 & 13846 & 227 & 327  & 0.0561 & 0 & 0.738 \\
    LC2022-aD2	& 1.78 & 29024 & 401 & 585  & 0.0540 & 0 & 0.733 \\
    LC2022-aE	  & 3.44 & 20638 & 134 & 220  & 0.0689 & 0 & 0.662 \\
    LC2022-aE2s & 3.96 & 23770 & 254 & 595  & 0.0829 & 0 & 0.498 \\
    LC2022-aF2	& 1.86 & 45788 & 560 & 824  & 0.0524 & 0 & 0.725 \\
    LC2022-aF2s & 1.94 & 47965 & 618 & 1038 & 0.0588 & 0 & 0.651 \\
  \end{tabular}
  \end{ruledtabular}
\end{table}

\begin{figure*}
  \includegraphics[width=\linewidth]{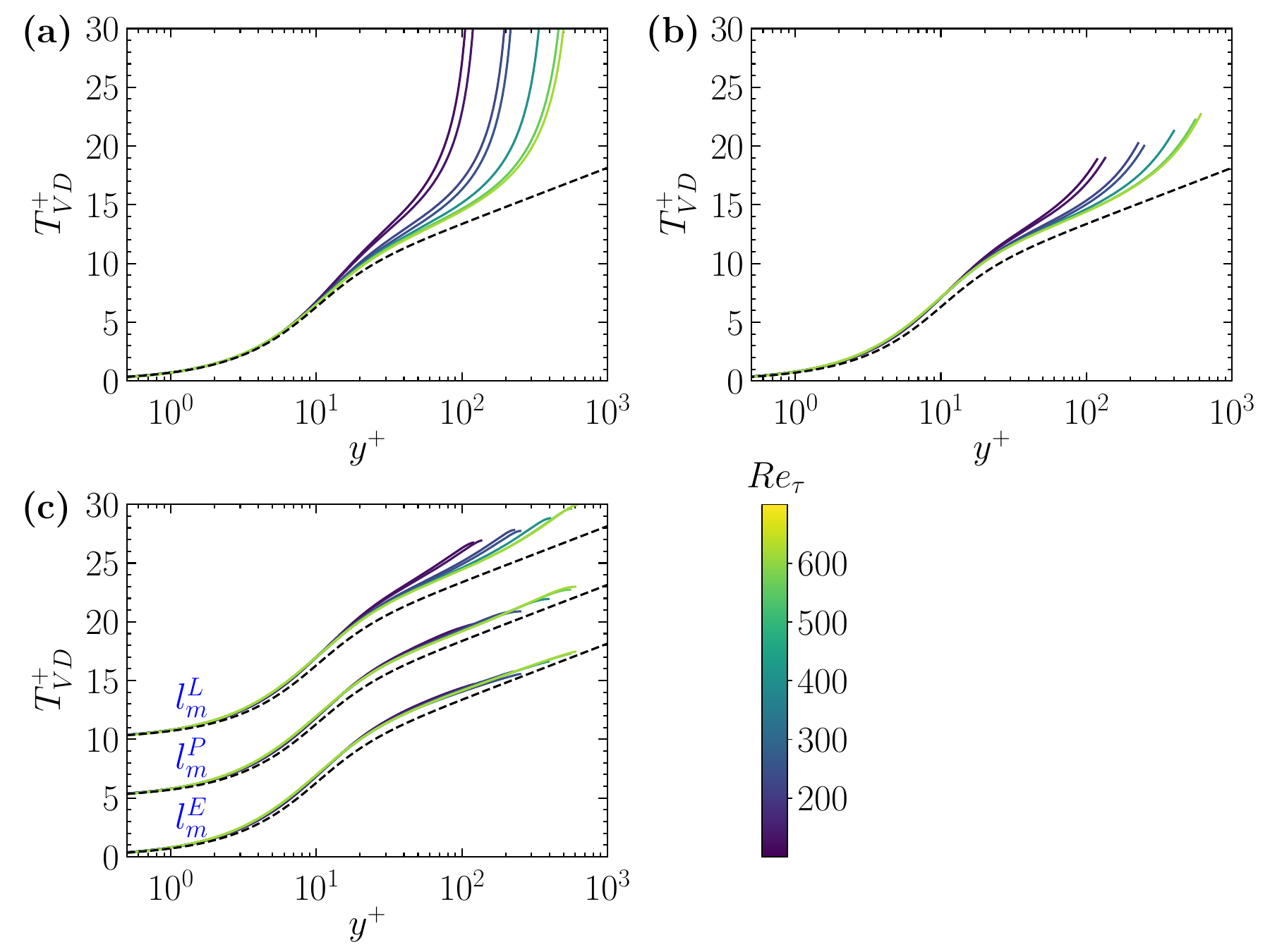}
  \caption{Temperature profiles above the adiabatic wall with mixed thermal configuration under the VD-type transformation of (a) \citet{Chen2022}, (b) \citet{Huang2023}, and (c) the present transformation given by Eq.~\eqref{eq:VD_Tplus}, using DNS data from \citet{Lusher2022}. All panels show the same 7 cases listed in Table \ref{table:DNS_LC2022_adiabatic_wall} and share the same color bar. In panel (c), results from \(l_m^P\) and \(l_m^L\) are shifted upward by 5 and 10 units, respectively. Black dashed lines: \(T^+_{LoW}\) from Eq.~\eqref{eq:IC_exLoW}.}
  \label{fig:LC2022_VD_Tplus_adiaWall}
\end{figure*}

\begin{figure*}
  \includegraphics[width=\linewidth]{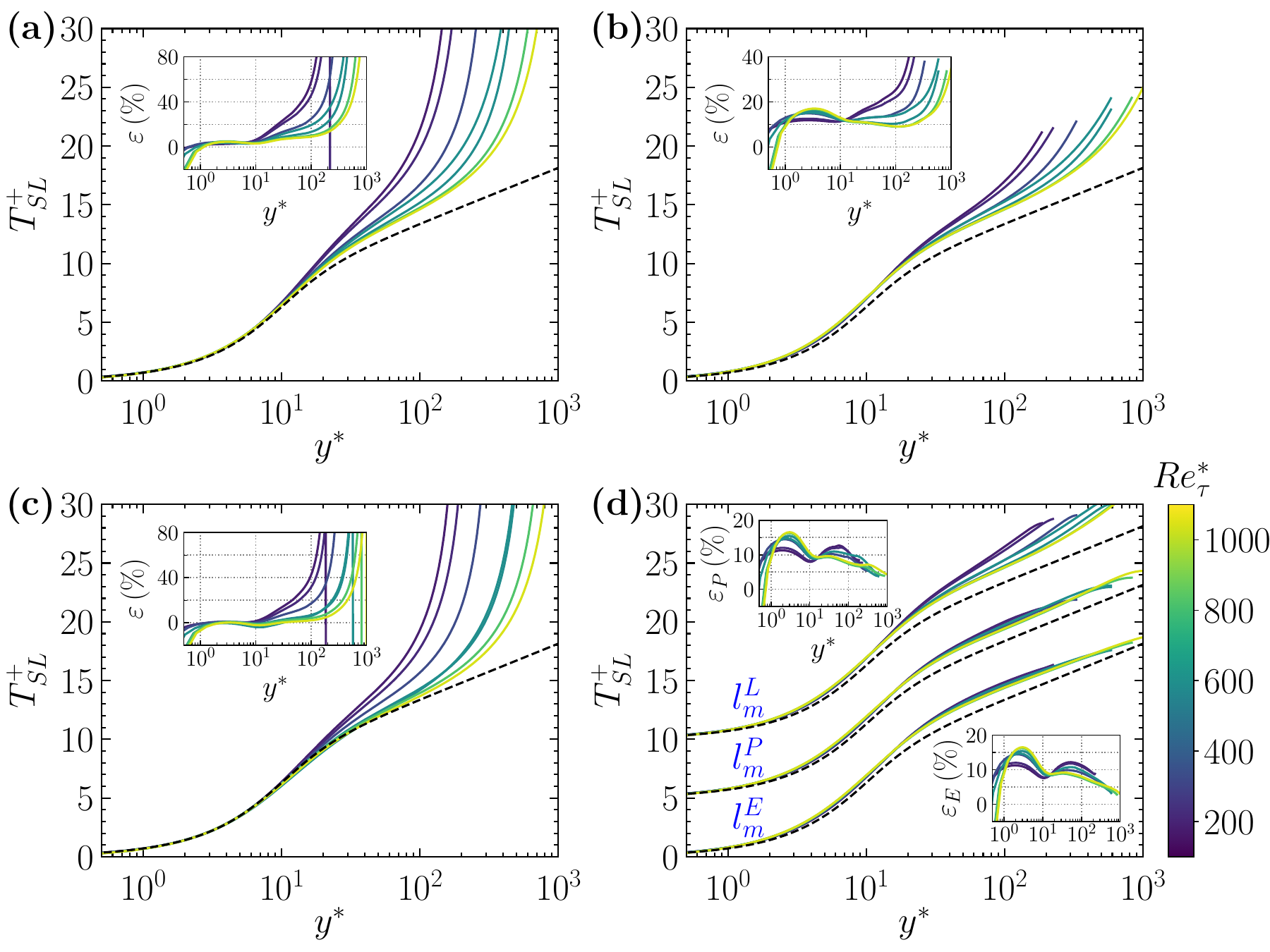}
  \caption{Temperature profiles above the adiabatic wall with mixed thermal configuration under the SL-type transformation of (a) \citet{Chen2022}, (b) \citet{Huang2023}, (c) \citet{Cheng2024b}, and (d) the present transformation given by Eq.~\eqref{eq:SL_Tplus}, using DNS data from \citet{Lusher2022}. All panels show the same 7 cases listed in Table \ref{table:DNS_LC2022_adiabatic_wall} and share the same color bar. In panel (d), results from \(l_m^P\) and \(l_m^L\) are shifted upward by 5 and 10 units, respectively. Black dashed lines: \(T^+_{LoW}\) from Eq.~\eqref{eq:IC_exLoW}. The insets show the local relative error.}
  \label{fig:LC2022_SL_Tplus_adiaWall}
\end{figure*}

Critical flow quantities on the adiabatic wall side are provided in Table \ref{table:DNS_LC2022_adiabatic_wall}. The temperature profiles under VD-type and SL-type transformations are presented in Figs.~\ref{fig:LC2022_VD_Tplus_adiaWall} and \ref{fig:LC2022_SL_Tplus_adiaWall}, respectively. For comparison, the transformations by \citet{Chen2022}, \citet{Huang2023} and \citet{Cheng2024b} are also included.

Unlike the isothermal boundary condition, temperature variation in the near-wall region above the adiabatic wall is minimal, suggesting only slight variations in density, dynamic viscosity, and thermal conductivity. As a result, the transformed temperature in the viscous sublayer and buffer layer collapses well under all these transformations. In the overlap layer, the performance of transformations by \citet{Chen2022} and \citet{Huang2023} improves as the Reynolds number increases. The transformation by \citet{Cheng2024b} yields similar behavior to those of \citet{Chen2022}, and shows closer agreement in magnitude. Our transformations based on \(l_m^P\) and \(l_m^E\) collapse the temperature profiles across the entire boundary layer. However, \(T^+_{SL}\) based on \(l_m^E\) presents smaller slope in the lower part of the outer layer, while \(l_m^P\) yields explicitly pronounced logarithmic profile. Although not shown here, this behavior is caused by the underprediction of mixing length by \(l_m^E\) in this region. 

The most pronounced difference for the adiabatic wall lies in the magnitude, which is noticeably larger than the incompressible temperature profile. The relative error of our transformation and that of \citet{Huang2023} exceed \(10\%\) below the overlap layer. Within our transformation framework, it is likely caused by an overprediction of \(l_m\) in this region. In addition, this discrepancy may also arise from differences in thermal boundary configurations, as the reference incompressible temperature profile is obtained from symmetric configuration rather than a mixed one \citep{Pirozzoli2016a}. 

An exception is the SL-type transformation by \citet{Cheng2024b}, which exhibits the correct magnitude with increasing Reynolds number. This is consistent with the results of their study.

\subsection{Integral mean error}\label{sec:integral_mean_error}
To evaluate the overall performance of the proposed transformation, we compute the integral mean error of \(T^+_{SL}\) over the entire half-channel for each case. Following the approach of \citet{Cheng2024b}, the integral mean error is defined as:
\begin{equation}\label{eq:mean_error_UT}
    \bar\varepsilon_P = \frac{\int_0^{h_e^+} \left|T^+_{SL} - T^+_{IC}\right|dy^*}{\int_0^{h_e^+} T^+_{IC} \, dy^*}, \quad \quad \\
    \bar\varepsilon_E = \frac{\int_0^{h_e^+} \left|T^+_{SL} - T^+_{LoW}\right|dy^*}{\int_0^{h_e^+} T^+_{LoW} \, dy^*}.
\end{equation}

Here, the subscript \(P\) and \(E\) denote the use of \(l_m^P\) and \(l_m^E\) for computing \(T^+_{SL}\). Regarding the compressible transformation, the ideal case is that the transformed temperature matches its incompressible counterpart across the entire boundary layer. Therefore, the upper integration bound extends well beyond the inner layer and approaches the boundary layer edge \((h^+_e)\). In contrast, \citet{Cheng2024b} evaluate the errors in viscous sublayer and log-layer, which is reasonable when the focus is on near-wall modeling or the agreement in the outer layer is far from satisfactory.

\begin{figure*}
  \includegraphics[width=\linewidth]{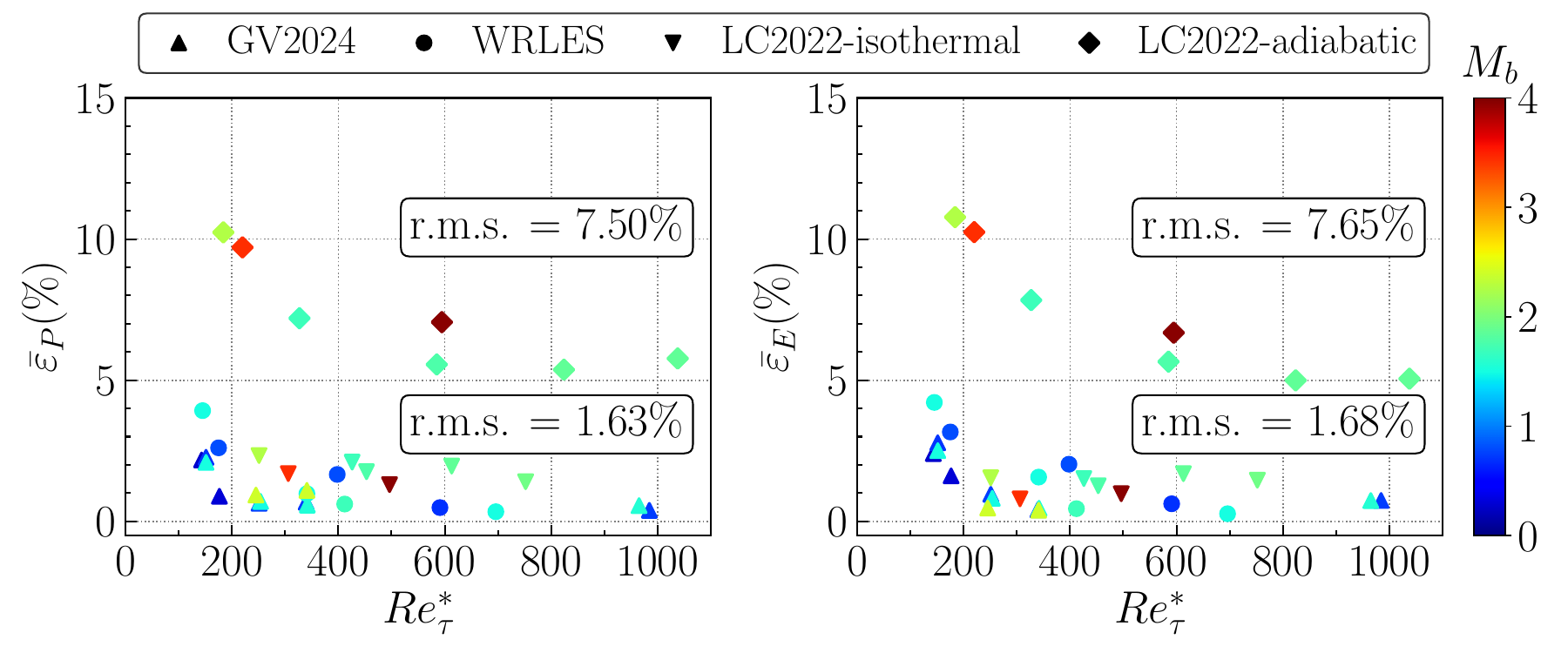}
  \caption{Integral mean error for \(T^+_{SL}\) based on \(l_m^P\) (a) and \(l_m^E\) (b) over the entire half-channel height. The two values in each panel represent the r.m.s. errors for isothermal and adiabatic walls, respectively.}
  \label{fig:integral_mean_error}
\end{figure*}

Fig.~\ref{fig:integral_mean_error} shows the computed results for the cases listed in Tables.~\ref{table:DNS_GV2024} to \ref{table:DNS_LC2022_adiabatic_wall}, covering 26 cases for the isothermal wall and 7 cases for the adiabatic wall. As shown, no significant differences are observed between \(\bar\varepsilon_P\) and \(\bar\varepsilon_E\). The integral mean error generally decreases with increasing \(Re^*_\tau\), consistent with the results of \citet{Cheng2024b}. For the isothermal wall, the mean error remains below \(2\%\) for most cases, with root-mean-square (r.m.s.) values of \(1.63\%\) and \(1.68\%\) for \(\bar\varepsilon_P\) and \(\bar\varepsilon_E\), respectively. In contrast, the adiabatic wall exhibits significantly higher errors, with r.m.s. values exceeding \(7.50\%\).

\subsection{Diagnostic function}\label{sec:disgnostic_function}
To assess the presence of logarithmic profile, the diagnostic is applied:
\begin{equation}\label{eq:diaganostic_fucntion_SL}
\Xi = y^+ \frac{dT^+_{VD}}{dy^+} \approx \text{const}
\ \ \ \text{or} \ \ \ 
\Xi = y^* \frac{dT^+_{SL}}{dy^*} \approx \text{const}.
\end{equation}

In principle, \(\Xi\) should remain constant (\(\Xi = {Pr_t}/{\kappa} \approx 2.073\)) within the logarithmic region. However, this requirement is quite strict under finite Reynolds number conditions \citep{Bernardini2014}, hence some deviation from constancy is expected.

\begin{figure*}
  \begin{tikzpicture}
    \node[anchor=south west, inner sep=0] (image) at (0,0) {\includegraphics[width=\linewidth]{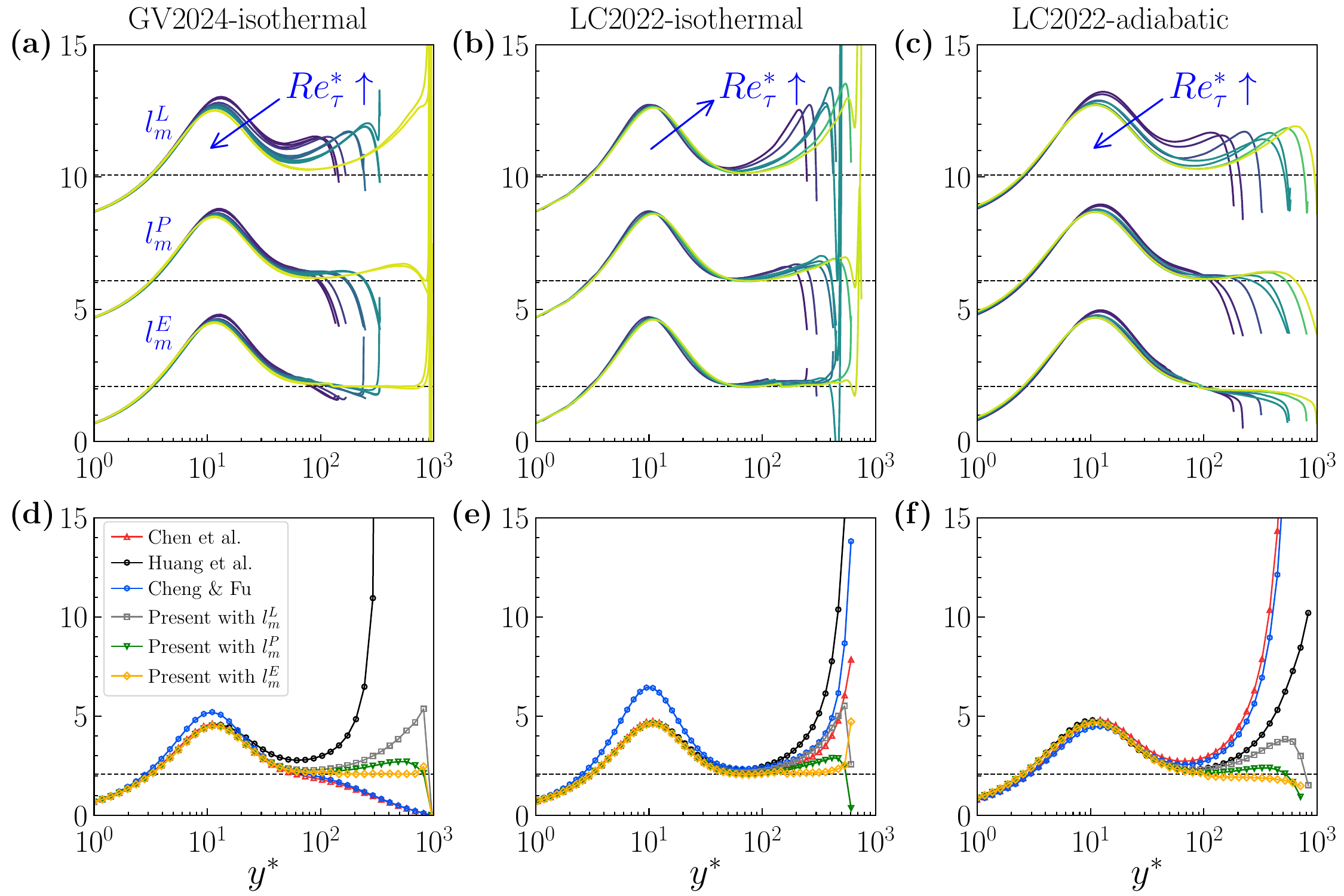}};
    \node[anchor=west, font=\fontsize{13}{11}\selectfont, rotate=90] at (0.4, 7.5) {\(\Xi\)};
    \node[anchor=west, font=\fontsize{13}{11}\selectfont, rotate=90] at (0.4, 2.5) {\(\Xi\)};
  \end{tikzpicture}
  \caption{Diagnostic function for different boundary conditions and various SL-type transformations. Panels (a, b, c) show the \(\Xi\) obtained using our SL-type transformation with three mixing length models. Cases are (a) GV2024-isothermal, (b) LC2022-isothermal, and (c) LC2022-adiabatic, respectively. Panels (d, e, f) present \(\Xi\) for various SL-type transformations under three flow conditions: (d) GV2024-MCLx1p50, (e) LC2022-iF2, and (f) LC2022-aF2. The dashed line represents the reference values \(\Xi = {Pr_t}/{\kappa} \approx 2.073, 6.073\), and \(10.073\).}
\label{fig:diagnostic_function}
\end{figure*}

Fig.~\ref{fig:diagnostic_function} shows the \(\Xi\) profile for \(T^+_{SL}\) under different boundary conditions and various SL-type transformations. As seen in panels (a, b, c), when using the parabolic model \(l_m^P\) and the enhanced model \(l_m^E\), \(\Xi\) collapses well in the inner layer across all three boundary conditions. Noticeable scatter is observed in the overlap layer for the linear model \(l_m^L\), as shown in panels (a) and (c). Compared to \(l_m^L\) and \(l_m^P\), the enhanced model \(l_m^E\) yields a broader range over which \(\Xi\) remains close to the reference value. As the Reynolds number increases, this region becomes more pronounced.

The kinks and bends near the channel center in panels (a) and (b) are caused by energy imbalance in the transformation, which will be discussed in Sec.~\ref{sec:influence_of_f1f2f3}. For the adiabatic wall, as shown in panel (c), the enhanced model results in relatively smaller \(\Xi\) than the reference. This is caused by the underprediction of the mixing length value.

Furthermore, for each boundary condition, we select one case with relatively high Reynolds number and compare the \(\Xi\) profile across different transformations, as shown in panels (d, e, f). Our transformation yields a significantly broader logarithmic region than the other three transformations, particularly when \(l_m^P\) and \(l_m^E\) are applied.

To summarize the previous sections, both the VD-type and SL-type transformations produce explicit logarithmic profile in the overlap region. The SL-type demonstrates superior performance in the viscous sublayer and buffer layer, thereby recovering the compressible LoW. \(T^+_{SL}\) based on \(l_m^P\) retains the wake profile, while \(T^+_{SL}\) based on \(l_m^E\) collapses throughout the entire outer layer, leading to extended logarithmic profile. Slight variations in the logarithmic intercept (\(B_T\)) are observed across different boundary conditions. For comparison, results under the transformations of \citet{Chen2022}, \citet{Huang2023}, and \citet{Cheng2024b} are presented, which demonstrate improved performance with increasing \( Re^*_\tau \).

\section{Simplified temperature transformations}\label{sec:simplified_temperature_transformations}
In Eqs.~\eqref{eq:VD_Tplus} and \eqref {eq:SL_Tplus}, three parameters (\(\psi_1\), \(\psi_2\), and \(\psi_3\)) are introduced, distinguishing these transformations from those of \citet{Huang2023}. In this section, we examine the roles of \(\psi_1\), \(\psi_2\), and \(\psi_3\) and demonstrate that their mechanisms stem from the damping effects in the  transformation. This analysis justifies the inclusion and exclusion of \(\psi_3\) depending on the thermal wall conditions. To remove the reliance on the high-order term in \(\psi_3\), a multi-layer structure of the turbulent TKE flux is identified, which further leads to approximate models for the turbulent TKE flux and simplified temperature transformations.

\subsection{Damping effects of \(\psi_1\), \(\psi_2\) and \(\psi_3\)}\label{sec:influence_of_f1f2f3}
\begin{figure*}
  \begin{tikzpicture}
    \node[anchor=south west, inner sep=0] (image) at (0,0) {\includegraphics[width=\linewidth]{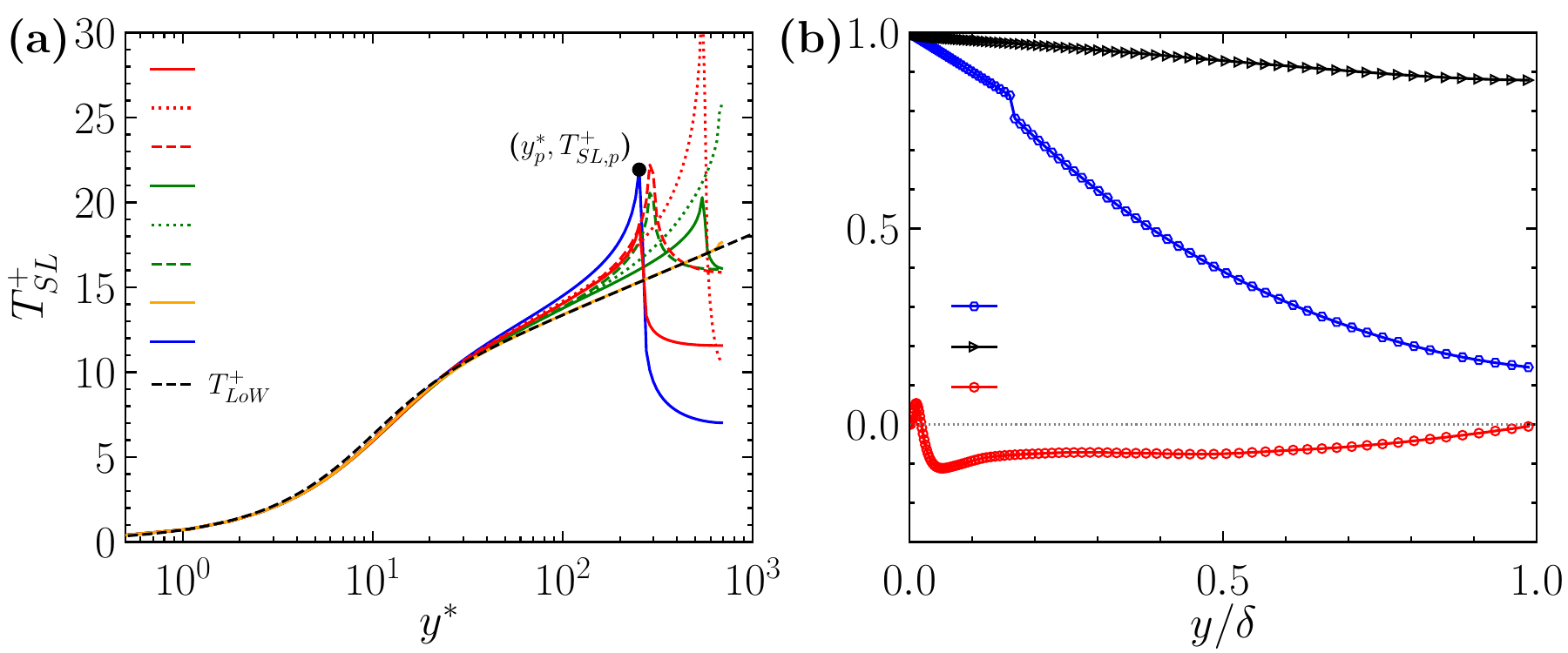}};
    \node[anchor=west, font=\fontsize{8}{11}\selectfont] at (2.0, 6.15) {with \(\psi_1\): (1)};
    \node[anchor=west, font=\fontsize{8}{11}\selectfont] at (2.0, 5.70) {with \(\psi_2\): (2)};
    \node[anchor=west, font=\fontsize{8}{11}\selectfont] at (2.0, 5.30) {with \(\psi_3\): (3)};
    \node[anchor=west, font=\fontsize{8}{11}\selectfont] at (2.0, 4.90) {with \( \psi_1, \psi_2 \): (4)};
    \node[anchor=west, font=\fontsize{8}{11}\selectfont] at (2.0, 4.50) {with \( \psi_2, \psi_3 \): (5)};
    \node[anchor=west, font=\fontsize{8}{11}\selectfont] at (2.0, 4.10) {with \( \psi_1, \psi_3 \): (6)};
    \node[anchor=west, font=\fontsize{8}{11}\selectfont] at (2.0, 3.70) {with \( \psi_1, \psi_2, \psi_3 \): (7)};
    \node[anchor=west, font=\fontsize{8}{11}\selectfont] at (2.0, 3.25) {Huang et al. : (8)};
    \node[anchor=west, font=\fontsize{8}{11}\selectfont] at (10.5, 3.65) {\(\psi_1\)};
    \node[anchor=west, font=\fontsize{8}{11}\selectfont] at (10.5, 3.20) {\(\psi_2\)};
    \node[anchor=west, font=\fontsize{8}{11}\selectfont] at (10.5, 2.75) {100 \(\psi_3\)};
    \node[anchor=west, font=\fontsize{13}{11}\selectfont, rotate=90] at (8.5, 3.5) {\( \psi \)};
  \end{tikzpicture}
  \caption{Influence of \(\psi_1\), \(\psi_2\), \(\psi_3\) for the case JXF-M1.5Re17000. (a) \( T^+_{SL} \) profile for different combinations of \(\psi_1\), \(\psi_2\), and \(\psi_3\). (b) Distributions of \(\psi_1\), \(\psi_2\), and \(\psi_3\). Each curve in (a) is labeled with a number. In the legend, the label "with \( \psi_i (i = 1, 2, 3) \)" indicates that the corresponding parameter is active. When any of \( \psi_i (i = 1, 2, 3) \) is inactive, the default values \( \psi_1 = 1.0 \), \( \psi_2 = 1.0 \), and \( \psi_3 = 0 \) are applied. Curve (8) represents the SL-type transformation of \citet{Huang2023}, given by Eq.~\eqref{eq:SL_Tplus_Huang2023}, where none of \(\psi_1\), \(\psi_2\), or \(\psi_3\) are active and all take their default values. The enhanced mixing length model \(l_m^E\), given by Eq.~\eqref{eq:lm_new}, is employed to compute \(\psi_1\) in curves (1), (4), (6), and (7). Black dashed lines: \(T^+_{LoW}\) from Eq.~\eqref{eq:IC_exLoW}.}
  \label{fig:f1f2f3_influence_wrles}
\end{figure*}

Fig.~\ref{fig:f1f2f3_influence_wrles}(a) presents the temperature profile under the SL-type transformation with different combinations of \(\psi_1\), \(\psi_2\), and \(\psi_3\) for the case JXF-M1.5Re17000. The profiles of \(\psi_1\), \(\psi_2\), and \(\psi_3\) are shown in Fig.~\ref{fig:f1f2f3_influence_wrles}(b). For clarity, one can directly follow the sequence \( (8)\rightarrow(1)\rightarrow(4)\rightarrow(7) \), which illustrates how \(T^+_{SL}\) transitions step by step from the transformation by \citet{Huang2023} to that of our new transformation by activating \(\psi_1\), \(\psi_2\), and \(\psi_3\). When all the three parameters are active, \(T^+_{SL}\) presents close agreement with the LoW, as shown in curve (7) and the black dashed line.

The influence of \(\psi_1\) is determined by comparing curve (5) with (7) in panel (a). Outside the buffer layer, curve (7) consistently presents a smaller slope and magnitude than curve (5). This behavior is attributed to the damping effect of \(\psi_1\). In the transformation of \citet{Huang2023}, \(\psi_1\) remains constant at 1.0 throughout the channel. However, as shown in panel (b), \(\psi_1\) in our transformation decreases from 1.0 at the wall to approximately 0.16 at the channel centerline, directly damping the transformation kernel and leading to a closer agreement with the reference. Similar behaviors are observed when comparing curves (2) with (4) and (3) with (6).

To illustrate the influence of \(\psi_2\), we examine curves (1), (3), (6), and (8), where \(\psi_2\) is inactive, resulting a  significant spike. This behavior is directly caused by the sign change in the denominator of Eqs.~\eqref{eq:VD_Tplus} and \eqref {eq:SL_Tplus}. To illustrate this, we focus on curve (8), which corresponds to the transformation by \citet{Huang2023}. Comparing Eqs.~\eqref{eq:log_layer_energy_balance} and \eqref{eq:log_layer_energy_balance_3}, we obtain:
\begin{equation}\label{eq:denominator_relation}
  \left(B_q + \psi_2 {(\gamma - 1)M_\tau^2 u^+} + \psi_3 \right)
  \propto 
  q_w - \left( \psi_2 \tau_w \tilde u - \overline{\rho v^{\prime\prime} \frac{1}{2}{u^{\prime\prime}_i u^{\prime\prime}_i}} \right).
\end{equation}

We define \(q_y = \psi_2 \tau_w \tilde u - \overline{\rho v^{\prime\prime} \frac{1}{2}{u^{\prime\prime}_i u^{\prime\prime}_i}} \). In our transformation, \(q_y\) includes the molecular and turbulent diffusion of kinetic energy across the \(y-\)plane, and the work of the external body force below it. However, in the transformation by \citet{Huang2023}, the total shear stress is assumed to be equal to the wall shear stress, and the effects of the body force and the turbulent TKE flux are neglected. As a result, the following relation holds:
\begin{equation}\label{eq:denominator_relation_Huang}
  \left(B_q + {(\gamma - 1)M_\tau^2 u^+}\right)
  \propto 
  \big ( q_w - \tau_w \tilde u \big ).
\end{equation}

This simplification indicates \(q_y = \tau_w \tilde u \). We point out that it is this simplification that directly leads to the peak point (\( y^*_p, T^+_{SL,p}\)) in curve (8). Here, \(y^*_p\) corresponds to the location where \(\tilde u = u_b\). Actually, for the density based body force, the overall energy balance of the whole channel satisfies \( q_w = \tau_w \, u_b \) \citep{Coleman1995}. For the volume based body force, \( q_w = \tau_w \, u_m \), where \(u_m\) is the mean velocity without density weighting. When the Mach number is not very high, \(u_b \approx u_m\). Consider the transformation of \citet{Huang2023}, applying \(q_y = \tau_w \, \tilde u \) results in \( \left | q_y \right | < \left | q_w \right | \) for \( y < y_p \), since \( \tilde u < u_b \), and \( \left | q_y \right | > \left | q_w \right |\) for \( y > y_p \), because \( \tilde u > u_b \). As a result, the sign of the denominator changes across this location, leading to the peak value \(T^+_{SL,p}\) in curve (8). The same reasoning applies to curves (1), (3), and (6).

As for the influence of \(\psi_3\), similar but significantly more narrow spikes are also observed in curves (2) and (4), where \(\psi_2\) is active but \(\psi_3\) is not. This behavior arises for the same reason as the spike caused by \(\psi_2\). Since the magnitude of \(\psi_2 {(\gamma - 1) M_\tau^2 u^+}\) is  typically much larger than \(\psi_3\), the spike caused by \(\psi_2\) generally affects a greater proportion of the channel, whereas the spike caused by \(\psi_3\) is considerably delayed and confined to a narrow region near the channel center. 

When both \(\psi_2\) and \(\psi_3\) are active, the spike is effectively eliminated, as shown in curves (5) and (7). The remaining small spikes result from the neglect of other high-order terms, which are not considered here. From Fig.~\ref{fig:f1f2f3_influence_wrles}(b), we observe that \(\psi_2\) decreases from \( 1 \) at the wall to approximately \( 0.85 \) at the channel centerline, effectively damping the second term \(\psi_2 {(\gamma - 1) M_\tau^2 u^+}\). In the outer layer, \(\psi_3\) is negative and has the same sign as \(B_q\), mitigating the sign change and providing an equivalent damping effect to the entire transformation kernel. Consequently, the overall behavior of \(T^+_{SL}\) stabilizes, and its distribution gets closer to the reference profile.

Near the channel center, the magnitude of both sides of Eq.~\eqref{eq:denominator_relation} decreases. Even a small energy imbalance can trigger the sign change in the denominator, consequently leading to a spike. This behavior is significantly different from that in velocity transformation, which also explains why developing an effective temperature transformation is more challenging than a velocity transformation.

A similar analysis can be applied to the SL-type transformation with different \(l_m\), the VD-type transformation, and the mixed isothermal/adiabatic configuration (results are not shown here for simplicity). Compared to the classical isothermal wall configuration, there are two distinct characteristics in mixed thermal condition. First, the thermal energy is removed exclusively from the isothermal wall side, making \( q_w \) approximately twice as large as that in the classical configuration. However, the shear stress and velocity in the channel do not increase proportionally. Consequently, the sign of denominator in the transformation of \citet{Huang2023} remains unchanged and no spikes are observed. Second, as shown in Fig.~\ref{fig:energy_budget}(b, c), the turbulent heat conduction component \(q^t_T\) makes a  significantly larger contribution to the total energy balance, and it does not decrease to zero near the channel center, in contrast to the classical configuration. As introduced in Sec.~\ref{sec:temperature_transformation}, \(q^t_k\) is significantly smaller than \(q^t_T\) under this configuration. Therefore, it is expected that \(\psi_3\) can be neglected accordingly. In addition, \(q_f\) is also smaller than \(q^t_T\) in the majority of the channel, suggesting a less significant role of \(\psi_2\). 

\begin{figure*}
  \begin{tikzpicture}
    \node[anchor=south west, inner sep=0] (image) at (0,0) {\includegraphics[width=\linewidth]{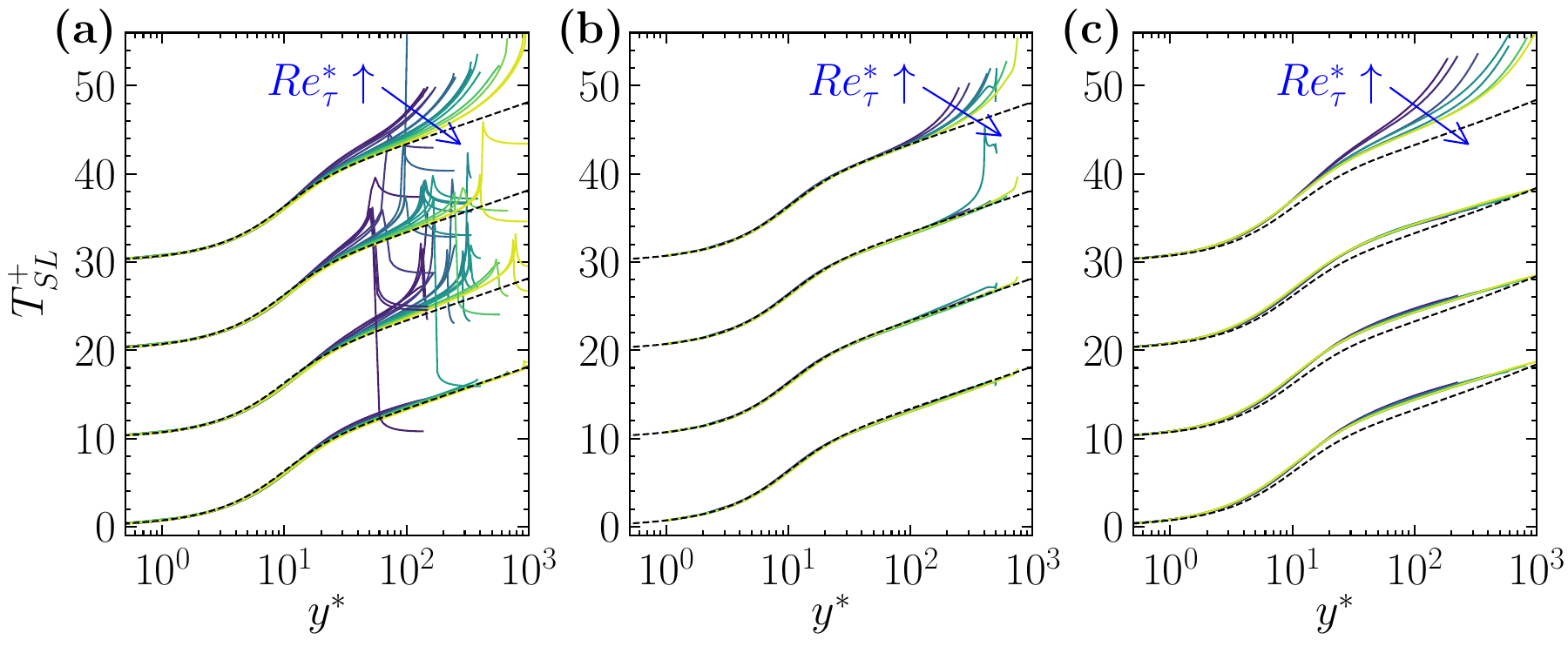}};
    \node[anchor=west, font=\fontsize{10}{11}\selectfont, rotate=22] at (1.4, 1.50) {\(\psi_1,\psi_2,\psi_3\)};
    \node[anchor=west, font=\fontsize{10}{11}\selectfont, rotate=22] at (1.4, 2.40) {\(\psi_1,\psi_2\)};
    \node[anchor=west, font=\fontsize{10}{11}\selectfont, rotate=22] at (1.4, 3.40) {\(\psi_1,\psi_3\)};
    \node[anchor=west, font=\fontsize{10}{11}\selectfont, rotate=22] at (1.4, 4.30) {\(\psi_2,\psi_3\)};
  \end{tikzpicture}
  \caption{Transformed temperature profiles in the absence of \(\psi_1\), \(\psi_2\), and \(\psi_3\) under SL-type transformation. (a) The classical isothermal wall configuration (19 cases listed in table~\ref{table:DNS_GV2024} and \ref{table:WRLES_JAXFluids}). (b) Isothermal wall side for the mixed thermal configuration (7 cases listed in table~\ref{table:DNS_LC2022_isothermal_wall}). (c) Adiabatic wall side (7 cases listed in table~\ref{table:DNS_LC2022_adiabatic_wall}). The first group of curves in the bottom of each panel indicates the complete transformation with all of \(\psi_1\), \(\psi_2\), and \(\psi_3\) activated. The other three groups of curves correspond to neglecting one of the three parameters, and are shifted upward by multiples of 10 units. Black dashed lines: \(T^+_{LoW}\) from Eq.~\eqref{eq:IC_exLoW}. }
\label{fig:SL_Tplus_f1f2f3_influence}
\end{figure*}

The above analysis can be validated through Fig.~\ref{fig:SL_Tplus_f1f2f3_influence}, which presents the transformed temperature profiles in the absence of \(\psi_1\), \(\psi_2\), and \(\psi_3\) under SL-type transformation using \(l_m^E\). The first group of curves in the bottom of each panel correspond to the complete transformation with all three parameters activated. As shown in panel (a), for the classical isothermal wall configuration, the exclusion of any of \(\psi_1\), \(\psi_2\), and \(\psi_3\) makes significant difference, suggesting that they are all important. Particularly, excluding either \(\psi_2\) or \(\psi_3\) would result in the kinks. In mixed thermal configuration, \(\psi_1\) is the most important parameter for both isothermal and adiabatic wall sides, while \(\psi_2\) and \(\psi_3\) make insignificant difference. However, this does not imply that the body force is unimportant in turbulent channel flow.

\subsection{Modeling the turbulent TKE flux}\label{sec:modeling_turbulent_tke_flux}
Analogous to the complete-form transformations by \citet{Chen2022} and \citet{Cheng2024b}, the inclusion of the turbulent TKE flux, \(q^t_k = -\overline{\rho v^{\prime\prime} \frac{1}{2}{u^{\prime\prime}_i u^{\prime\prime}_i}} \), adds complexity to the proposed transformation, which is a weakness. A practical challenge also arises because many open-source DNS databases do not provide this high-order statistics, making the validation of our transformations difficult on such datasets.

As demonstrated in Fig.~\ref{fig:SL_Tplus_f1f2f3_influence}, the turbulent TKE flux can be safely neglected in mixed thermal wall configuration, while should be retained in classical isothermal wall setup. To eliminate the dependence on \(q^t_k\), two approaches have been applied in previous studies. \citet{Huang2023} and \citet{Cheng2024b} used \(q_y = \tau_w \tilde u \) based on the constant stress assumption. \citet{Chen2022} applied \(q_y = \tau_{xy} \tilde u \) to derive a simplified version of their transformation, where \( \tau_{xy} = (\bar\mu + \bar\mu_t) d{\tilde u}/dy \) represents the total shear stress. Both methods neglect the influence of the external body force. In the present study, we introduce a third approach that accounts for the turbulent TKE flux \(q^t_k\) rather than \(q_y\).

\begin{figure*}[hb]
  \includegraphics[width=\linewidth]{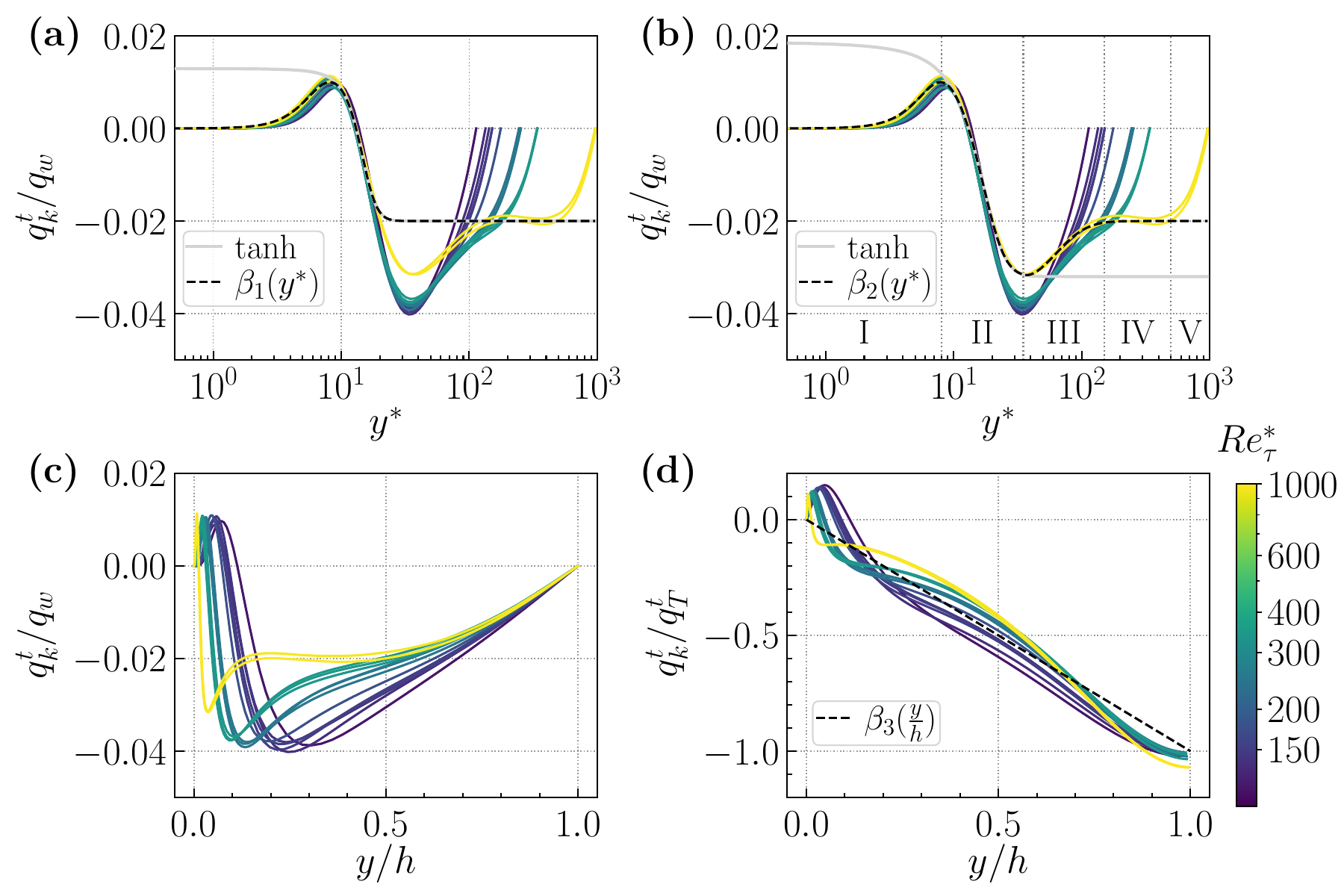}
  \caption{Distributions of the turbulent TKE flux in compressible turbulent channel flow with the classical isothermal configuration. Panels (a, b) \(q^t_k/q_w\) versus \( y^* \), (c) \(q^t_k/q_w\) versus \( y/h \), and (d) \(q^t_k/q^t_T\) versus \( y/h \). DNS data listed in Table~\ref{table:DNS_GV2024} are used. I-V in (b) indicate the multi-layer structure of \(q^t_k/q_w\) at high Reynolds numbers.}
\label{fig:q_vTKE_profile}
\end{figure*}

Fig.~\ref{fig:q_vTKE_profile} shows the distribution of the turbulent TKE flux in both inner and outer coordinates. Panels (a) and (b) depict \(q^t_k/q_w\) as a function of \(y^*\). The distribution is mainly influenced by \(Re^*_\tau\). At high Reynolds numbers, \(q^t_k/q_w\) exhibits a pronounced multi-layer structure, denoted as \textbf{I - V} in panel (b).

\textbf{Layer I} spans approximately \(0 < y^* < 8\), where the turbulent motion is strongly suppressed by the wall. At \(y^* \approx 8\), \(q^t_k/q_w\) reaches a maximum of about 0.012. \textbf{Layer II} covers \(8 < y^* < 35\), where \(q^t_k/q_w\) decreases continuously until reaching a minimum at \(y^* \approx 35\). This minimum value exhibits a moderate reduction at lower \(Re^*_\tau\). Within this layer, the turbulent TKE flux changes direction at \(y^*_k \approx 13.5\): toward the wall for \(y^* < y^*_k\) and towards the channel centerline for \(y^* > y^*_k\). In \textbf{Layer III}, \(q^t_k/q_w\) recovers, approaching an asymptotical plateau at \(y^* \approx 150\). \textbf{Layer IV} maintains an approximately constant \(q^t_k/q_w\), and its lower and upper bound may expand slightly with increasing \(Re^*_\tau\). For the cases GV2024-MCLx1p50 and GV2024-MCLx0p81, the outer bound is around \(y^* = 500\), approximately \(50\%\) of the corresponding \(Re^*_\tau\). \textbf{Layer V} represents the wake region, where the distribution of \(q^t_k/q_w\) is not well collapsed in inner coordinate. However, panel (c) suggests that an outer coordinate may be more appropriate. This observation is consistent with the study of \citet{She2017}, although their focus was on the mixing length distribution. 

Based on the above analysis of panels (a), (b), and (c), we proposed the following model for \(q^t_k/q_w\). The basic idea is to use a base model that captures the overall distribution, with additional corrections applied to improve the accuracy in different layers. Here, we employ a hyperbolic tangent function:
\begin{equation}
  F_k(y^*; F_{\min}, F_{\max}, A_1, A_2) =
  -\frac{F_{\max} - F_{\min}}{2}\,
  \tanh\!\left[A_1\,(y^* - A_2)\right]
  + \frac{F_{\max} + F_{\min}}{2}
\end{equation}

Here, \(F_{\min}\) and \(F_{\max}\) denote the minimum and maximum values of the tanh function, which may differ slightly from the lower and upper bounds of \(q^t_k/q_w\). The constants \(A_1\) and \(A_2\) control the steepness and the position of the transition, respectively. The thick gray lines in panels (a) and (b) illustrate two examples of this base distribution.

Building upon the base distribution, we introduce two models for high-Reynolds-number flows. Note that \textit{ad hoc} procedures are required to determine the values of free parameters. The black dashed line in panel (a) demonstrates the first type, which is expressed as:
\begin{equation}\label{eq:qvTKE_beta1}
  \begin{gathered}
  \frac{q^t_k}{q_w} = \beta_{1}(y^*) = F_{k}\left (y^*; -0.02, 0.013, 0.23, 14 \right ) \, D_1, \\[1mm]
  D_1 = 1 - \exp\!\left[- \left(y^*/6\right)^3\right].
  \end{gathered}
\end{equation}

Here, a damping function \(D_1\) is applied to improve the representation in \textbf{Layer I}. The asymptotical distribution in \textbf{Layer IV} is maintained through imposing \(F_{\min} = -0.02\). However, distribution in the outer part of \textbf{Layer II} and in \textbf{Layer III} is overpredicted.

To address this limitation, we introduce the second type model that captures the V-shaped distribution, as indicated by the black dashed line in panel (b). The model is expressed as:
\begin{equation}
  \begin{gathered}\label{eq:qvTKE_beta2}
  \frac{q^t_k}{q_w} = \beta_{2}(y^*) = F_{k}(y^*; -0.032, 0.02, 0.12, 15) \, D_2 + H(y^* - 35) \, B(y^*), \\
  D_2 = 1 - \exp\!\left[- \left(y^*/6.5\right)^3\right], \,\,\,
  B(y^*) = 0.012 \, \left[1 - \exp\!\left(- (y^* - 35)/27\right) \right]^2.
  \end{gathered}
\end{equation}

Here, we employs \(F_{\min} = -0.032\) for the base tanh function to capture the valley. A correction term \(B(y^*)\) is included to recover the asymptotic value in \textbf{Layer IV}. \(H(y^* - 35)\) is the Heaviside step function, which activates the correction only for \(y^* > 35\), corresponding to \textbf{Layers III-IV}.

One may introduce additional model to describe the valley as a function of \(Re^*_\tau\). Additionally, the distribution in the wake region can be further corrected based on \(\beta_{2}(y^*)\) by employing the outer coordinate \(y/h\). Here, we introduce another model that is entirely formulated in terms of the outer coordinate. In Sec.~\ref{sec:temperature_transformation}, we point out that the turbulent diffusion of thermal energy (\(q^t_T\)) should serve as the basis when evaluating the relative importance of each term in Eq.~\eqref{eq:global_energy_balance_2} within the overlap layer. This motivates the normalization by \(q^t_T\) shown in Fig.~\ref{fig:q_vTKE_profile}(d). The black dashed line exhibits good overall collapse in the outer layer. The model can be approximately expressed as:
\begin{equation}\label{eq:qvTKE_beta3}
  \frac{q^t_k}{q^t_T} = \beta_3\left(\frac{y}{h}\right) \approx -\frac{y}{h}.
\end{equation}

The model \(\beta_1\) is expected to perform well at high Reynolds numbers, while \(\beta_2\) captures the V-shaped distribution. Compared with \(\beta_1\) and \(\beta_2\), the expression of \(\beta_3\) is simpler. However, it does not capture the positive peak in the buffer layer.

\subsection{Derivation and application of simplified temperature transformations}\label{sec:derivation_and_application_simplified_transformations}
Using the above models for the turbulent TKE flux, simplified temperature transformations can be obtained. By invoking \(\psi_{3,\beta_1} = -\beta_1 B_q\) and \(\psi_{3,\beta_2} = -\beta_2 B_q\) in Eq.~\ref{eq:energy_f1f2f3}, we obtain the simplified SL-type transformations:
\begin{equation}\label{eq:SL_Tplus_modeled_1}
  T^+_{SL,\beta_1} = \int_0^{\theta^+} 
  \frac{\psi_1}{(1 - \beta_1) B_q + \psi_2 {(\gamma - 1)M_\tau^2 u^+}}
  \sqrt{\rho^+}
  \left(1 + \frac{1}{2}\frac{y^+}{\rho^+}\frac{d\rho^+}{dy^+} - \frac{y^+}{\mu^+}\frac{d\mu^+}{dy^+}\right)
  {d\theta^+}.
\end{equation}
\begin{equation}\label{eq:SL_Tplus_modeled_2}
  T^+_{SL,\beta_2} = \int_0^{\theta^+} 
  \frac{\psi_1}{(1 - \beta_2) B_q + \psi_2 {(\gamma - 1)M_\tau^2 u^+}}
  \sqrt{\rho^+}
  \left(1 + \frac{1}{2}\frac{y^+}{\rho^+}\frac{d\rho^+}{dy^+} - \frac{y^+}{\mu^+}\frac{d\mu^+}{dy^+}\right)
  {d\theta^+}.
\end{equation}

Similarly, substituting \(\beta_3\) into Eq.~\eqref{eq:log_layer_energy_balance} and following the same approach as in Sec.~\ref{sec:temperature_transformation} yields:
\begin{equation}\label{eq:SL_Tplus_modeled_3}
  T^+_{SL,\beta_3} = \int_0^{\theta^+} 
  \frac{\psi_1 (1 + \beta_3)}{B_q + \psi_2 {(\gamma - 1)M_\tau^2 u^+}}
  \sqrt{\rho^+}
  \left(1 + \frac{1}{2}\frac{y^+}{\rho^+}\frac{d\rho^+}{dy^+} - \frac{y^+}{\mu^+}\frac{d\mu^+}{dy^+}\right)
  {d\theta^+}.
\end{equation}

The corresponding simplified VD-type transformations can be derived following a similar approach. However, they are not presented here for simplicity. Compared to the complete forms in Eqs.~\eqref{eq:VD_Tplus} and \eqref{eq:SL_Tplus}, the effect of \(\psi_3\) is captured through (\(1 - \beta_1\)) and (\(1 - \beta_2\)) in Eqs.~\eqref{eq:SL_Tplus_modeled_1} and \eqref{eq:SL_Tplus_modeled_2}, and through (\(1 + \beta_3\)) in Eq.~\eqref{eq:SL_Tplus_modeled_3}. These modifications remove the direct reliance on the high-order turbulent TKE flux term while still retaining its effects.

\begin{figure*}
  \includegraphics[width=\linewidth]{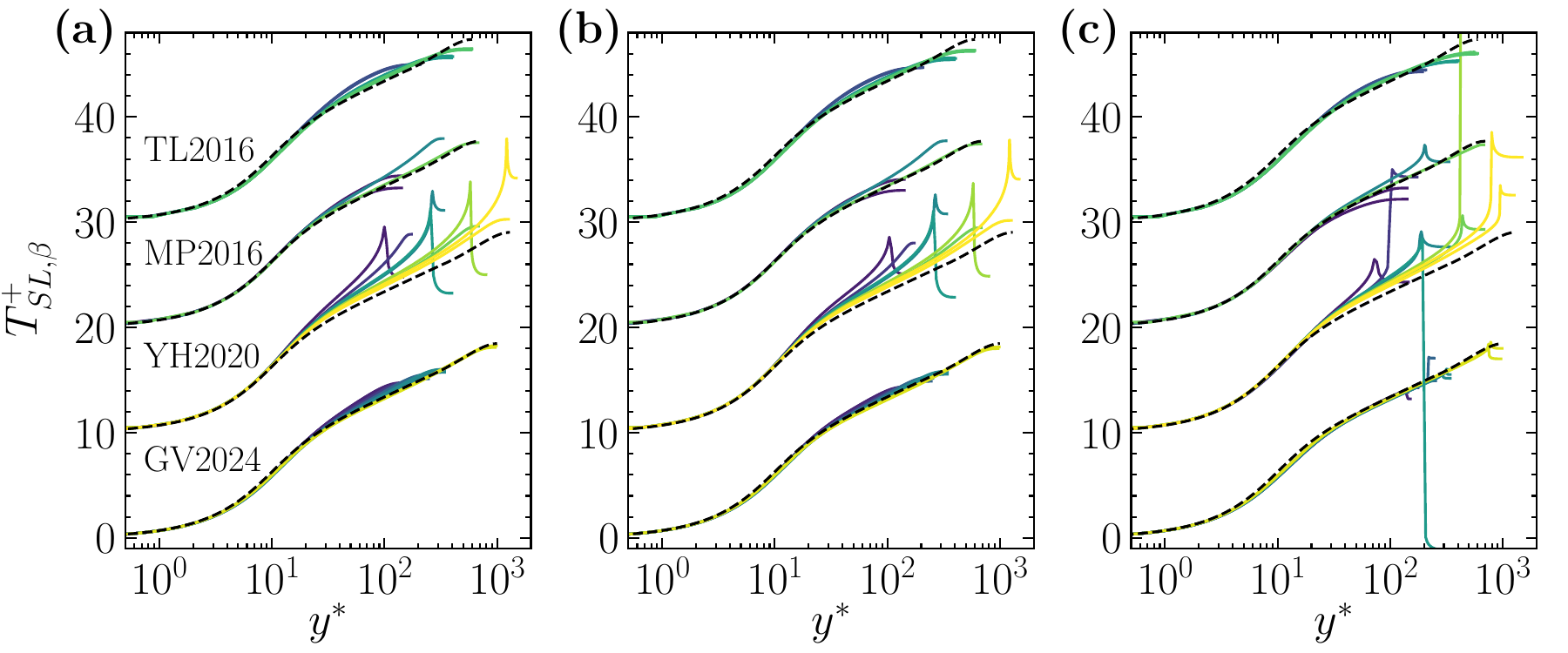}
  \caption{Transformed temperature profiles obtained using the simplified SL-type transformation: (a) \(T^+_{SL,\beta_1}\), (b) \(T^+_{SL,\beta_2}\), and (c) \(T^+_{SL,\beta_3}\). DNS data from \citet{Trettel2016}, \citet{Modesti2016}, \citet{Yao2020}, and \citet{Gerolymos2023, Gerolymos2024a, Gerolymos2024b} are employed and are labeled as "TL2016", "MP2016", "YH2020", and "GV2024", respectively. The parabolic mixing length model \(l_m^P\) is applied to compute \(\psi_1\). For clarity, the profiles are shifted upward by multiple 10 units. Black dashed lines denote the analytical incompressible profile \(T^+_{IC}\) corresponding to one representative case from each dataset.}
\label{fig:Tplus_beta_lmP}
\end{figure*}

\begin{figure*}
  \includegraphics[width=\linewidth]{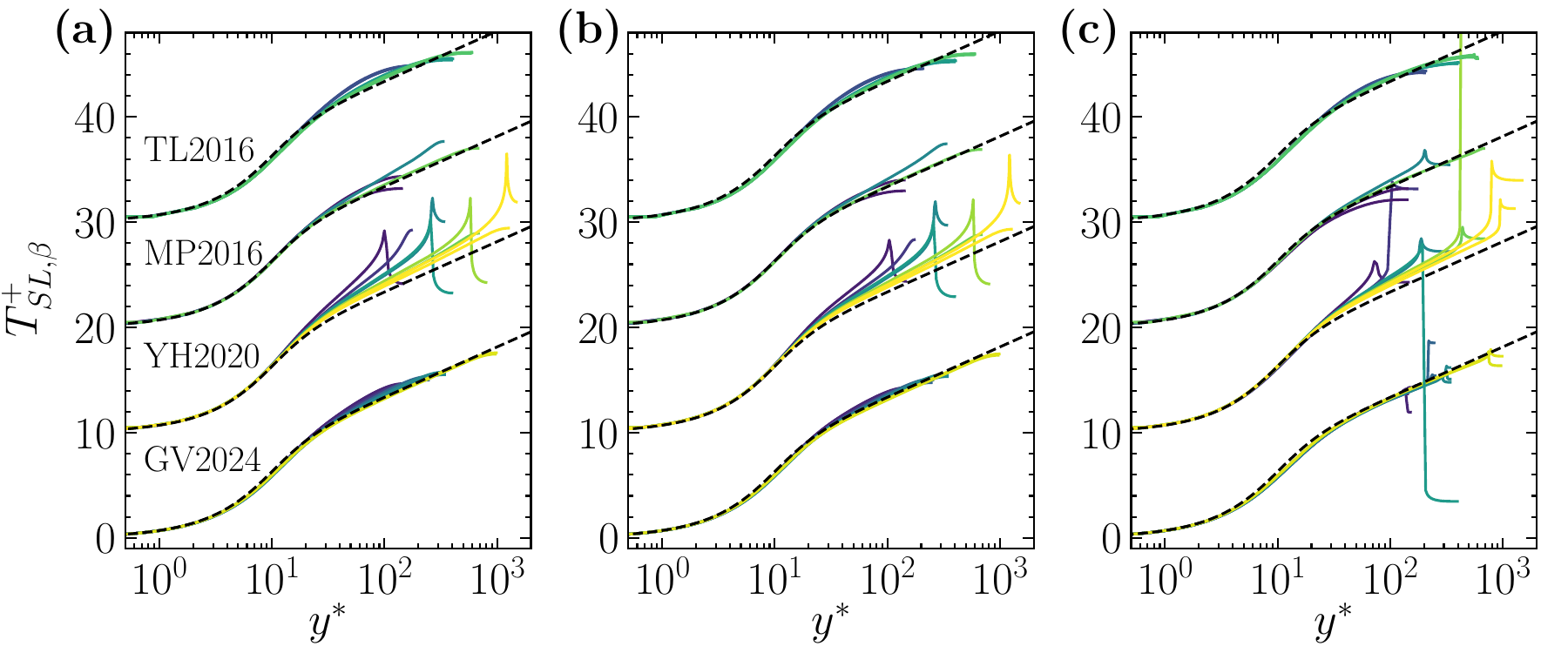}
  \caption{The same to Fig.~\ref{fig:Tplus_beta_lmP}, except that the enhanced mixing length model \(l_m^E\) is used to compute \(\psi_1\), and the black dashed lines represent \(T^+_{LoW}\) from Eq.~\eqref{eq:IC_exLoW}.}
\label{fig:Tplus_beta_lmE}
\end{figure*}

Since the publicly available DNS datasets of \citet{Trettel2016}, \citet{Modesti2016}, and \citet{Yao2020} do not include the turbulent TKE flux, they cannot be used to validate the transformations in Eqs.~\eqref{eq:VD_Tplus} and \eqref{eq:SL_Tplus}. Nevertheless, these datasets are suitable for validating the simplified transformations. Figs.~\ref{fig:Tplus_beta_lmP} and \ref{fig:Tplus_beta_lmE} present the transformed temperature profiles using \(l_m^P\) and \(l_m^E\), respectively. Results from the DNS of \citet{Gerolymos2023, Gerolymos2024a, Gerolymos2024b} are also included.

As shown, all three simplified SL-type transformations perform fairly well, especially at high Reynolds numbers. For most of the boundary layers, \(T^+_{SL,\beta}\) agrees closely with the analytical incompressible profile \(T^+_{IC}\) and \(T^+_{LoW}\) when using \(l_m^P\) and \(l_m^E\), respectively. However, both \(T^+_{SL,\beta_1}\) and \(T^+_{SL,\beta_2}\) are slightly underpredicted in the wake region. This is directly caused by the underprediction of \(\beta\) in this region, as shown in Fig.~\ref{fig:q_vTKE_profile} (a, b). Using \(\beta_3\) introduces slight kinks in the datasets "GV2024" and "MP2016". As explained in Sec.~\ref{sec:influence_of_f1f2f3}, this is due to the energy imbalance caused by removal of \(\psi_3\) from the denominator. However, including \((1 + \beta_3)\) in the numerator significantly mitigates these kinks.

An exception is the dataset of \citet{Yao2020}, where noticeable discrepancies and kink appear. We note that the temperature profiles (\(\tilde T/\tilde T_w\)) from \citet{Yao2020} are slightly higher than those from \citet{Modesti2016} at the same Mach and Reynolds numbers. This is connected to the overprediction of transformed temperature in Figs.~\ref{fig:Tplus_beta_lmP} and \ref{fig:Tplus_beta_lmE}. Additionally, it may also be responsible for the pronounced kinks.

\begin{figure*}
  \includegraphics[width=\linewidth]{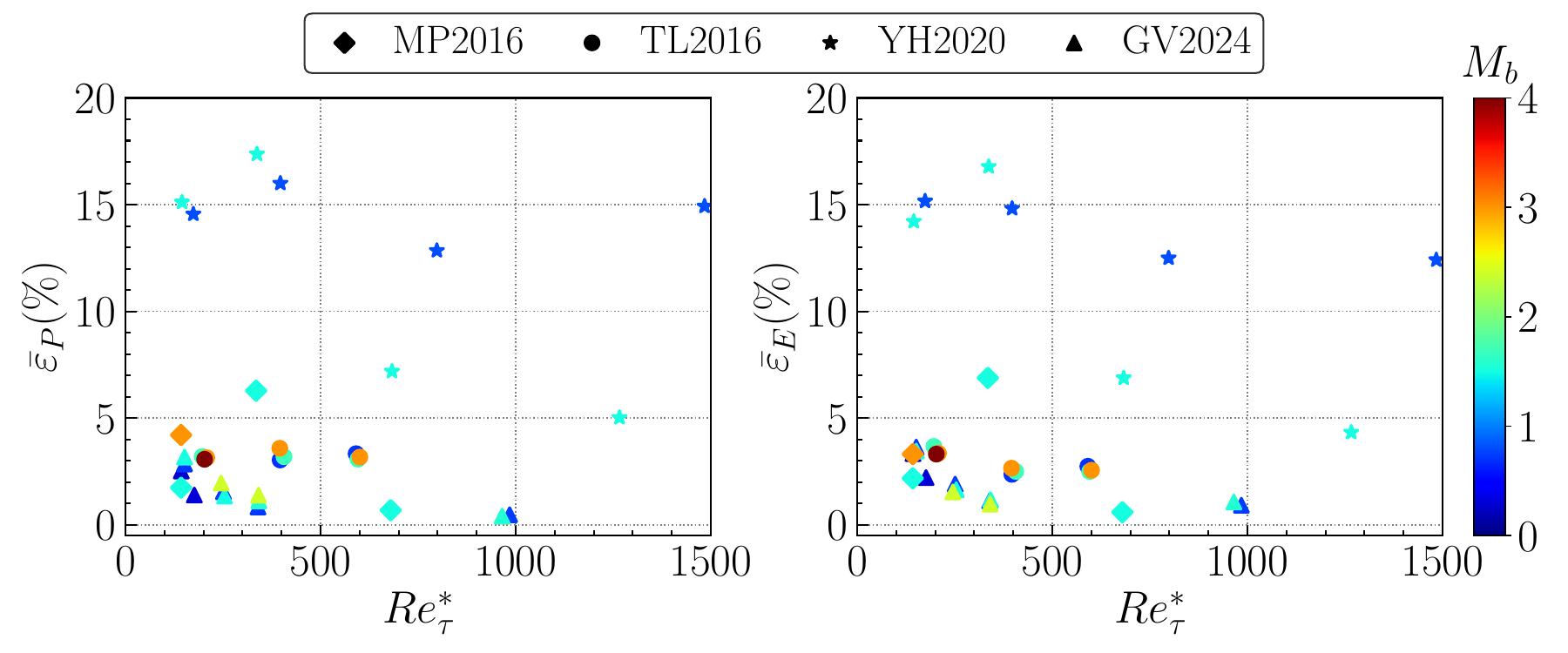}
  \caption{Integral mean error of \(T^+_{SL}\) over the entire half-channel height using \(\beta_2\). Panels show (a) \(l_m^P\) and (b) \(l_m^E\). Symbols denote different DNS datasets \citep{Modesti2016,Trettel2016,Yao2020,Gerolymos2024a}.}
\label{fig:integral_mean_error_model_v2}
\end{figure*}

Fig.~\ref{fig:integral_mean_error_model_v2} presents the integral mean error computed from Eq.~\eqref{eq:mean_error_UT} over the entire half-channel height. Only results obtained with the \(\beta_2\) model are shown. Except for the DNS dataset of \citet{Yao2020}, both \(\varepsilon_P\) and \(\varepsilon_E\) remain below \(4\%\) for most of the tested cases, with r.m.s. values of \(2.77\%\) and \(2.78\%\) for \(\varepsilon_P\) and \(\varepsilon_E\), respectively. These errors are slightly higher than those obtained using the complete transformations shown in Fig.~\ref{fig:integral_mean_error}.

For the mixed thermal configuration, there is no significant difference among the complete transformation, the simplified transformation, and the transformation with \(\psi_3 = 0\) for both isothermal and adiabatic sides (results are not shown here). In addition, the model for the turbulent TKE flux are also used to derive the simplified transformation for compressible turbulent pipe flows, as given in  Eqs.~\eqref{eq:SL_Tplus_pipe_modeled_1} - \eqref{eq:SL_Tplus_pipe_modeled_3} in Appendix \ref{sec:app_A}.

To summarize, the simplified transformations given in Eqs.~\eqref{eq:SL_Tplus_modeled_1} to \eqref{eq:SL_Tplus_modeled_3} apply to both isothermal and adiabatic walls. For mixed thermal configuration, \(\psi_3 = 0\) can also be employed, equivalent to \(\beta = 0\). The above results demonstrate the effectiveness of the proposed approximations for the turbulent TKE flux and the resulting simplified temperature transformations. 

\section{Potential Applications and Limitations}\label{sec:application_and_limitation}
\subsection{Applications to near-wall modeling}\label{sec:application}
Over the years, various approaches have been developed for near-wall modeling of turbulent flows \citep{Larsson2016, Bose2018}. However, many of these methods encounter limitations when applied to compressible turbulent boundary layers. Knowledge from compressibility transformations, including both velocity and temperature transformations, can be leveraged to improve the performance of existing near-wall modeling techniques for compressible flows or to extend models originally developed for incompressible flows to compressible ones.

For example, in the studies by \citet{Hendrickson2022, Hendrickson2023}, the incompressible eddy viscosity is corrected using the velocity transformation kernel to model the compressible eddy viscosity. Following this idea, the temperature transformation could similarly be implemented to improve the modeling of compressible thermal eddy diffusivity. In the study of \citet{Chen2025}, the temperature transformation proposed by \citet{Cheng2024b} is applied to remove the reliance on boundary-layer-edge quantities in the temperature-velocity relation. Additionally, \citet{Modesti2024} demonstrated that the velocity and temperature transformations can be applied inversely to reconstruct compressible mean profiles from their incompressible counterparts. Although their study focuses on low-Mach-number flows with variable properties, the same approach can be explored in compressible flows. Furthermore, wall models directly based on the log-law have been applied to incompressible wall modeling, such as the algebraic wall models \citep{Schumann1975, Groetzbach1987, Piomelli1989, Piomelli-Balaras} and the control-based approaches \citep{Nicoud2001, Templeton2006}. It will be of great significance to develop wall models for compressible flows based on the log-law of transformed velocity and temperature profiles.

Based on the above analysis, the potential applications of the proposed temperature transformation are briefly discussed below.

First, in the case of large eddy simulation, the high-order term may be partially resolved. Accordingly, the transformation can be readily employed for wall-modelled large eddy simulation following the control-based approach proposed by \citet{Nicoud2001}. The main difference is that both velocity and temperature transformations, along with a more advanced control strategy, are required for compressible flows. The simplified transformations may also be used for this purpose.

Second, the proposed transformation can be applied inversely to construct the mean profiles and key flow quantities of compressible flows, following a similar approach as used by \citet{Modesti2024}. More specifically, given the \(M_b\) and \(Re_b\) of the turbulent channel flow, the inverse transformation outputs the mean profiles of velocity, temperature, density, as well as key quantities including \(Re_\tau\), \(Re^*_\tau\), \(C_f\), \(B_q\), and \(\tilde T_c/\tilde T_w\). In this application, reference "incompressible" velocity and temperature profiles should be prescribed. For example, the analytical incompressible profiles for velocity \citep{Nagib2008, She2017} and temperature \citep{Pirozzoli2023,Pirozzoli2024} can be used. A critical issue is the treatment of the high-order term in the transformation, which is unknown \emph{a priori}. In this regard, the simplified transformations given in Eqs.~\eqref{eq:SL_Tplus_modeled_1}, \eqref{eq:SL_Tplus_modeled_2}, and \eqref{eq:SL_Tplus_modeled_3} can be readily applied. In addition, it may be practical to set \(\psi_3 \approx 0\) in the inverse transformation. Despite the observed discrepancies in Fig.~\ref{fig:f1f2f3_influence_wrles}, \ref{fig:Tplus_beta_lmP}, and \ref{fig:Tplus_beta_lmE} associated with these simplifications, our tests suggest that the predicted mean profiles and key quantities show reasonable agreement with the DNS results. Implementation details are outside the scope of the present study and will be reported in a separate manuscript.

Third, the transformation can also be incorporated in the classical wall-stress models in LES following the approach of \citet{Griffin2023}. As shown in Fig.~\ref{fig:SL_Tplus_f1f2f3_influence}, when \(\psi_3 \approx 0\) is applied, the transformed temperature profiles still approximately agree with the LoW at high Reynolds numbers within the inner layer. In such cases, \(\psi_3 \approx 0\) might be reasonable when applied inversely in the wall model \citep{Griffin2023}. Moreover, as shown in Fig.~\ref{fig:Tplus_beta_lmP} and \ref{fig:Tplus_beta_lmE}, the simplified transformations demonstrate reasonable agreement with the incompressible distribution within the typical wall-modelled layer. Therefore, the simplified transformations may also be applied. Applications of these approaches in wall-modeled LES will be the subject of future investigation.

\subsection{Limitation and future direction}\label{sec:limitation_and_future_direction}
The proposed temperature transformation shows good performance for compressible turbulent channel flows. Its application in compressible turbulent pipe flows is provided in Appendix~\ref{sec:app_A}. Despite these success, further improvements are possible.

First, the approximate models \(\beta_1\), \(\beta_2\), and \(\beta_3\) given in Eqs.~\eqref{eq:qvTKE_beta1}, \eqref{eq:qvTKE_beta2}, and \eqref{eq:qvTKE_beta3} are only phenomenological approximations. \textit{ad hoc} procedures are involved to specify the values of free parameters. A more physically fundamental analysis is required to further improve the models. Additionally, the multi-layer structure of \(q^t_k/q_w\) is not fully aligned with the classical boundary layer structure of the mean velocity.

Second, similar to the transformations by \citet{Chen2022} and \citet{Huang2023}, the transformed temperature above the adiabatic wall is overpredicted. This may be due to an overprediction of \(l_m\) in the near wall region. 

Third, the derivation is entirely based on the momentum and energy balance equations in the overlap layer. The linear law is realized through imposing additional restrictions on \(\psi_1\) (or \(l_m\)). This may introduce potential physical inconsistency, as it requires the \(l_m\) to remain undamped, in contrast with the common practice of using a damping function in the viscous sublayer, such as in previous transformations \citep{Hasan2023, Huang2023, Modesti2024} and near-wall modeling \citep{Larsson2016, Yang2018, Chen2022b}. Actually, a \(l_m \approx \kappa \, y\) effectively removes the explicit influence of \(l_m\) on \(\psi_1\) in the viscous sublayer, leaving only \(\tau^+_{tot}\). Consequently, \(\psi_1\) acts as a modulating factor that explicitly accounts for the turbulent diffusion in the overlap layer and implicitly account for the viscous effects within the viscous sublayer. In addition, similar to the velocity transformation, a fundamental requirement for the temperature transformation is to maintain \(dT^+_I/dY^+_I = Pr \, q_y/q_w \approx Pr\) within the viscous sublayer. Here, \(Y^+_I\) and \(T^+_I\) denote the transformed "incompressible" coordinate and temperature, and \(q_y\) is the local heat flux. According to Eq.~\eqref{eq:sublayer_Tplus}, the SL-type transformation gives \(dT^+_{SL}/dy^* = Pr \, \psi_1\). Using \(l_m \approx \kappa \, y\) ensures \(\psi_1 \approx 1\), and thus \(dT^+_{SL}/dy^* \approx Pr\). One may also follow the approach of \citet{Hasan2023}, \citet{Modesti2024}, \citet{Huang2023}, and more recently \citet{Zhu2025c} to include both molecular and turbulent diffusions in the viscous sublayer. Nevertheless, in the final transformations of \citet{Huang2023}, the molecular and turbulent Prandtl numbers are not included, making their transformations a special case of Eqs.~\eqref{eq:VD_Tplus} and ~\eqref{eq:SL_Tplus} with \(\psi_1 = 1\), \(\psi_2 = 1\), and \(\psi_3 = 0\). In our transformations, introducing \(\psi_1\), \(\psi_2\), and \(\psi_3\) accounts for the influence of the mixing length model, the work of body fore, and the turbulent TKE flux in the outer layer. More importantly, it provides the flexibility to develop more advanced transformations in the future.

Lastly, the proposed approach is mainly designed for compressible turbulent channel flow. The application to compressible turbulent pipe flows is illustrated in Appendix~\ref{sec:app_A}. The proposed approach may be extended to more general configurations that exhibit similar inner layer structure. However, such an extension would require more advanced models for \(\psi_1\), \(\psi_2\), and \(\psi_3\). 

For example, in zero-pressure-gradient boundary layer flows, the constant stress assumption approximately holds in the inner layer, indicating \(\tau^+_{tot} \approx 1\) in Eq.~\eqref{eq:params_f1f2f3}. One may also use \(\tau^+_{tot} = 1 - (y/\delta_e)^{1.5}\) \citep{Sun1973, Chen2016} for the entire boundary layer, where \(\delta_e\) denotes the boundary layer thickness. At high Reynolds numbers, the commonly used linear model given in Eq.~\eqref{eq:lm_linear} can be readily applied. Nevertheless, we retain the flexibility to employ alternative models for \(\psi_1\) when necessary. The influence of the body force term may potentially be neglected. Regarding the turbulent TKE flux in \(\psi_3\), it remains to be validated whether the proposed models \(\beta_1\), \(\beta_2\), and \(\beta_3\) still hold in boundary layers. The influence of other high-order terms should also be carefully evaluated for different thermal wall conditions. Another challenge arises from the non-monotonic temperature profile in the cold wall configuration of compressible turbulent boundary layer flows. A local maximum appears in the buffer layer \citep{Zhang2018, Szajnecki2025}, with \(d\theta^+ < 0\) below this position and \(d\theta^+ > 0\) above it. Our tests indicate that including only the turbulent TKE flux is not sufficient as the denominator \(\left(B_q + \psi_2 {(\gamma - 1)M_\tau^2 u^+} + \psi_3 \right)\) still changes sign within the buffer layer due to energy imbalance. More importantly, the location of this sign change does not coincide with that of the maximum temperature. This mismatch causes the transformed temperature to exhibit a non-monotonic profile in the buffer layer. In this regard, it is considerably more challenging to develope the temperature transformation than in the turbulent channel flow. 

Addressing the above limitations would significantly improve the temperature transformation. However, this is beyond the scope of the present study and is left for future research.

\section{Conclusion}\label{sec:Conclusion}
Compressible temperature transformations have received increasing attention in recent years. In this study, we propose new VD-type and SL-type temperature transformations for compressible turbulent channel flow. Three parameters (\(\psi_1\), \(\psi_2\), and \(\psi_3\)) are introduced to account for the effects of the mixing length model, the work of body force, and the turbulent TKE flux, respectively. The complete transformations are given in Eqs.~\eqref{eq:VD_Tplus} and \eqref{eq:SL_Tplus}, while the simplified forms are provided in Eqs.~\eqref{eq:SL_Tplus_modeled_1}, \eqref{eq:SL_Tplus_modeled_2}, and \eqref{eq:SL_Tplus_modeled_3}. Corresponding transformations for compressible turbulent pipe flow are given in Appendix.~\ref{sec:app_A}.

To evaluate the performance, both classical isothermal and mixed isothermal/adiabatic wall configurations are considered. The SL-type transformation yields better data collapse than the VD-type in the viscous sublayer and buffer layer, consistent with previous studies on temperature and velocity transformations \citep{Patel2017,Chen2022,Huang2023,Cheng2024b, Patel2016,Trettel2016,Griffin2021}. For the isothermal wall, the linear model \(l_m^L\) reproduces the log-law only at high Reynolds numbers. The SL-type \(T^+_{SL}\) based on \(l_m^P\) and \(l_m^E\) shows good agreement with the analytical incompressible \(T^+_{IC}\) and \(T^+_{LoW}\), respectively. The relative errors are generally higher in the buffer layer than in the viscous sublayer and outer layer. At high reynolds numbers for the isothermal wall, most of the outer layer has a local relative error of \(\varepsilon < 2\%\) for the classical configuration and \(\varepsilon < 3.5\%\) for the mixed thermal configuration. The integral mean error over the entire boundary layer is generally below \(2\%\) for most tested cases, with r.m.s. values of \(1.63\%\) and \(1.68\%\) for \(\bar\varepsilon_P\) and \(\bar\varepsilon_E\), respectively. Regarding the log-law, we obtain \(B_T \approx 3.65 \) with \(\kappa = 0.41\) and \(Pr_t = 0.85\) under the SL-type transformation at \(Re^*_\tau \approx 1000\). As \(Re^*_\tau\) increases, \(B_T\) is expected to decrease slightly. For the adiabatic wall, both \(T^+_{VD}\) and \(T^+_{SL}\) collapse well. However, their magnitudes are systematically higher than the reference profiles. Future improvements are needed to address this limitation.

The present study demonstrates the feasibility of transforming compressible temperature onto their incompressible counterparts or onto the extended LoW, which is particularly significant for near-wall modeling and inverse transformations. It also highlights the damping effects of \(\psi_1\), \(\psi_2\), and \(\psi_3\) in the proposed temperature transformations. For the mixed isothermal/adiabatic wall configuration, \(\psi_2 = 1\) and \(\psi_3 = 0\) are valid approximations. In contrast, for the classical isothermal wall configuration, accounting for the turbulent TKE flux in \(\psi_3\) is essential. The present study identifies the multi-layer structure of \(q^t_k/q_w\) for the classic isothermal wall configuration. The proposed models \(\beta_1\), \(\beta_2\), and \(\beta_3\) avoid direct reliance on the high-order term in \(\psi_3\) while still retaining its effect. Based on these models, simplified transformations are obtained, which demonstrate overall good performance.

The proposed transformation may be extended to more general configurations. Such an extension would require more advanced models for \(\psi_1\), \(\psi_2\), and \(\psi_3\). We hope this study contributes to a deeper understanding of compressible temperature transformations and motivates further research on near-wall modeling of compressible turbulent flows.

\begin{acknowledgments}
The first author gratefully acknowledges financial support from the China Scholarship Council (No. 202006320042). He also acknowledges insightful discussions with Dr. Deniz A. Bezgin and Dr. Aaron B. Buhendwa regarding simulations of turbulent channel flow using JAX-Fluids. Special thanks are extended to Gary N. Coleman (NASA Langley Research Center) for his critical suggestions on the implementation of the DNS data.
\end{acknowledgments}

\section*{AUTHOR DECLARATIONS}
\subsection*{Conflict of Interest}
The authors have no conflicts to disclose.

\subsection*{Author Contributions}
\textbf{Youjie Xu}: Conceptualization (lead); Data curation (lead); Formal analysis (lead); Investigation (lead); Methodology (lead); Software (lead); Validation (lead); Visualization (lead); Writing-original draft (lead). \textbf{Steffen J. Schmidt}: Funding acquisition (equal); Project administration (equal); Resources (equal); Supervision (equal); Writing-review \& editing (equal). \textbf{Nikolaus A. Adams}: Funding acquisition (equal); Project administration (equal); Resources (equal); Supervision (equal); Writing-review \& editing (equal).

\section*{Data Availability Statement}
The DNS data that used in this study are available in the cited literature. The wall-resolved large eddy simulation data are available from the corresponding author upon reasonable request.

\section*{Author ORCID.}
Youjie Xu \href{https://orcid.org/0009-0006-8445-3200}{https://orcid.org/0009-0006-8445-3200};

Steffen J. Schmidt \href{https://orcid.org/0000-0001-6661-4505}{https://orcid.org/0000-0001-6661-4505};

Nikolaus A. Adams \href{https://orcid.org/0000-0001-5048-8639}{https://orcid.org/0000-0001-5048-8639}.

\appendix
\section{Temperature transformation for compressible turbulent pipe flow}\label{sec:app_A}
Compared to the compressible turbulent channel flow, a slight difference appears in the energy balance equation due to the circular geometry of pipe flow. We define \(\eta = y/\delta\), where \(y\) is the distance to the wall, and \(\delta\) represents the pipe radius. The momentum and energy balance equations between the wall and position \(\eta\) can be expressed as:
\begin{equation}\label{eq:pipe_global_momentumn_balance}
    \bar\mu \frac{d\tilde u }{dy} - 
    \bar\rho \widetilde{u^{\prime\prime} v^{\prime\prime}} = \tau_w (1 - \eta),
\end{equation}
\begin{equation}\label{eq:pipe_global_energy_balance_2}
    \bar\lambda \frac{d\tilde T}{dy}
    -\overline{\rho c_p v^{\prime\prime} T^{\prime\prime}}
    + \bar\mu \frac{d\tilde u}{dy} \tilde u
    -\overline{\rho v^{\prime\prime} u^{\prime\prime}}\tilde u
    -\overline{\rho v^{\prime\prime} \frac{1}{2} u^{\prime\prime}_i u^{\prime\prime}_i}
    + \frac{2\tau_w}{1 - \eta} \int_0^{\eta} (1 - \eta^\prime) \, \tilde u \, d{\eta^\prime}
    = \frac{q_w}{1 - \eta}.
\end{equation}

Following the approach in Sec.~\ref{sec:temperature_transformation}, we obtain the temperature transformation:
\begin{equation}\label{eq:pipe_VD_Tplus}
  T^+_{VD} = \int_0^{\theta^+} 
  \frac{\psi_1}{B_q/(1 - \eta) + \psi_2 {(\gamma - 1)M_\tau^2 u^+} + \psi_3}
  \sqrt{\rho^+}
  {d\theta^+}, 
\end{equation}
\begin{equation}\label{eq:pipe_SL_Tplus}
  T^+_{SL} = \int_0^{\theta^+} 
  \frac{\psi_1}{B_q/(1 - \eta) + \psi_2 {(\gamma - 1)M_\tau^2 u^+} + \psi_3}
  \sqrt{\rho^+}
  \left(1 + \frac{1}{2}\frac{y^+}{\rho^+}\frac{d\rho^+}{dy^+} - \frac{y^+}{\mu^+}\frac{d\mu^+}{dy^+}\right)
  {d\theta^+}.
\end{equation}

\begin{equation}\label{eq:pipe_params_f1f2f3}
    \psi_1 = \frac{l_m\sqrt{\tau^+_{tot}}}{\kappa y}, \quad
    \psi_2 = \tau^+_{tot} + \frac{\tilde u^i_b}{\tilde u} \eta, \quad
    \psi_3 = \frac {-\overline{\rho v^{\prime\prime} u^{\prime\prime}_i u^{\prime\prime}_i/2}}{\bar\rho_w c_p u_\tau \tilde T_w},
\end{equation}
\begin{equation}\label{eq:pipe_Ui_b_ratio}
    \frac{\tilde u^i_b}{\tilde u} = \frac {2}{(1 - \eta) \eta \tilde u} \int_0^{\eta} (1 - \eta^\prime) \tilde u \, d{\eta^\prime}.
\end{equation}

Here, \(\lambda\), \(B_q\), \(M_\tau\), and \(\theta^+\) are the same as those in Sec.~\ref{sec:temperature_transformation}. The total shear stress also similar to that in channel flow, \(\tau^+_{tot} = 1 - \eta\). Following Appendix~\ref{sec:app_B}, \({\tilde u^i_b}/{\tilde u} = 1\) at the wall for pipe flow.

Compared to Eqs.~\eqref{eq:VD_Tplus} and ~\eqref{eq:SL_Tplus} for turbulent channel flow, the factors \((1 - \eta)\) and \(2/(1 - \eta)\) are introduced into \(B_q\) and \({\tilde u^i_b}/{\tilde u}\), respectively, to account for the circular geometry.
Similarly, using the models \(\beta_1\), \(\beta_2\), and \(\beta_3\) given in Sec.~\ref{sec:modeling_turbulent_tke_flux}, we obtain the simplified SL-type transformations:
\begin{equation}\label{eq:SL_Tplus_pipe_modeled_1}
  T^+_{SL,p,\beta_1} = \int_0^{\theta^+} 
  \frac{\psi_1}{[1/(1 - \eta) - \beta_1] B_q + \psi_2 {(\gamma - 1)M_\tau^2 u^+}}
  \sqrt{\rho^+}
  \left(1 + \frac{1}{2}\frac{y^+}{\rho^+}\frac{d\rho^+}{dy^+} - \frac{y^+}{\mu^+}\frac{d\mu^+}{dy^+}\right)
  {d\theta^+}.
\end{equation}
\begin{equation}\label{eq:SL_Tplus_pipe_modeled_2}
  T^+_{SL,p,\beta_2} = \int_0^{\theta^+} 
  \frac{\psi_1}{[1/(1 - \eta) - \beta_2] B_q + \psi_2 {(\gamma - 1)M_\tau^2 u^+}}
  \sqrt{\rho^+}
  \left(1 + \frac{1}{2}\frac{y^+}{\rho^+}\frac{d\rho^+}{dy^+} - \frac{y^+}{\mu^+}\frac{d\mu^+}{dy^+}\right)
  {d\theta^+}.
\end{equation}
\begin{equation}\label{eq:SL_Tplus_pipe_modeled_3}
  T^+_{SL,p,\beta_3} = \int_0^{\theta^+} 
  \frac{\psi_1 (1 + \beta_3)}{B_q/(1 - \eta) + \psi_2 {(\gamma - 1)M_\tau^2 u^+}}
  \sqrt{\rho^+}
  \left(1 + \frac{1}{2}\frac{y^+}{\rho^+}\frac{d\rho^+}{dy^+} - \frac{y^+}{\mu^+}\frac{d\mu^+}{dy^+}\right)
  {d\theta^+}.
\end{equation}

\begin{figure*}
  \includegraphics[width=\linewidth]{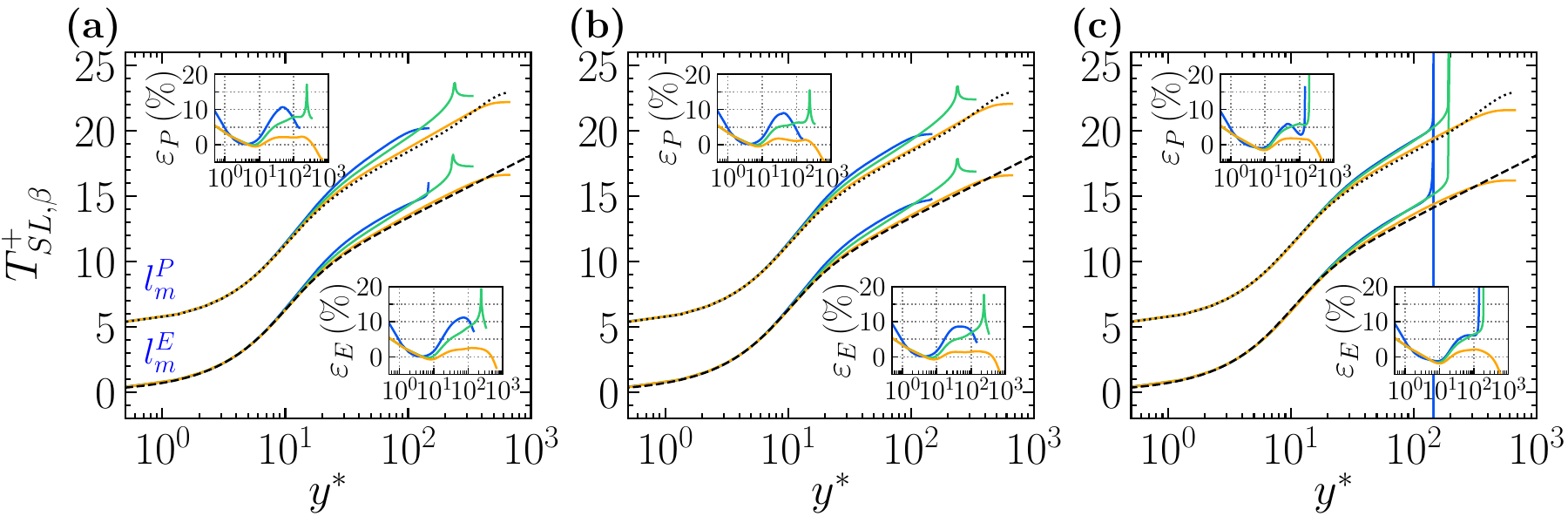}
  \caption{Transformed temperature profiles for compressible pipe flows obtained using the simplified SL-type transformations: (a) \(T^+_{SL,p,\beta_1}\), (b) \(T^+_{SL, p,\beta_2}\), and (c) \(T^+_{SL,p,\beta_3}\). DNS data from \citet{Modesti2019} are employed. For clarity, the profiles using \(l_m^P\) are shifted upward by 5 units. Black dotted lines: the analytical incompressible profile \(T^+_{IC}\) corresponding to one representative case. Black dashed lines: \(T^+_{LoW}\) from Eq.~\eqref{eq:IC_exLoW}.}
\label{fig:Tplus_beta_pipe}
\end{figure*}

Fig.~\ref{fig:Tplus_beta_pipe} presents the transformed temperature profiles, where the DNS data of compressible turbulent pipe flow from \citet{Modesti2019} are employed. The considered three cases are \(M_b = 1.5, Re^*_\tau = 334\); \(M_b = 1.5, Re^*_\tau = 667\), and \(M_b = 3.0, Re^*_\tau = 147\). As shown, the transformed temperature show reasonable agreement with the reference profiles \(T^+_{IC}\) and \(T^+_{LoW}\). Note that the \(T^+_{IC}\) is obtained using different parameter values for pipe flow \citep{Pirozzoli2024}. For the high-Reynolds-number case (\(M_b = 1.5, Re^*_\tau = 667\)), the local relative error is below \(2.5\%\) for most of the boundary layer. The wake region is underpredicted, which may be caused by the employed models for \(l_m\) and the turbulent TKE flux, as these models are obtained based on channel configurations. The studies of \citet{She2017} and \citet{Pirozzoli2024} indicate that channel and pipe flows are often characterized by different parameter values.

\section{Value of \(\tilde u^i_b / \tilde u\) at the wall}\label{sec:app_B}
When implementing the transformations in Eqs.~\eqref{eq:VD_Tplus} and \eqref{eq:SL_Tplus}, we need to compute \( {\tilde u^i_b}/{\tilde u} \) in \(\psi_2\). Its value at the wall is derived below.

Recall the definition in Eq.~\ref{eq:ub_i}, we obtain:
\begin{equation}\label{eq:u_limit}
  \lim_{y \to 0} \frac{{\tilde u_b^i}(y)}{\tilde u(y)} 
  = \lim_{y \to 0} \frac{1}{y \tilde{u}(y)} \int_0^y \tilde{u}(\eta) \, d\eta.
\end{equation}

In the viscous sublayer, the velocity follows a linear profile, i.e., \(\tilde u = C \, y \), where C is a constant. Thus, we get:
\begin{equation}\label{eq:u_intg}
  \int_0^y \tilde{u}(\eta) \, d\eta = \int_0^y C \eta \, d\eta = C \frac{y^2}{2}.
\end{equation}

Finally, we obtain \( {\tilde u^i_b}/{\tilde u} \) at the wall:
\begin{equation}\label{eq:u_limit_result}
  \left. \frac{{\tilde u_b^i}}{\tilde u} \right|_w
  = \lim_{y \to 0} \frac{{\tilde u_b^i}(y)}{\tilde u(y)}
  = \lim_{y \to 0} \frac{C {y^2}/{2}}{C y^2} = \frac{1}{2}.
\end{equation}

Similarly, for density-based body force, we use the following equation to approximate the value at the wall:
\begin{equation}\label{eq:u_limit_result_rho}
  \left. \frac{{\tilde u_b^{i,\rho}}}{\tilde u} \right|_w
  = \lim_{y \to 0} \frac{{\tilde u_b^i}(y)}{\tilde u(y)}
  = \frac{1}{2} \frac{\bar \rho_w}{\rho_b}.
\end{equation}

\begin{figure*}
  \includegraphics[width=\linewidth]{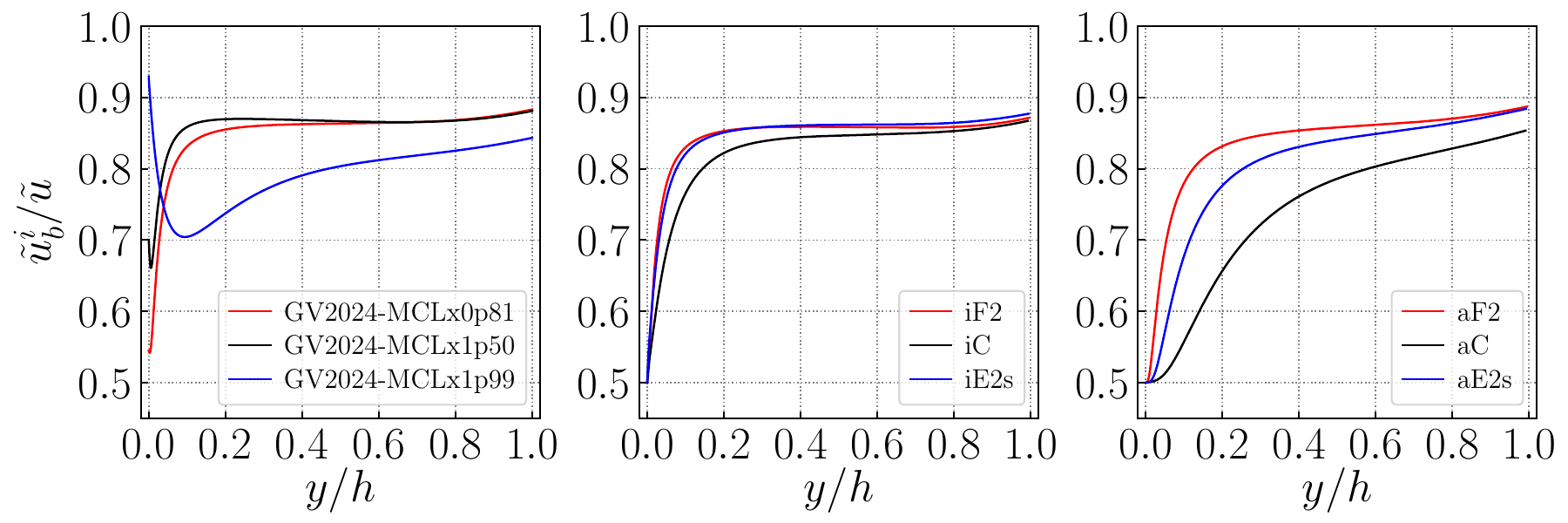}
  \caption{Profiles of \(\tilde u^i_b/\tilde u\) under different body force configurations. (a) Density-based body force, computed using Eq.~\eqref{eq:ub_i_new} and DNS data from \citet{Gerolymos2023, Gerolymos2024a, Gerolymos2024b}.
  (b) Volume-based body force, computed using Eq.~\eqref{eq:ub_i} and DNS data from \citet{Lusher2022}, isothermal wall side. (c) Volume-based body force, computed using Eq.~\eqref{eq:ub_i} and DNS data from \citet{Lusher2022}, adiabatic wall side.}
\label{fig:ub_i}
\end{figure*}

Representative distributions of \(\tilde u^i_b/\tilde u\) under different body force and thermal configurations are presented in Fig.~\ref{fig:ub_i}. It can be observed that \(\tilde u^i_b/\tilde u = 0.5\) at \(y = 0\) for the volume-based body force. In contrast, for the density-based body force, the value at the wall increases with increasing Mach numbers. In the vicinity of channel center, the values are similar across the three configurations. Above the adiabatic wall, the ratio increases more slowly than above the isothermal wall.

\section{Temperature transformations in the literature}\label{sec:app_C}
The VD-type and SL-type transformations by \citet{Chen2022}, originally presented in Eq.~(5.1) and (5.2) of their paper, are rewritten here for convenience.
\begin{equation}\label{eq:VD_Tplus_Chen2022}
    T^{+,Chen}_{VD} = \int_0^{\theta^{Chen}}  \frac {d\theta^{Chen}}{\theta^*_{\tau,c}}.
\end{equation}
\begin{equation}\label{eq:SL_Tplus_Chen2022}
    T^{+,Chen}_{SL} = \int_0^{\theta^{Chen}} 
    \left( 1 + \frac{y}{Re^*_{\tau,Chen}} \frac{dRe^*_{\tau,Chen}}{dy} \right)
    \frac {d\theta^{Chen}}{\theta^*_{\tau,c}}.
\end{equation}

Here, \( \theta^{Chen} = \tilde T_w - \tilde T \), \( Re^*_{\tau,Chen} = {\bar\rho \sqrt{\tau_w/\bar \rho} h}/{\bar\mu} \), \( \theta^*_{\tau,c} = {(q_w^{Chen} + \bar q^{Chen})}/{(\bar \rho c_p u^*_\tau)}\), \( u^*_\tau = \sqrt{\tau_w/\bar\rho} \), 
\( \bar q^{Chen} = \overline{\sigma_{i2} u_i} 
+ \overline{\sigma^\prime_{i2} u^\prime_i}
- \overline{\rho v^{\prime\prime}u^{\prime\prime}_i}\tilde u_i
- \overline{\rho v^{\prime\prime} \frac{1}{2}{u^{\prime\prime}_i u^{\prime\prime}_i}} \), and \( q_w^{Chen} \) represents the wall heat flux in the same direction of wall-normal coordinate. 
\\ \\

The VD-type and SL-type transformations by \citet{Huang2023}, originally presented in Eq.~(4.6) and (5.6) of their paper, are rewritten here for convenience.
\begin{equation}\label{eq:VD_Tplus_Huang2023}
    {T}^{+,Huang}_{VD} = \int_0^{\theta^{Huang}} 
    \frac{1}{B_q + (\gamma - 1)M_\tau^2u^+} \sqrt {\frac{\bar\rho} {\bar\rho_w}} \, d{\theta^{Huang}}.
\end{equation}
\begin{equation}\label{eq:SL_Tplus_Huang2023}
    T^{+,Huang}_{SL} = \int_0^{\theta^{Huang}} 
    \frac {1} {B_q + ({\gamma}-1) M_\tau^2 u^+}
    \sqrt {\frac{\bar \rho} {\bar\rho_w}}
    \left( 1 + \frac {1}{2} \frac {y^+}{\bar \rho} \frac{\partial \bar \rho }{\partial y^+} - \frac{y^+}{\bar \mu } \frac{\partial \bar \mu}{\partial y^+} \right) \, 
    d{\theta^{Huang}}.
\end{equation}

Here, \( \theta^{Huang} = (T_w - T)/T_w \). The definition of \( B_q \) and \( M_\tau \) are the same as those in Sec.~\ref{sec:compressible_law_of_the_wall}. 
\\ \\

\citet{Cheng2024b} originally proposed two types of semi-local transformations depending on the high-order term is neglected or not, see Eq.~(22) and (30) of their paper. For comparison, we select the complete form and rewrite it in the following equation:
\begin{equation}\label{eq:SL_Tplus_Cheng2024}
  T^{+,Cheng}_{SL} = \int_0^{\theta^{+,Cheng}}
  \frac{\sqrt{\bar \rho^+}}{1 - \frac{q^{Cheng}}{q_w}}
  d\theta^{+,Cheng}.
\end{equation}

Here, \( \theta^{+,Cheng} = (T - \bar T_w)/\theta_\tau \) with \( \theta_\tau = q_w/(\bar \rho_w c_p u_\tau) \), \( \bar \rho^+ = {\bar\rho}/{\bar \rho_w} \), and 
\( \bar q^{Cheng} = \overline{u \tau_{xy}} 
- \overline{\rho \tilde u u^{\prime\prime} v^{\prime\prime}}
- \overline{\rho v^{\prime\prime} \frac{1}{2}{u^{\prime\prime} u^{\prime\prime}}} \). Note that \( q_w^{Cheng} \) represents the heat flux removed from the wall.

\section{Analytical incompressible temperature profile}\label{sec:app_D}
\citet{Pirozzoli2024} proposed an analytical expression for the temperature difference \(\varTheta = T - T_w\) in incompressible turbulent channel and pipe flows. It consists of two parts: the inner layer and wake region distributions. In the present study, this analytical profile is denoted as \(T^+_{IC}\). 

The inner part is expressed as:
\begin{equation}\label{eq:Theta_plus}
  \begin{aligned}
  \varTheta_i^+(y^+, Pr) &= \frac{1}{2 k_\theta \zeta_0 (2 + 3 Pr \zeta_0)} 
  \Bigg\{ 
  \frac{2 \big( 2 \zeta_0 + 3 Pr^2 C_\theta^2 \zeta_0 + Pr (C_\theta^2 + 2 \zeta_0^2) \big)}{\Delta} \\ 
  &\quad \times \Big[ 
  \arctan \Big( \frac{1 + Pr \zeta_0}{\Delta} \Big) - \arctan \Big( \frac{1 + Pr (2 \zeta + \zeta_0)}{\Delta} \Big)
  \Big] \\
  &\quad + 2 Pr (C^2 + \zeta_0^2) \log \Big( 1 - \frac{\zeta}{\zeta_0} \Big) + \\
  &\quad \left( Pr (2 \zeta_0^2 - C_\theta^2) + 2 \zeta_0 \right) 
  \log \frac{Pr \zeta^2 + (1 + Pr \zeta_0)(\zeta + \zeta_0)}{\zeta_0 (1 + Pr \zeta_0)} 
  \Bigg\}
  \end{aligned},
\end{equation}

where \(\zeta = k_\theta y^+\), \(\Delta = \sqrt{3 Pr^2 \zeta_0^2 + 2 Pr \zeta_0 - 1}\). The parameter \(\zeta_0\) is given by:
\begin{equation}\label{eq:Theta_plus_1}
    \begin{gathered}
        \zeta_0 = \frac{1}{3 Pr} \Big(-1 + \frac{1}{z} + z \Big), \,
        z = \left[ \frac{1}{2} \Big( -2 - 27 Pr^2 C_\theta^2 + \sqrt{-4 + (2 + 27 Pr^2 C_\theta^2)^2} \Big) \right]^{1/3}.
    \end{gathered}
\end{equation}

The wake region distribution is given by:
\begin{equation}\label{eq:Theta_plus_wake}
  \begin{gathered}
      \varTheta^+_e - \varTheta^+_c = C_w (1 - \eta)^2, \\
      \varTheta^+_e = \varTheta^+_i (\eta^* \delta_t^+, Pr) + C_w (1 - \eta^*)^2.
  \end{gathered}
\end{equation}

For turbulent channel flow with uniform internal heat source, \(C_\theta = 10\), \(C_w = 5.48\), \(\eta^* = 0.274\), and \(\delta_t\) is the half-channel height. For turbulent pipe flow with uniform internal heat source, \(C_\theta = 10\), \(C_w = 6.00\), \(\eta^* = 0.238\), and \(\delta_t\) is the pipe radius.

\bibliography{reference}

\end{document}